\documentclass[preprint,11pt]{elsarticle}

\usepackage{amssymb}
\usepackage{amsfonts}
\usepackage{graphicx}
\usepackage{amsmath}
\usepackage{latexsym}
\usepackage{amssymb}
\usepackage{graphics}
\usepackage{mathrsfs}
\usepackage{fancyhdr}
\usepackage{subfigmat}
\usepackage{multirow}

\usepackage{longtable}
\usepackage{hyperref}
\usepackage{natbib}

\usepackage[total={16.5cm,24cm},           
            top=2cm,                        
            left=2.5cm,                     
            ]{geometry}                     

\newcommand{\bftheta}{\mbox{\boldmath{$\theta$}}}

\newcommand{\bfP}{\mbox{\boldmath{$P$}}}
\newcommand{\bfx}{\mbox{\boldmath{$x$}}}
\newcommand{\bff}{\mbox{\boldmath{$f$}}}
\newcommand{\bfm}{\mbox{\boldmath{$m$}}}
\newcommand{\bfv}{\mbox{\boldmath{$v$}}}

\newcommand{\bfJ}{\mbox{\boldmath{$J$}}}
\newcommand{\bfq}{\mbox{\boldmath{$q$}}}

\newcommand{\M}[1]{\boldsymbol{\mathsf{#1}}}
\newcommand{\de}{\,{\mathrm d}}
\newcommand{\trn}{^\mathsf{T}}

\usepackage{xcolor}

\journal{International Journal of Heat and Mass Transfer}

\begin{document}

\begin{frontmatter}

\title{Hygro-thermo-mechanical analysis of spalling in concrete walls at high temperatures
as a moving boundary problem}

\author[k101]{Michal Bene\v{s}}

\author[k133]{Radek \v{S}tefan\corref{cor1}}
\cortext[cor1]{Corresponding author. Tel.: +420 22435 4633; fax: +420 23333 5797}
\ead{radek.stefan@fsv.cvut.cz}

\address[k101]{Department of Mathematics,}
\address[k133]{Department of Concrete and Masonry Structures,\\Faculty of Civil Engineering,\\
Czech Technical University in Prague,\\
Th\'{a}kurova 7, 166 29 Prague 6, Czech Republic}

\begin{abstract}
A mathematical model allowing coupled hygro-thermo-mechanical analysis of spalling in concrete walls at high temperatures by means of the moving boundary problem is presented. A simplified mechanical approach to account for effects of thermal stresses and pore pressure build-up on spalling is incorporated into the model. The numerical algorithm based on finite element discretization in space and the semi-implicit method for discretization in time is presented. The validity of the developed model is carefully examined by a comparison between the experimentally determined data stated in literature and the results obtained from the numerical simulation.
\end{abstract}


\end{frontmatter}


\section{Introduction}
\label{Intro}

{
Mathematical modelling seems to be an effective and powerful tool to simulate the heated concrete behaviour (see e.g. \citep{Gawin2012}).}
Several models based on more or less general physical background have been developed to simulate transport processes in heated concrete (see \citep{Gawin2011b} and references therein). All developed models build on a system of conservation of mass and energy, but differ in the complexity of phase description of state of pore water as well as chemical reactions and different physical mechanisms of coupled transport processes in a pore system. A descriptive phenomenological approach was used to explore hygro-thermal processes in concrete exposed to temperatures exceeding $100~^{\circ}$C starting with the Ba\v{z}ant \& Thonguthai model \citep{BaTh1978}. Here, the liquid water and water vapour are treated as a single phase, moisture, and the evaporable water is assumed to be formed by the capillary water only. The main advantages for usage of this approach is relative simplicity and small number of parameters that can be obtained from experiments.
However, in such a way it is impossible to consider the effects of phase changes of water and the applicability of the single-fluid-phase models for temperatures above the critical point of water is disputed. These deficiencies led to the development of more detailed multi-phase description, see e.g. the works of \citet{Gawin1999} and \citet{Davie2010} for specific examples. Coupled multi-phase models reflect the multi-phase structure of concrete, interactions between phases, phase changes of fluids and solids and non-linear couplings between thermal, hygral and mechanical processes. However, such increase in complexity comes at the expense of a large number of model parameters, which determination can be hardly obtained directly from experiments. Moreover, multi-phase models are computationally expansive. Despite rapid progress in computer technologies, complex models still exceed capabilities of recently developed numerical algorithms and computational hardware. Therefore, in this paper we will adopt a pragmatic concept and consider the simplified model obtained directly from the complex multi-phase description. Relative importance of thermodynamic fluxes will be quantitatively evaluated at the level of material point and these results will allow us to neglect less important transport phenomena without lost of capability to realistically predict behavior of concrete at extremely high temperatures.

High-temperature exposure of concrete can lead to the risk of concrete spalling and, consequently, to the damage of the entire structure (see e.g. \citep{Gawin2006} and references therein, for the examples see \citep{Ju1998,Sch2002,Shekarchi2003}).
It is generally accepted that the spalling process in rapidly heated concrete is caused by two main processes -- increase of pore pressure end development of thermal stresses -- that may act separately or, which is more likely, in a combined way (see e.g. \citep[Section~4]{KhouryAnderberg2000}).

In literature, we can find many criteria to asses the risk and, in some cases, also the amount of concrete spalling (for a brief summary of some of these criteria, see e.g. \citep{Gawin2006,KhouryAnderberg2000}). In our previous work \citep{Benes2013}, we have employed a heuristic engineering approach, originally proposed by \citet[Section~3.3]{DwaiKod2009}, in which the spalling is supposed to occur if the effective pore pressure exceeds the temperature dependent tensile strength of concrete. In the present paper, we extend this criterion in order to take into account not only the pore pressure (which seems to be not the dominant mechanism of spalling, as observed by recent experimental investigations \citep{Jansson2010,Mindeguia2013} and numerical simulations \citep{Msaad2007}) but also the thermal stress as a driving force of spalling.

The paper is organized as follows. In Section~\ref{sec:basic_cons_eqs}, we specify general thermodynamical and mechanical assumptions on concrete as a porous multi-phase medium to obtain a reasonably simple but still realistic model to predict hygro-thermal behavior of concrete at very high temperatures. This Section is concluded by presentation of conservation of mass and thermal energy, carefully derived in \ref{parameter_analysis} by quantitative parameter analysis. In Sections~\ref{sec:const_rel} and \ref{BoundaryConditions}, constitutive relationships are discussed in details and suitable boundary conditions for description of transport processes through the surfaces of the concrete wall are presented, respectively. Section~\ref{sec:spalling} deals with the problem of moving boundary for spalling simulation. Our approach is based on the combined effect of pore pressure and thermal stresses in concrete under high temperature exposure. In Section~\ref{sec:complete_form}, the complex problem is formulated as a fully coupled system of highly nonlinear partial differential equations (PDE's) supplemented with appropriate boundary and initial conditions. Based on the full FEM (in space) and semi-implicit (in time) discretization of the mentioned system of PDE's, an in-house research MATLAB code has been developed for the solution of the system of nonlinear algebraic equations. The numerical algorithm is presented in Section~\ref{sec:full_discr}. Section~\ref{Sec_MaterialProperties} brings the complete list of material properties of moist concrete at high temperatures used in the numerical model. Section~\ref{Sec_Examples} is the key part of the paper. Here we present validation of the model by comparison of the numerical results with experiments reported in literature by means of three examples.
First two examples, performed by \citet{Kalifa2000} and, more recently, \citet{Mindeguia2009}, examine the comparison of the measurements of pore pressure and temperature distributions with the predicted numerical simulations based on the present model under specific conditions excluding spalling phenomena. Finally, we apply the present model to investigate the surface spalling of high strength concrete prismatic specimen under unidirectional heating by the ISO~834 fire curve and compare the numerical results with experimental observations reported by \citet{Mindeguia2009} and \citet{Mindeguia2009a,Mindeguia2013}. The summary of the outcomes achieved in this paper as well as the general conclusions and recommendations for future research appear in Section~\ref{Sec_Conclusions}.

\section{Basic conservation equations}
\label{sec:basic_cons_eqs}

\subsection{General assumptions}
A number of assumptions have been adopted to develop the coupled hygro-thermo-mechanical model for concrete spalling due to high temperatures exposure. Some of them of particular interest are as follows:

\begin{itemize}
\item concrete is assumed as a multi-phase system consisting of different phases and components: solid skeleton (composed of various chemical compounds and chemically bound water), liquid phase (combined capillary and adsorbed water), gas phase (a mixture of dry air and water vapour) (see e.g. \citep{Gawin1999});

\item the diffusive mass flux of water vapour is neglected and the adsorbed water diffusion is assumed to be expressed by the liquid water relative permeability term, $K_{rw}$, instead of a separate term (see \citep{ChungCon2005,Davie2010});

\item the effects of variations of pressure of dry air is neglected. The mass of dry air in concrete is considered to be much smaller than the mass of liquid water and vapour and, consequently, the vapour pressure plays the crucial role (when compared to dry air) in spalling phenomena (see \citep{DwaiKod2009});

\item above the critical point of water, only the vapour contribution to the mass transport is taken into account (cf. \citep{Gawin2002});

\item the spalling of concrete in a heated wall is assumed to be caused by a combination of the hygro-thermal stress due to the pore pressure build-up and the thermo-mechanical stress resulting from the restrained thermal dilatation (see e.g. \citep{KhouryAnderberg2000});

\item the wall is considered as fully mechanically restrained in the plane perpendicular to the wall thickness (see \citep{Msaad2007,Zeiml2008});

\item the stresses resulting from the external mechanical load as well as the self-weight of the wall are not accounted for since their effect on the potential spalling may be considered to be negligible compared with the effect of mechanical restraint (see \citep{Msaad2007,Zeiml2008});

\item the hygro-thermal and the mechanical problems are coupled in the analysis.
\end{itemize}

\subsection{Conservation laws}
The mathematical model of moisture and heat transfer in concrete consists of the balance equations governing the conservation of mass (moisture) and thermal energy (cf. equations \eqref{balance_moisture} and \eqref{app:balance_energy}).

\emph{The mass balance equation of moisture (liquid water and vapour):}
\begin{equation}\label{moisture}
\frac{\partial }{\partial t}(\eta_{w}\rho_{w}+\eta_v\rho_v)
+
\frac{\partial }{\partial x} \left( \eta_w \rho_w v_{w} + \eta_v \rho_v v_v \right)
=
\frac{\partial m_{d}}{\partial t};
\end{equation}

\emph{energy conservation equation for moist concrete as the multi-phase system:}
\begin{equation}\label{energy temp}
(\rho c_p)
\frac{\partial\theta}{\partial t}
=
-
\frac{\partial q_c}{\partial x}
-
(\rho c_p  v)
\frac{\partial \theta}{\partial x}
-
\frac{\partial m_{e}}{\partial t} h_{e}
-
\frac{\partial m_{d}}{\partial t} h_{d},
\end{equation}
where
\begin{eqnarray}
(\rho c_p)
&=&
c_p^{w}\, \rho_{w}\,\eta_{w}
+
c_p^g \rho_g \, \eta_{g}
+
c_p^{s}\, \rho_{s}\,\eta_{s},
\label{HeatCapacity}
\\
(\rho c_p  v)
&=&
c^w_p\rho_w\eta_w  v_{w}
+
c_p^g \rho_g \eta_g  v_g
\end{eqnarray}
and $c_p^g \rho_g = c_p^{v}\rho_{v}
+ c_p^{a}\rho_{a}$.
Governing equations \eqref{moisture} and \eqref{energy temp} are carefully derived in \ref{parameter_analysis} and the meanings of all symbols are explained in a well arranged way in \ref{nomenclature}. It should be underlined that material coefficients of concrete and fluids that appear in governing equations \eqref{moisture} and \eqref{energy temp} depend in non-linear manner upon the primary unknowns -- temperature and pore pressure, which completely describe the state of concrete under thermal loading.

\section{Constitutive relationships}
\label{sec:const_rel}

Balance equations \eqref{moisture} and \eqref{energy temp} are supplemented by an appropriate set of constitutive equations.

\paragraph{Moisture flux}

As the constitutive equations for fluxes of fluid phases (liquid water and vapour) the \emph{Darcy's law} is applied 
\\
\begin{eqnarray}
\eta_w\rho_{w} v_{w}  &=& - \rho_{w}
\frac{K K_{rw}}{\mu_{w}} \frac{\partial P_w}{\partial x} ,
\quad  P_w = P - P_c,
\label{flux_w}
\\
\eta_v\rho_{v}  v_v &=& - \rho_{v}
\frac{K K_{rg}}{\mu_g} \frac{\partial P}{\partial x} ,
\label{flux_v}
\end{eqnarray}
where $P_w$ [Pa] is the pressure of liquid water, $P_c$ [Pa] is the capillary pressure and $P$ [Pa] represents the pore pressure due to the water vapour. Further, $K~[\rm{m^2}]$ represents the intrinsic permeability, $K_{rw}~[-]$ and $K_{rg}~[-]$ are the relative permeability of liquid water and relative permeability of gas, $\mu_{w}$ [Pa s] and $\mu_g$ [Pa s] represent the liquid water and gas dynamic viscosity.

\bigskip

The equilibrium state of the capillary water with the water vapour is expressed in the form of the \emph{Kelvin equation}
\begin{equation}\label{equilibrium meniscus}
P_c = - \rho_w  \frac{\theta R}{M_w} \ln \left(
\frac{P}{P_{s}} \right),
\end{equation}
where $P_c$ denotes the capillary pressure and the water vapour saturation pressure $P_{s}$ [Pa] can be calculated from the following formula as a function of temperature $\theta$ [K]
\begin{equation}\label{watervapoursaturationpressure}
P_{s}(\theta) =  \exp  \left( 23,5771
- \frac{4042,9}{\theta-37,58} \right).
\end{equation}

\bigskip

Water vapour is considered to behave as perfect gas, therefore, \emph{Clapeyron equation} \citep{Gawin1999}
\begin{equation}\label{clapeyron}
 \rho_{v} = \frac{P M_w}{\theta R}
\end{equation}
is assumed as the state equation, where  $R~\rm{[J\,mol^{-1}\,K^{-1}]}$ is the gas constant $(8.3145~\rm{J\,mol^{-1}\,K^{-1}})$ and $M_w~\rm{[kg\,mol^{-1}]}$ represents molar mass of water vapour.

\paragraph{Heat flux} For the heat flux induced by conduction the Fourier's law is applied in the form
\begin{equation}
q_c = - \lambda_c \frac{\partial  \theta}{\partial x},
\end{equation}
where $\lambda_c~\rm{[W\,m^{-1}\,K^{-1}]}$ represents the temperature and saturation dependent effective thermal conductivity of moist concrete.

\paragraph{Evaporation}
In order to determine the amount of heat due to evaporation or, reversely, condensation processes, the water vapour conservation equation
\begin{equation}\label{balance_vapour}
\frac{\partial (\eta_v\rho_v)}{\partial t}
+
\frac{\partial  \left( \eta_v \rho_v v_v \right)}{\partial x}
=
\frac{\partial m_{e}}{\partial t}
\end{equation}
needs to be incorporated into the energy conservation equation \eqref{energy temp} which will be handled in Section \ref{sec:mod_energy}.

\paragraph{Dehydration}

Following \citet{PontEhrlacher2004,Feraille-Fresnet:2003}, the evolution of mass source term  $m_d$ $[{\rm kg\, m^{-3}}]$ related to the dehydration process is considered through the following evolution law
\begin{equation}\label{dehydration}
\frac{\partial m_{d}}{\partial t} = -\frac{1}{\tau}(m_{d}-m_{d,eq}(\theta)),
\end{equation}
where $m_{d,eq}$ $[{\rm kg\, m^{-3}}]$ is the mass of water released at the equilibrium according to temperature $\theta$ and $\tau$ $[s]$ is the characteristic time of mass loss governing the asymptotic evolution of the dehydration process.

\section{Boundary conditions}
\label{BoundaryConditions}

To describe coupled transport processes through the surfaces of the wall, one should prescribe the appropriate boundary conditions across the boundary. Homogeneous Neumann conditions
\begin{eqnarray}
h_e \rho_v\eta_v  v_v  - \lambda_c \frac{\partial \theta}{\partial x}  &=& 0,
\label{neumann a}
\\
\rho_w\eta_w v_{w\overline{}} + \rho_v\eta_v v_v &=& 0
\label{neumann b}
\end{eqnarray}
are usually applied on the insulated surface of the wall. Boundary conditions of the form
\begin{eqnarray}
\left(
h_e \rho_v\eta_v v_v  - \lambda_c \frac{\partial \theta}{\partial x}
\right) n_x
 &=&
 \alpha_c(\theta-\theta_{\infty})
 +
 e\sigma_{SB} (\theta^4-\theta^4_{\infty}),
 \label{radiative a}
\\
(\rho_w\eta_w v_w + \rho_v\eta_v v_v ) n_x
&=&
\beta_c (\rho_{v} - \rho_{v \infty})
\label{radiative b}
\end{eqnarray}
($n_x=\pm 1$)
are of importance on the exposed side of the wall, where the terms on the right hand side of \eqref{radiative a} represent the heat energy dissipated by convection and radiation to the surrounding medium and the term on the ride hand side of \eqref{radiative b} represents a water vapour dissipated into the surrounding medium.

\section{Moving boundary for spalling simulation in concrete walls}
\label{sec:spalling}

\subsection{Spalling criterion}
Let us assume a concrete wall of a thickness $\ell$ exposed to fire on boundary $x = \ell$. The spalling at position $x \in (0,\ell)$ and time $t$ occurs if (cf. \citep[p.~613]{Msaad2007}; \citep[Section~2.3]{Phan2012})
\begin{equation}\label{spalling_criterion}
F (f_c(\theta),f_t(\theta),\sigma_{ht}(P,\theta),\sigma_{tm}(\theta)) > 1,
\end{equation}
where $F$ is a dimensionless failure parameter (failure function), $f_c$ and $f_t$ are the temperature dependent uniaxial compressive and tensile strengths of concrete, respectively, $\sigma_{ht}$ and $\sigma_{tm}$ are the actual stresses in concrete caused by hygro-thermal and thermo-mechanical processes, respectively, and $P$ and $\theta$ are the pore pressure and temperature in concrete at position $x$ and time $t$, respectively.

Note that the strengths of concrete (both compressive and tensile) are assumed to be positive values while the strains and stresses are taken as positive in tension and negative in compression.

\subsection{Hygro-thermal stress}
Hygro-thermal stress in heated concrete is a tensile stress caused by the pore pressure build-up. There are several approaches to determine the hygro-thermal stress. These approaches may differ both in the definition of the pore pressure and also in the manner in which the pore pressure is converted into the hygro-thermal stress. The pore pressure may be assumed to be equal to the vapour pressure (e.g. \citep{DwaiKod2009}), to the gas pressure, or can be calculated as the Bishop's stress \citep{Davie2006,Gawin1999,Msaad2007,Phan2012}. The hygro-thermal stress (i.e. the effective pore pressure) can be determined from the pore pressure by the hollow spherical model \citep{Ichikawa2004}, or multiplying respectively by a Biot's coefficient \citep{Msaad2007,Phan2012} or by a concrete porosity \citep{DwaiKod2009}.

Here, we adopt the approach proposed by \citet{DwaiKod2009}, in which
\begin{equation}\label{hygro_thermal_stress}
\sigma_{ht}(P,\theta) = P \phi(\theta),
\end{equation}
where $P$ is the pore pressure due to water vapour and $\phi$ is the temperature dependent concrete porosity.

\subsection{Thermo-mechanical stress}\label{Section_Thermo_mechanical_stress}
The thermo-mechanical stress in a heated concrete wall generally depends on the external mechanical load applied on the wall, on its geometry (wall thickness, load eccentricity), material properties (both hygro-thermal and mechanical), and boundary conditions (heating, mechanical restraint).

In our approach, we follow a conservative assumption, also adopted by e.g. \citet{Msaad2007,Zeiml2008}, in which the wall is supposed to be fully mechanically restrained in the plane perpendicular to the wall thickness. On the other hand, the stresses resulting from the external mechanical load as well as the self-weight of the wall are not accounted for since their effect on the potential spalling may be considered to be negligible when compared with the effect of mechanical restraint (cf. \citep{Msaad2007,Zeiml2008}).

In order to asses the thermal stress arising from the mechanical restraint, we have to focus on the stress-strain conditions in the heated wall. It is widely accepted that the total strain in concrete subjected to high temperatures may be  decomposed into several parts which differ in their physical meaning. Hence, we can write (\citep[eq.~(1)]{Anderberg2008}; \citep[eq.~(4.2)]{Anderberg1976}; \citep[eq.~(1)]{LiPurkiss2005})
\begin{equation}\label{total_strain}
\epsilon_{tot} = \epsilon_{\theta}(\theta) + \epsilon_{\sigma}(\sigma,\theta) + \epsilon_{cr}(\sigma,\theta,t) + \epsilon_{tr}(\sigma,\theta),
\end{equation}
where $\epsilon_{tot}$ is the total strain, $\epsilon_{\theta}$ is the free thermal strain, $\epsilon_{\sigma}$ is the instantaneous stress-related strain (which can be divided into the elastic part and the plastic part, $\epsilon_{e}$ and $\epsilon_{p}$, respectively (see e.g. \citep{Bratina2005}), $\epsilon_{cr}$ is the creep strain, $\epsilon_{tr}$ is the transient strain, $\sigma$ is the stress, and $t$ is the time.

As stated in e.g. \citep{GernayFransen2010}, the creep strain, $\epsilon_{cr}$, is usually neglected. Moreover, the stress-dependent strains can be assumed together (with or without $\epsilon_{\sigma}$ or $\epsilon_{cr}$) as the mechanical strain (see \citep{GernayFransen2010}) or as the so called load induced thermal strain -- LITS (see e.g. \citep{Anderberg2008,Kukla2010,Terro1998,Zeiml2008}).

Here, we adopt a constitutive law proposed by \citet{Eurocode2-1-2}, in which the stress is expressed in terms of the total mechanical strain, $\epsilon_m$, that includes the transient strain implicitly, while the creep strain is omitted \citep{Anderberg2008,Carstensen2011,GernayFransen2010}, and hence, we can write
\begin{equation}\label{total_strain_Eurocode}
\epsilon_{tot} = \epsilon_{\theta}(\theta) + \epsilon_{m}(\sigma,\theta).
\end{equation}

Since the wall is supposed to be fully mechanically restrained, the total strain is equal to zero and the mechanical strain can be expressed as
\begin{equation}\label{mechanical_strain}
\epsilon_{m} = -\epsilon_{\theta}(\theta).
\end{equation}

As mentioned above, the constitutive law given by \citet{Eurocode2-1-2} has the form
\begin{equation}\label{stress_Eurocode}
\sigma = \mathcal{L} (\epsilon_m, \theta).
\end{equation}

In our case (see equation \eqref{mechanical_strain}), the stress can be expressed directly as a function of temperature (the formulas provided by \citet{Eurocode2-1-2} for $\epsilon_{\theta}(\theta)$ and
$\mathcal{L} (\epsilon_m, \theta)$ are stated in Section~\ref{Sec_MaterialProperties}). It should be noted that the stress in concrete determined by \eqref{stress_Eurocode} belongs to the uniaxial conditions. For the plane stress conditions, we get
\begin{equation}\label{thermo_mechanical_stress}
\sigma_{tm}(\theta) = \frac{1}{1-\nu(\theta)} \, \sigma(\theta),
\end{equation}
where $\nu$ is the Poisson's ratio (see Section~\ref{Sec_MaterialProperties}), and $\sigma$ is the uniaxial stress given by \eqref{stress_Eurocode} (with $\epsilon_m$ determined by \eqref{mechanical_strain}).

\subsection{Resulting stress conditions}
\begin{figure}[h]
  \centering
  {\includegraphics[angle=0,width=7cm]{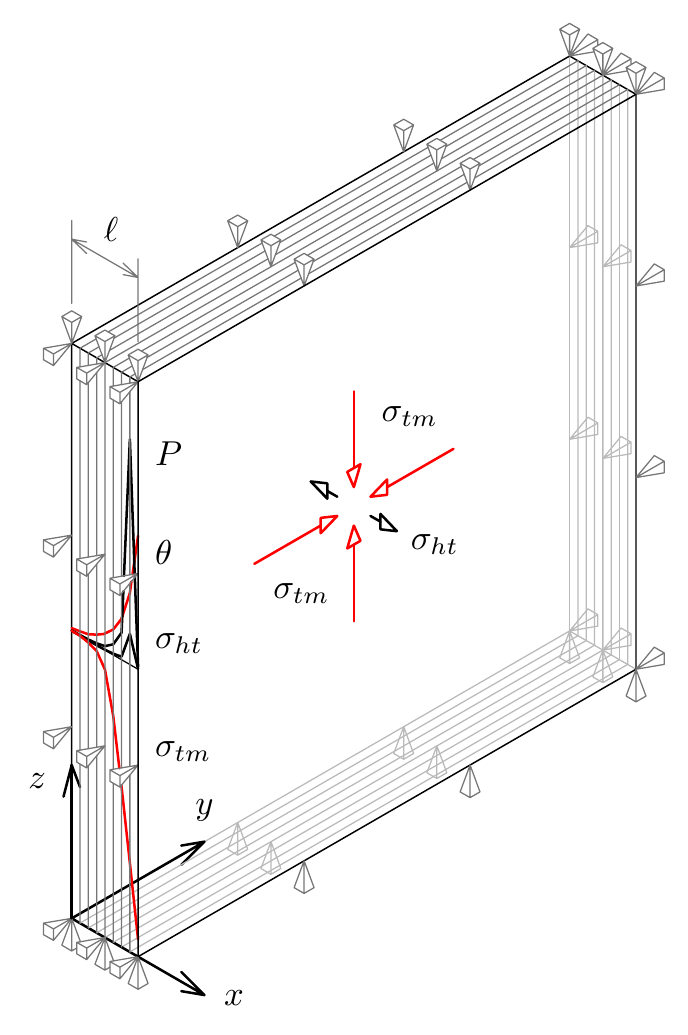}}
  \caption{Stress conditions in the wall}
  \label{Fig_Theory_Wall}
\end{figure}

Let us assume that the wall thickness, $\ell$, which is much smaller than the other dimensions of of the wall, is parallel to the $x$-axis, and the wall surfaces (at positions of $x = 0$ and $x = \ell$) are parallel to the plane $y$-$z$ (cf. \citep{Msaad2007}). As mentioned above and as stated in \citep{Msaad2007}, we can suppose that the hygro-thermal stress acts as a tensile stress perpendicular to the wall surface (i.e. in the $x$-direction) and the thermo-mechanical stress acts as a compressive plane stress in the planes parallel to the wall surface (i.e. in the directions of $y$ and $z$), see Figure~\ref{Fig_Theory_Wall}. Hence, in terms of the principal stresses, ${\sigma}_1$, ${\sigma}_2$, ${\sigma}_3$, and the normal stresses, ${\sigma}_x$, ${\sigma}_y$, ${\sigma}_z$, we may write \citep{Msaad2007}
\begin{equation}
\sigma_1 = \sigma_x = \sigma_{ht}(P,\theta),
\end{equation}
\begin{equation}
\sigma_2 = \sigma_3 = \sigma_y = \sigma_z = \sigma_{tm}(\theta).
\end{equation}

\subsection{Failure function}
\label{Section_Failure_Function}
As a failure function, we adopt the Men\'{e}trey--Willam triaxial failure criterion defined as \citep[eq.~(6)]{MenetreyWillam1995}
\begin{equation}\label{failure criterion}
F =
\left( \sqrt{1.5} \frac{\rho}{f_c} \right)^2 +
\left( 3 \frac{f_c^2 - f_t^{2}}{f_c f_t} \frac{e_F}{e_F+1} \right)
\left( \frac{\rho}{\sqrt{6}f_c} r(\vartheta,e_F) + \frac{\xi}{\sqrt{3}f_c} \right),
\end{equation}
where $e_F$ is the dimensionless eccentricity ($0.5 < e_F \leq 1.0$), which influences the shape of the failure function (see e.g. \citep[Fig.~3]{MenetreyWillam1995}) and can be expressed as a function of $f_c$, $f_t$ and $f_b$, with $f_b$ being the biaxial compressive strength of concrete (see \citep[eq.~(21.28)]{JirasekBazant2002}). Further, function $r(\vartheta,e_F)$ is given by \citet[eq.~(1)]{MenetreyWillam1995} in the form
\begin{equation}\label{function_r}
r(\vartheta,e_F) = \frac
{4(1-e_F^2)\cos^2{\vartheta}+(2e_F-1)^2}
{2(1-e_F^2)\cos{\vartheta}+(2e_F-1)\sqrt{4(1-e_F^2)\cos^2{\vartheta}+5e_F^2-4e_F}},
\end{equation}
and $\xi$, $\rho$ and $\vartheta$ are the Haigh--Westergaard coordinates (see e.g. \citep{JirasekBazant2002,MenetreyWillam1995}).

\begin{figure}[h]
\centering
\begin{minipage}[t]{.47\textwidth}
  \centering
  {\includegraphics[angle=0,width=6.4cm]{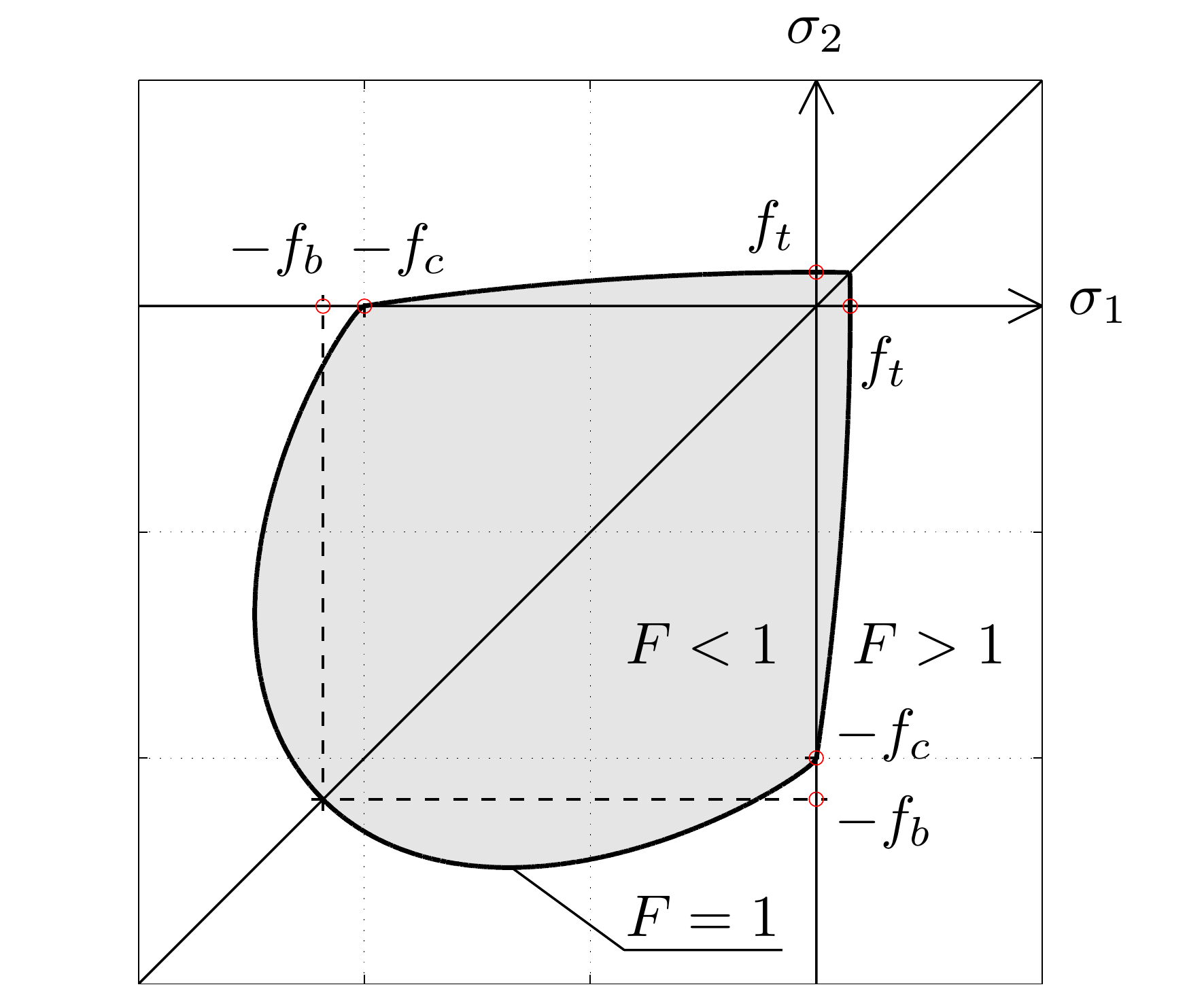}}
  \caption{Men\'{e}trey--Willam failure function \citep{MenetreyWillam1995}}
  \label{Fig_Theory_Failure_Function}
\end{minipage}%
\hspace{.01\textwidth}
\begin{minipage}[t]{.47\textwidth}
  \centering
  {\includegraphics[angle=0,width=6.4cm]{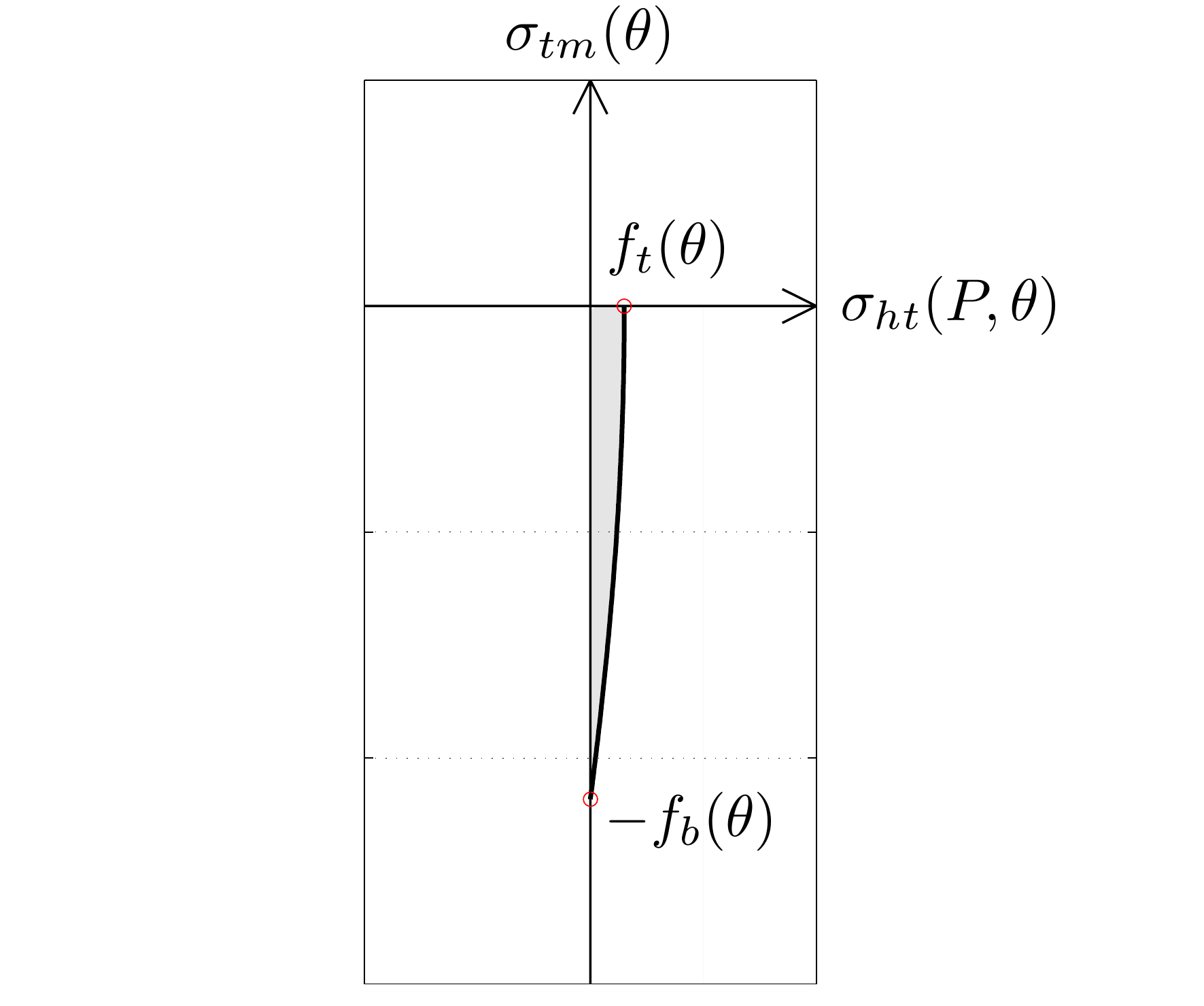}}
  \caption{Failure function used in our model}
  \label{Fig_Theory_Failure_Function_Simplified}
\end{minipage}
\end{figure}

The Men\'{e}trey--Willam failure function for the plane stress conditions (i.e. for $\sigma_3 = 0$) is shown in Figure~\ref{Fig_Theory_Failure_Function}.

In our case, in which $\sigma_1 = \sigma_{ht}$ and $\sigma_2 = \sigma_3 = \sigma_{tm}$, we can write
\begin{eqnarray}
\xi = \frac{1}{\sqrt{3}}(\sigma_{ht} + 2\sigma_{tm}),
\\
\rho = \sqrt{\frac{2}{3} (\sigma_{ht} - \sigma_{tm})^2},
\\
\cos{(3 \vartheta)} = \frac{   \sigma_{ht} - \sigma_{tm}    }{      \sqrt{(\sigma_{ht} - \sigma_{tm})^2}       }.
\end{eqnarray}

Since $ \sigma_{tm} \leq 0 < \sigma_{ht}$, and $0.5 < e_F \leq 1.0$, it is obvious that $\vartheta = 0$, and $r = 1/e_F$. Substituting this in \eqref{failure criterion} and assuming the strengths of concrete as temperature dependent, we get the resulting failure function used in our model (cf. \citep[p.~613]{Msaad2007})
\begin{equation}\label{failure criterion_simplified}
F = \left( \frac{\sigma_{ht}(P,\theta)-\sigma_{tm}(\theta)}{f_c(\theta)} \right) ^2 +
\frac{f_c(\theta)^2-f_t(\theta)^2}{f_c(\theta)^2 f_t(\theta)} \left( \sigma_{ht}(P,\theta) + \frac{2e_F-1}{e_F+1} \sigma_{tm}(\theta) \right),
\end{equation}
which is illustrated in Figure~\ref{Fig_Theory_Failure_Function_Simplified}.

\subsection{Evolution law for moving boundary due to spalling}
In our regularized approach, the instantaneous spalling of concrete is approximated by rapid continuous process in such a way that sheet of concrete is continuously removed from the wall and the receding outer surface forms the moving boundary. In general, in one dimension, the external loading to the concrete wall according to the boundary conditions \eqref{radiative a} and \eqref{radiative b} is prescribed on the unknown ablating boundary $x=\ell(t)$.
For the spalling of the wall originally occupying the space $0 \leq x \leq \ell_0$, we propose the governing equation of the form
\begin{equation}\label{spalling_process}
\frac{d \ell}{d t} = - \frac{\ell}{\gamma} \, [ \max(F) - 1]^+,
\qquad \ell(0)=\ell_0,
\end{equation}
where the position of the moving boundary $\ell$ has to be determined as a part of the solution. Further, $x^+ = \max(0,x)$, $\gamma$ [s] represents the characteristic time of spalling process and $\max(F)$ is the maximal value of failure function \eqref{failure criterion_simplified} achieved within the wall $(0,\ell)$ at actual time $t$.

\section{Formulation of the problem}
\label{sec:complete_form}

\subsection{Modification of mass balance equation}
Let us start with modification of the equation \eqref{moisture}. Denote by $m = \eta_{w}\rho_{w}+\eta_v\rho_v$ the total mass of moisture (including liquid water and vapour). Incorporating the constitutive relations \eqref{flux_w} and \eqref{flux_v} into the mass balance equation \eqref{moisture} yields
\begin{equation}\label{mass_balance_11}
\frac{\partial m}{\partial t}
=
\frac{\partial}{\partial x}\left[ \left(
\rho_{w}\frac{K K_{rw}}{\mu_{w}} + \rho_{v} \frac{K K_{rg}}{\mu_{g}} \right)
\frac{\partial P}{\partial x}
-
\rho_{w}\frac{K K_{rw}}{\mu_{w}}\frac{\partial P_{c}}{\partial x} \right]
+
\frac{\partial m_{d}}{\partial t}.
\end{equation}
Here $P_c=P_c(P,\theta)$ via equation \eqref{equilibrium meniscus}. After additional modification, the equation \eqref{mass_balance_11} can be written in a general form
\begin{equation}\label{mass_balance_12}
\frac{\partial m}{\partial t}
-
\frac{\partial m_{d}}{\partial t}
=
\frac{\partial}{\partial x}\left(
K_{mP}
\frac{\partial P}{\partial x}
+
K_{m\theta} \frac{\partial \theta}{\partial x}
\right),
\end{equation}
where
\begin{eqnarray}
K_{mP} &=& \rho_{w}\frac{K K_{rw}}{\mu_{w}}\left( 1-\frac{\partial P_c}{\partial P} \right)
+
\rho_{v} \frac{K K_{rg}}{\mu_{g}},
\label{kmp}
\\
K_{m\theta}
&=&
-
\rho_{w}\frac{K K_{rw}}{\mu_{w}}\frac{\partial P_c}{\partial \theta}.
\label{kmt}
\end{eqnarray}

\subsection{Modification of energy conservation equation}
\label{sec:mod_energy}
Incorporating water vapour conservation equation into the term corresponding to the latent heat of evaporation leads to the following modified energy balance equation
\begin{equation}\label{energy temp_10}
(\rho c_p) \frac{\partial\theta}{\partial t}
+
h_{e}\frac{\partial \left(\eta_{v}\rho_{v}\right)}{\partial t}
+
h_{d}\frac{\partial m_{d}}{\partial t}
=
\frac{\partial}{\partial x}
\left(
\lambda_c \frac{\partial \theta}{\partial x}
\right)
-
h_{e} \frac{\partial\left(\eta_v \rho_v v_{v}\right)}{\partial x}
-
\left( c^w_p\rho_w\eta_w v_{w} + c^g_p\rho_g\eta_g v_g \right)
\frac{\partial\theta}{\partial x}.
\end{equation}
Simple calculation yields
\begin{eqnarray}\label{sup_eq_10}
h_{e} \frac{\partial\left(\eta_v \rho_v v_{v}\right)}{\partial x}
&=&
\frac{\partial \left(h_{e} \eta_v \rho_v v_{v}\right)}{\partial x}
-
\left(\eta_v \rho_v v_{v}\right) \frac{\partial h_{e}}{\partial x}
\nonumber
\\
&=&
-\frac{\partial}{\partial x}
\left(
h_{e}\rho_{v}
\frac{K K_{rg}}{\mu_{g}}
\frac{\partial P}{\partial x}
\right)
+
\rho_{v}
\frac{K K_{rg}}{\mu_{g}}
\frac{\partial h_{e}}{\partial \theta}
\frac{\partial P}{\partial x}
\frac{\partial \theta}{\partial x}.
\end{eqnarray}
Incorporating the equation \eqref{sup_eq_10} into \eqref{energy temp_10} reads
\begin{equation}\label{energy temp_11}
M_{\theta P} \frac{\partial P}{\partial t}
+
M_{\theta \theta} \frac{\partial \theta}{\partial t}
+
h_{d}\frac{\partial m_{d}}{\partial t}
=
\frac{\partial}{\partial x}
\left(
K_{\theta P}
\frac{\partial  P}{\partial x}
+
K_{\theta\theta} \frac{\partial \theta}{\partial x}
\right)
+
\left(
C_{\theta P}\frac{\partial P}{\partial x}
+
C_{\theta \theta} \frac{\partial\theta}{\partial x}
\right)
\frac{\partial\theta}{\partial x},
\end{equation}
where
\begin{eqnarray}
M_{\theta P}
&=&
h_{e}\frac{\partial \left(\eta_{v}\rho_{v}\right)}{\partial P},
\label{mtp}
\\
M_{\theta \theta}
&=&
(\rho c_p)
+
h_{e}\frac{\partial \left(\eta_{v}\rho_{v}\right)}{\partial \theta},
\label{mtt}
\\
{K}_{\theta P}
&=&
h_{e}\rho_{v}
\frac{K K_{rg}}{\mu_{g}},
\label{ktp}
\\
{K}_{\theta \theta} &=&  \lambda_c,
\label{ktt}
\\
C_{\theta P}
&=&
c_p^w \rho_{w}\frac{K K_{rw}}{\mu_{w}}\left( 1-\frac{\partial P_c}{\partial P} \right)
+
\left( \left( c_p^v - \frac{\partial h_{e}}{\partial \theta} \right)\rho_{v} + c_p^a \rho_{a} \right)
\frac{K K_{rg}}{\mu_{g}},
\label{ctp}
\\
C_{\theta \theta}
 &=&
-
c_p^w \rho_{w}\frac{K K_{rw}}{\mu_{w}}\frac{\partial P_c}{\partial \theta}.
\label{ctt}
\end{eqnarray}

\subsection{Resulting model}
The full mathematical model consists of the balance equations for moisture and energy, state equation of pore water (moisture), governing equation for dehydration, evolution equation for moving boundary due to spalling, the set of appropriate boundary and initial conditions specifying the fields of pore pressure, temperature, mass of dehydrated water and initial thickness of the wall.

\bigskip

\noindent
\emph{Moisture conservation equation:}
\begin{equation}
\label{eq:model_moisture}
\frac{\partial m}{\partial t}
-
\frac{\partial m_{d}}{\partial t}
=
\frac{\partial}{\partial x} \left(
K_{mP}
\frac{\partial P}{\partial x}
+
K_{m\theta} \frac{\partial\theta}{\partial x}
\right);
\end{equation}
\emph{energy conservation equation:}
\begin{equation}
M_{\theta P} \frac{\partial P}{\partial t}
+
M_{\theta \theta} \frac{\partial \theta}{\partial t}
+
h_{d}\frac{\partial m_{d}}{\partial t}
=
\frac{\partial}{\partial x}
\left(
K_{\theta P}
\frac{\partial P}{\partial x}
+
K_{\theta\theta}\frac{\partial\theta}{\partial x}
\right)
+
\left(
C_{\theta P}\frac{\partial P}{\partial x}
+
C_{\theta\theta} \frac{\partial\theta}{\partial x}
\right)
\frac{\partial\theta}{\partial x};
\end{equation}
\emph{state equation of moisture:}
\begin{equation}
m = \eta_{w}\rho_{w}+\eta_v\rho_v;
\end{equation}
\emph{governing equation of dehydration:}
\begin{equation}
\frac{\partial m_{d}}{\partial t} = -\frac{1}{\tau}(m_{d}-m_{d,eq}(\theta));
\end{equation}
\emph{evolution law for moving boundary due to spalling:}
\begin{equation}\label{res_model_moving_bound}
\frac{d \ell}{d t} = - \frac{\ell}{\gamma} \, [ \max(F) - 1]^+  ;
\end{equation}
\emph{boundary conditions (at $x=0$ and $x=\ell$):}
\begin{eqnarray}
-\left(
K_{mP}\frac{\partial P}{\partial x}
+
K_{m\theta} \frac{\partial\theta}{\partial x}
\right) n_x
&=&
\beta_c (\rho_{v} - \rho_{v \infty}),
\\
-\left(
K_{\theta P}
\frac{\partial P}{\partial x}
+
K_{\theta\theta} \frac{\partial\theta}{\partial x}
\right) n_x
&=&
\alpha_c(\theta-\theta_{\infty})
+
e\sigma_{SB} (\theta^4-\theta^4_{\infty})
\end{eqnarray}
and \emph{initial conditions (at $t=0$):}
\begin{eqnarray}\label{eq102}
P = P_0,
\\
\theta   = \theta_0,
\\
\ell =  \ell_0,
\\
m_d = 0.
\label{eq:ini_md}
\end{eqnarray}

The unknowns in the model are the moisture content $m$, pore pressure $P$, temperature $\theta$, mass of dehydrated water $m_{d}$ and the actual thickness of the wall $\ell$. Transport coefficients $K_{mP}$, $K_{m\theta}$, $M_{\theta P}$, $M_{\theta \theta}$, $K_{\theta P}$, $K_{\theta \theta}$, $C_{\theta P}$, $C_{\theta \theta}$, defined by \eqref{kmp} and \eqref{kmt} and \eqref{mtp}--\eqref{ctt}, depend in non-linear manner upon temperature $\theta$ and pressure $P$.
Material non-linearities are described in detail in Section \ref{Sec_MaterialProperties}.

\section{FEM formulation and solution strategy}
\label{sec:full_discr}
Let $0=t_0<t_1<\dots<t_N=T$ be an equidistant partitioning of time interval $[0,T]$ with step $\Delta t$. Set a fixed integer $n$  such that $0\leq n <N$. In what follows we abbreviate $f(x,t_n)$ by $f^n$ for any function $f$. The time discretization of the continuous model is accomplished through a semi-implicit difference scheme
\begin{eqnarray}
\frac{ m^{n+1} - m^n}{\Delta t}
& = &
\frac{\partial}{\partial x} \left(
K_{mP}^n
\frac{\partial P^{n+1}}{\partial x}
+
K_{m\theta}^n \frac{\partial \theta^{n+1}}{\partial x}
\right)
\nonumber
\\
& & +
\frac{m_{d}^{n+1} - m_{d}^{n}}{\Delta t},
\\
M_{\theta P}^n \frac{P^{n+1} - P^n}{\Delta t}
+
M_{\theta \theta}^n \frac{\theta^{n+1} - \theta^n}{\Delta t}
& = &
\frac{\partial}{\partial x}
\left(
K_{\theta P}^n
\frac{\partial  P^{n+1}}{\partial x}
+
K_{\theta\theta}^n \frac{\partial \theta^{n+1}}{\partial x}
\right)
\nonumber
\\
& & +
\left(
C_{\theta P}^n
\frac{\partial P^n}{\partial x}
+
C_{\theta\theta}^n \frac{\partial \theta^n}{\partial x}
\right)
\frac{\partial \theta^n}{\partial x}
\nonumber
\\
& & -
h_{d}^n \frac{m_{d}^{n+1} - m_{d}^{n}}{\Delta t},
\\
m^{n+1}
& = &
\eta_{w}^{n+1}\rho_{w}(\theta^{n+1})+\eta_v^{n+1}\rho_v(\theta^{n+1},P^{n+1}),
\\
\frac{m_{d}^{n+1} - m_{d}^{n}}{\Delta t}
& = &
-\frac{1}{\tau}(m_{d}^n-m_{d,eq}(\theta^n)),
\label{dehydr_discr}
\\
\frac{\ell^{n+1} - \ell^{n}}{\Delta t}
& = &
- \frac{\ell^{n+1}}{\gamma} \, [ \max(F^{n+1}) - 1]^+,
\label{spall_discr}
\\
%
-n_x\left(
K_{mP}^{n}
\frac{\partial P^{n+1}}{\partial x}
+
K_{m\theta}^{n}
\frac{\partial \theta^{n+1}}{\partial x}
\right)\Bigg|_{x=0,\;x=\ell^{n+1}}
& = &
\beta_c
\left(
\rho_v(P^{n+1},\theta^{n+1}) - \rho_v (P_{\infty}^{n+1},\theta_{\infty}^{n+1})
\right),
\\
-n_x\left(
K_{\theta P}^{n}
\frac{\partial P^{n+1}}{\partial x}
+
K_{\theta\theta}^{n}
\frac{\partial \theta^{n+1}}{\partial x}
\right)\Bigg|_{x=0,\;x=\ell^{n+1}}
& = &
\alpha_c(\theta^{n+1}-\theta^{n+1}_{\infty})
\nonumber
\\
& & +
e\sigma_{SB} \left( (\theta^{n+1})^4-(\theta_{\infty}^{n+1})^4 \right),
\end{eqnarray}
where $\eta_{w}^{n+1}=\eta_{w}(\theta^{n+1},P^{n+1})$ and $\eta_v^{n+1}=\eta_v(\theta^{n+1},P^{n+1})$.
Further, $n_x=+1$ for $x=\ell^{n+1}$ and $n_x=-1$ for $x=0$.

Applying the Galerkin procedure to the mass and energy conservation equations leads to the system of non-linear algebraic equations
\begin{equation}\label{eq:discrete_system}
\frac{1}{\Delta t} \M{M}^n \left( \bfx^{n+1} - \bfx^n \right) + \M{K}^n \bfx^{n+1} +  \bff^{n+1}( \bfx^{n+1} )  = {\bf0} ,
\end{equation}
where
$\bfx^{n+1}
=
\left(
{{\bfm}}^{n+1} , {{\bftheta}}^{n+1} , {{\bfP}}^{n+1}
\right)^T$,
$m^{n+1}(x) = \M{N}(x) {{\bfm}}^{n+1}$,
$\theta^{n+1}(x) = \M{N}(x) {{\bftheta}}^{n+1}$
and
$P^{n+1}(x) = \M{N}(x) {{\bfP}}^{n+1}$,
stores the unknown nodal values of moisture, temperature and pore pressure at time $t_{n+1}$, respectively. The constant matrices in \eqref{eq:discrete_system} exhibit a block structure
\begin{equation}\label{eq:block_structure_linear}
\M{M}^n = \left[
\begin{array}{ccc}
\M{M}_{mm}^n  & \M{0}  & \M{0}
\\
\M{0}  & \M{M}_{\theta\theta}^n  &   \M{M}_{\theta P}^n
\\
\M{0}   &  \M{0}   &  \M{0}
\end{array}
\right],
\qquad
\M{K}^n =
\left[
\begin{array}{ccc}
  \M{0}   &  \M{K}_{m\theta}^n & \M{K}_{mP}^n
\\
  \M{0}   &  \M{K}_{\theta\theta}^n  &   \M{K}_{\theta P}^n
\\
  \M{0}   &  \M{0}   &  \M{0}
\end{array}
\right]
\end{equation}
and the non-linear term reads as
\begin{equation}
\label{eq:block_structure_non_linear}
\bff^{n+1}( \bfx^{n+1} )
=
\left(
\begin{array}{c}
{\bff}^{n+1}_m( \bfx^{n+1} )
 \\
{\bff}^{n+1}_{\theta}( \bfx^{n+1} )
\\
{\bff}^{n+1}_{P}( \bfx^{n+1} )
\end{array}
\right).
\end{equation}
The non-linear system \eqref{eq:discrete_system} is solved iteratively using the Newton's method (see \citep{Reddy2004}).

The individual matrices $\M{M}_{\bullet}^{\bullet}$ can be expressed as
%
\begin{eqnarray}
\M{M}^{n}_{mm} & = & \int_0^{\ell^{n+1}} \M{N}(x)\trn \M{N}(x) \de x,
\\
\M{M}^{n}_{\theta \theta} & = & \int_0^{\ell^{n+1}}
\M{N}(x)\trn
\left[
\left(
(\rho c_p)
+
h_{e}\frac{\partial \left(\eta_{v}\rho_{v}\right)}{\partial \theta}
\right)
\M{N}(x)
\right]
\de x,
\\
\M{M}^{n}_{\theta P} & = & \int_0^{\ell^{n+1}}
\M{N}(x)\trn
\left[
h_{e}\frac{\partial \left(\eta_{v}\rho_{v}\right)}{\partial P}
\M{N}(x)
\right]
\de x,
\end{eqnarray}
%
whereas the blocks
$\M{K}_{\bullet}^{\bullet}$ attain the
form
%
\begin{eqnarray}
\M{K}_{m\theta}^n
& = &
\int_0^{\ell^{n+1}}
\left( \frac{\partial }{\partial x} \M{N}(x) \right) \trn
\left[
-
\rho_{w}\frac{K K_{rw}}{\mu_{w}}\frac{\partial P_c}{\partial \theta}
\frac{\partial }{\partial x} \M{N}(x)
\right]
\de x,
\\
\M{K}_{mP}^n
& = &
\int_0^{\ell^{n+1}}
\left( \frac{\partial }{\partial x} \M{N}(x) \right) \trn
\left[
\left(
\rho_{w}\frac{K K_{rw}}{\mu_{w}}\left( 1-\frac{\partial P_c}{\partial P} \right)
+
\rho_{v} \frac{K K_{rg}}{\mu_{g}}
\right)
\frac{\partial }{\partial x} \M{N}(x)
\right]
\de x,
\nonumber
\\
\\
\M{K}^n_{\theta\theta}
& = &
\int_0^{\ell^{n+1}}
\left( \frac{\partial }{\partial x} \M{N}(x) \right) \trn
\left[
\lambda_c
\frac{\partial }{\partial x} \M{N}(x)
\right]
\de x,
\\
\M{K}^n_{\theta P}
& = &
\int_0^{\ell^{n+1}}
\left( \frac{\partial }{\partial x} \M{N}(x) \right) \trn
\left[
h_{e}\rho_{v}
\frac{K K_{rg}}{\mu_{g}}
\frac{\partial }{\partial x} \M{N}(x)
\right]
\de x.
\end{eqnarray}
%
%
%
The non-linear terms  $\bff^\bullet$ are provided by
\begin{eqnarray}
\bff^{n+1}_m(\bfx^{n+1})
& = &
\int_0^{\ell^{n+1}}
-\frac{m_{d}^{n+1} - m_{d}^{n}}{\Delta t}
\M{N}(x)\trn
\de x
\nonumber
\\
&&
+
\Bigg[
\left( \beta_c
\left(
\rho_v(P^{n+1},\theta^{n+1}) - \rho_v (P_{\infty}^{n+1},\theta_{\infty}^{n+1})
\right) \right)
 \M{N}(x)\trn
\Bigg]^{x=\ell^{n+1}}_{x=0},
\\
\bff^{n+1}_{\theta}(\bfx^{n+1})
& = &
\int_0^{\ell^{n+1}}
h_{d}^n \frac{m_{d}^{n+1} - m_{d}^{n}}{\Delta t}
\M{N}(x)\trn
\de x
\nonumber
\\
&&
-
\int_0^{\ell^{n+1}}
\left(
C_{\theta P}^n
\frac{\partial P^n}{\partial x}
+
C_{\theta\theta}^n \frac{\partial \theta^n}{\partial x}
\right)
\frac{\partial \theta^n}{\partial x}
\M{N}(x)\trn
\de x
\nonumber
\\
&&
+
\Bigg[
\left(\alpha_c(\theta^{n+1}-\theta^{n+1}_{\infty})
+
 e\sigma_{SB} \left( (\theta^{n+1})^4-(\theta_{\infty}^{n+1})^4 \right)\right)
 \M{N}(x)\trn
\Bigg]^{x=\ell^{n+1}}_{x=0},
\\
\bff^{n+1}_{P}(\bfx^{n+1})
& = &
\bfm^{n+1}
-
\eta_{w}^{n+1}(\bftheta^{n+1},\bfP^{n+1})\rho_w(\bftheta^{n+1})
\nonumber
\\
&&
-
\eta_v^{n+1}(\bftheta^{n+1},\bfP^{n+1})\rho_v(\bftheta^{n+1},\bfP^{n+1}).
\end{eqnarray}

\clearpage

\paragraph{Numerical algorithm} The described numerical algorithm becomes:
\paragraph{Step 1} Set $T$, $\Delta t$, $N = T/\Delta t$

\paragraph{Step 2} Set initial values $P_0$, $\theta_0$, $\ell_0$, $m_d(0) = 0$

\paragraph{Step 3}

\begin{itemize}

\item[] For $n=0, \dots, N-1$

\begin{itemize}

\item[] $t_n = n \Delta t$

\item[] update $m_{d,eq}(\theta^n)$, $F^{n+1}$

\item[] solve \eqref{spall_discr} to get $\ell^{n+1}$

\item[] solve \eqref{dehydr_discr} to get $m_{d}^{n+1}$

\item[] update $P_{\infty}^{n+1}$, $\theta_{\infty}^{n+1}$

\item[] update  $\M{M}^n$, $\M{K}^n$, $\bff^{n+1}$

\item[] solve \eqref{eq:discrete_system} by Newton's iteration procedure to get $\bfx^{n+1}$

\end{itemize}

\item[] end $n$

\end{itemize}

\section{Material data for concrete at high temperatures}
\label{Sec_MaterialProperties}

{
In our approach, concrete is assumed to be a homogeneous multi-phase system. Hence, most of its material properties (such as free thermal strain, thermal conductivity, etc.) are treated as so called effective or smeared characteristics. It means that for the solid skeleton, we do not distinguish between the aggregates and the cement paste. Such macroscopic approach has been widely accepted by the scientific community for investigation of hygro-thermo-mechanical and spalling behaviour of heated concrete (see e.g. the works by \citet{DwaiKod2009,Gawin2006,Witek2007}). It is clear that by turning to a mesoscale level, one can obtain more realistic results, especially when investigating the spalling phenomenon (see e.g. the recent works by \citet{Le2011,Zhao2014}). However, such approach is out of scope of the present paper.
}

Based on the literature review, the material properties of concrete and its components subjected to high temperatures are assumed as follows.

The free thermal strain of concrete, $\epsilon_{\theta}~[-]$, is defined in the temperature range of $293.15~{\rm{K}}$ to $1473.15~{\rm{K}}$ as \citep[Section~3.3.1]{Eurocode2-1-2}

\medskip

for siliceous aggregates concrete:
\begin{equation}\label{thermal_strain_siliceous}
\epsilon_{\theta}(\theta) = \left\{ \begin{array}{lll}
\displaystyle -1.8 \times 10^{-4} + 9 \times 10^{-6} \, \theta + 2.3 \times 10^{-11} \, \theta^3
& {\rm{for}} & \theta \leq 973.15~{\rm{K}},
\\
\displaystyle 14 \times 10^{-3}
& {\rm{for}} & \theta > 973.15~{\rm{K}},
\end{array} \right.
\end{equation}

\medskip

for calcareous aggregates concrete:
\begin{equation}\label{thermal_strain_calcareous}
\epsilon_{\theta}(\theta) = \left\{ \begin{array}{lll}
\displaystyle -1.2 \times 10^{-4} + 6 \times 10^{-6} \, \theta + 1.4 \times 10^{-11} \, \theta^3
& {\rm{for}} & \theta \leq 1078.15~{\rm{K}},
\\
\displaystyle 12 \times 10^{-3}
& {\rm{for}} & \theta > 1078.15~{\rm{K}}.
\end{array} \right.
\end{equation}

\begin{figure}[h]
\centering
\begin{minipage}[t]{.47\textwidth}
  \centering
  {\includegraphics[angle=0,width=6.4cm]{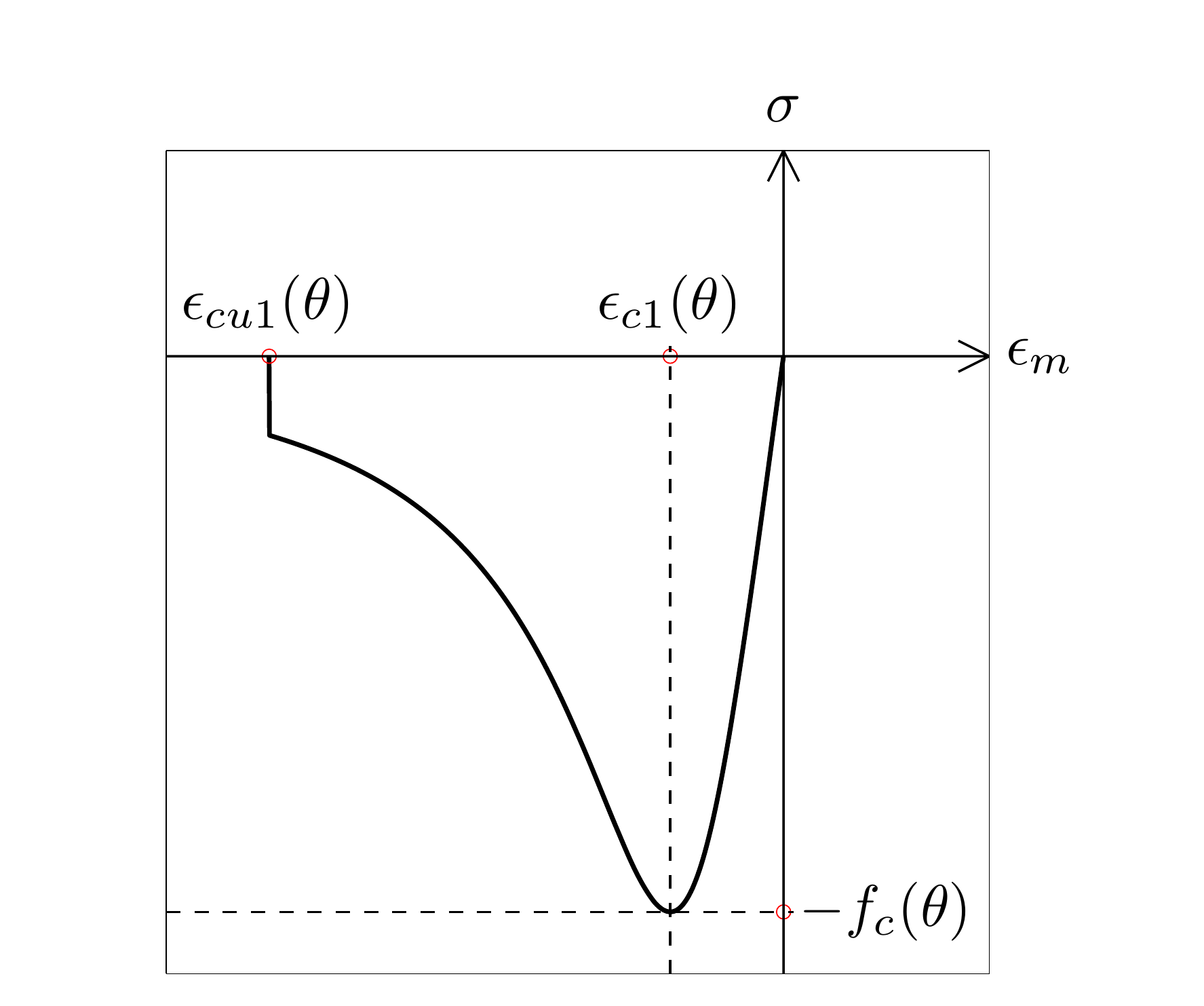}}
  \caption{Constitutive law for concrete in compression given by \citet[Figure~3.1]{Eurocode2-1-2}}
  \label{Fig_Theory_Eurocode_Constitutive_Law}
\end{minipage}%
\hspace{.01\textwidth}
\begin{minipage}[t]{.47\textwidth}
  \centering
  {\includegraphics[angle=0,width=6.4cm]{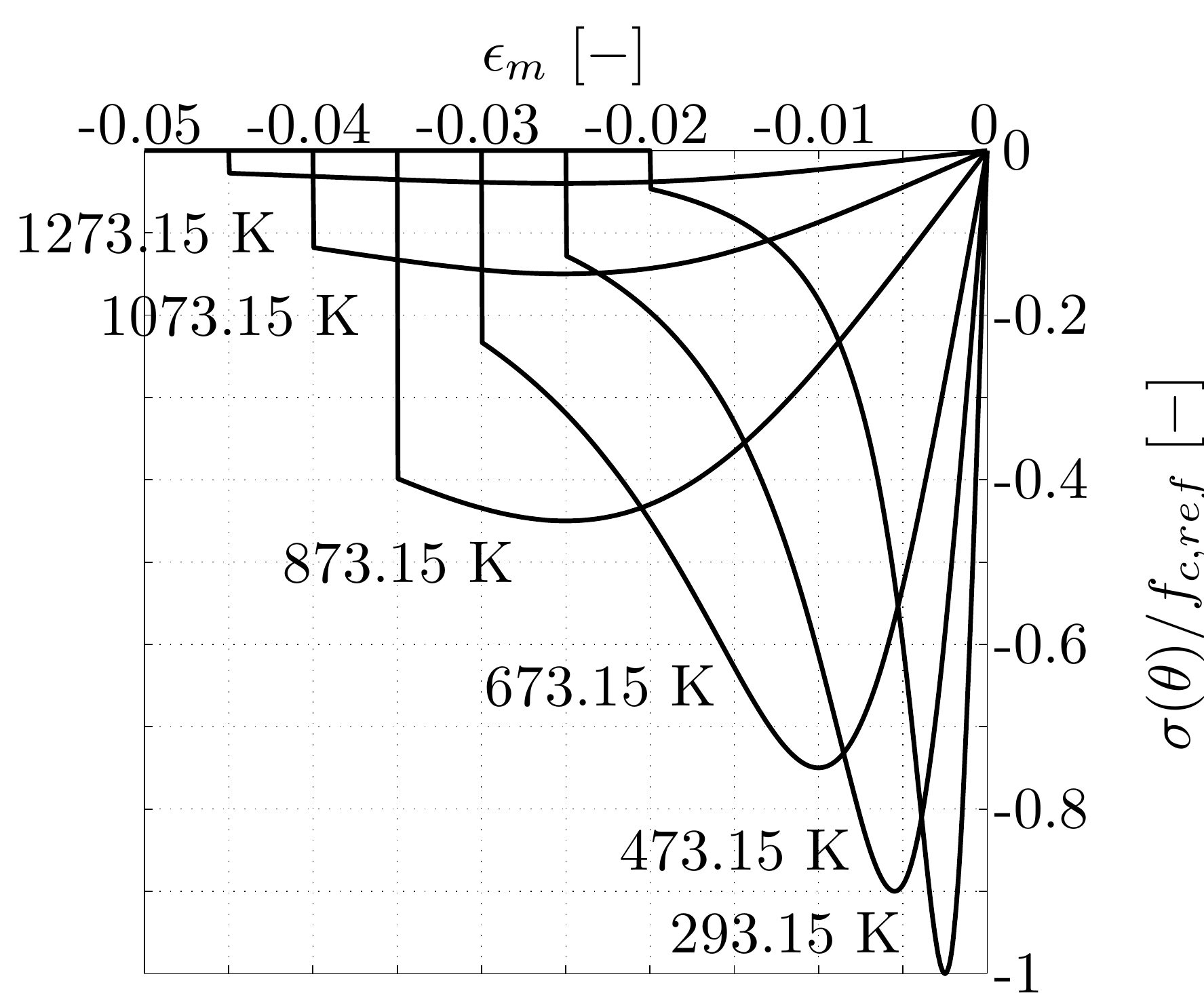}}
  \caption{An example of the constitutive law for concrete in compression at high temperatures \citep{Eurocode2-1-2}}
  \label{Fig_Theory_Eurocode_Constitutive_Law_Temp}
\end{minipage}
\end{figure}

The constitutive law for concrete in compression, $\sigma = \mathcal{L} (\epsilon_m, \theta)$, is adopted from Eurocode~2 in the form \citep[Section~3.2.2.1]{Eurocode2-1-2}
\begin{equation}\label{concrete_constitutive_law}
\sigma(\epsilon_m, \theta) = \left\{
 \begin{array}{lll}
 \displaystyle - \frac{3 \epsilon_m f_c(\theta)}{\epsilon_{c1}(\theta) \left[ 2 + \left( \displaystyle\frac{\epsilon_m}{\epsilon_{c1}(\theta)}\right)^3\right]}
 & {\rm{for}} & \epsilon_{cu1}(\theta) < \epsilon_m \leq 0,
 \\
 0
 & {\rm{for}} & \epsilon_m  \leq \epsilon_{cu1}(\theta),
 \end{array} \right.
\end{equation}
where $\epsilon_m$ is the total mechanical strain, $f_c$ is the compressive strength of concrete, $\epsilon_{c1}$ is the strain corresponding to $f_c$, and $\epsilon_{cu1}$ is the ultimate strain, see Figure~\ref{Fig_Theory_Eurocode_Constitutive_Law} and Figure~\ref{Fig_Theory_Eurocode_Constitutive_Law_Temp}.

\begin{figure}[h]
\centering
\begin{minipage}[t]{.47\textwidth}
  \centering
  {\includegraphics[angle=0,width=6.4cm]{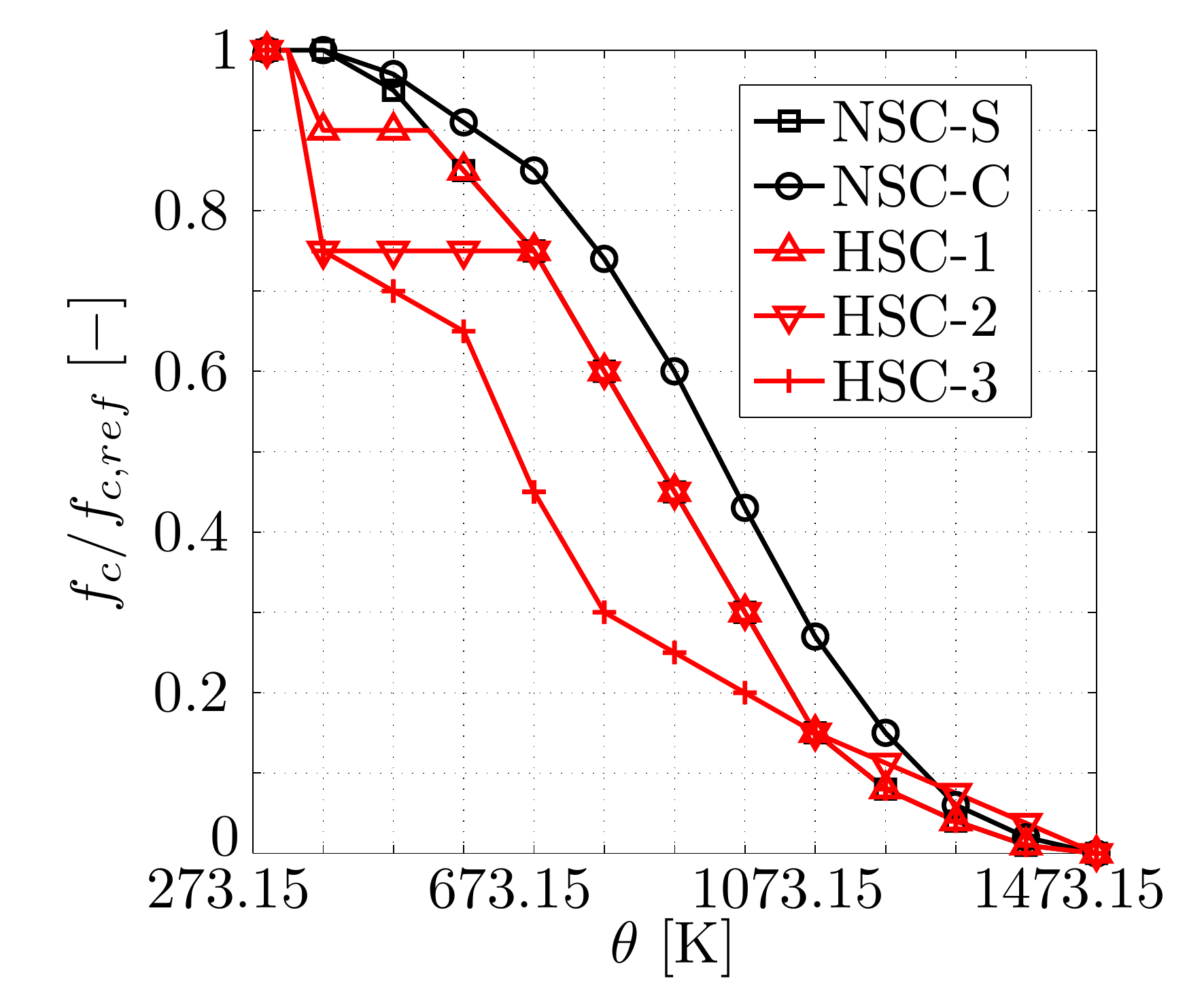}}
  \caption{Reduction of the compressive strength of NSC and HSC given by \citet{Eurocode2-1-2}}
  \label{Fig_Theory_Eurocode_Strength}
\end{minipage}%
\hspace{.01\textwidth}
\begin{minipage}[t]{.47\textwidth}
  \centering
  {\includegraphics[angle=0,width=6.4cm]{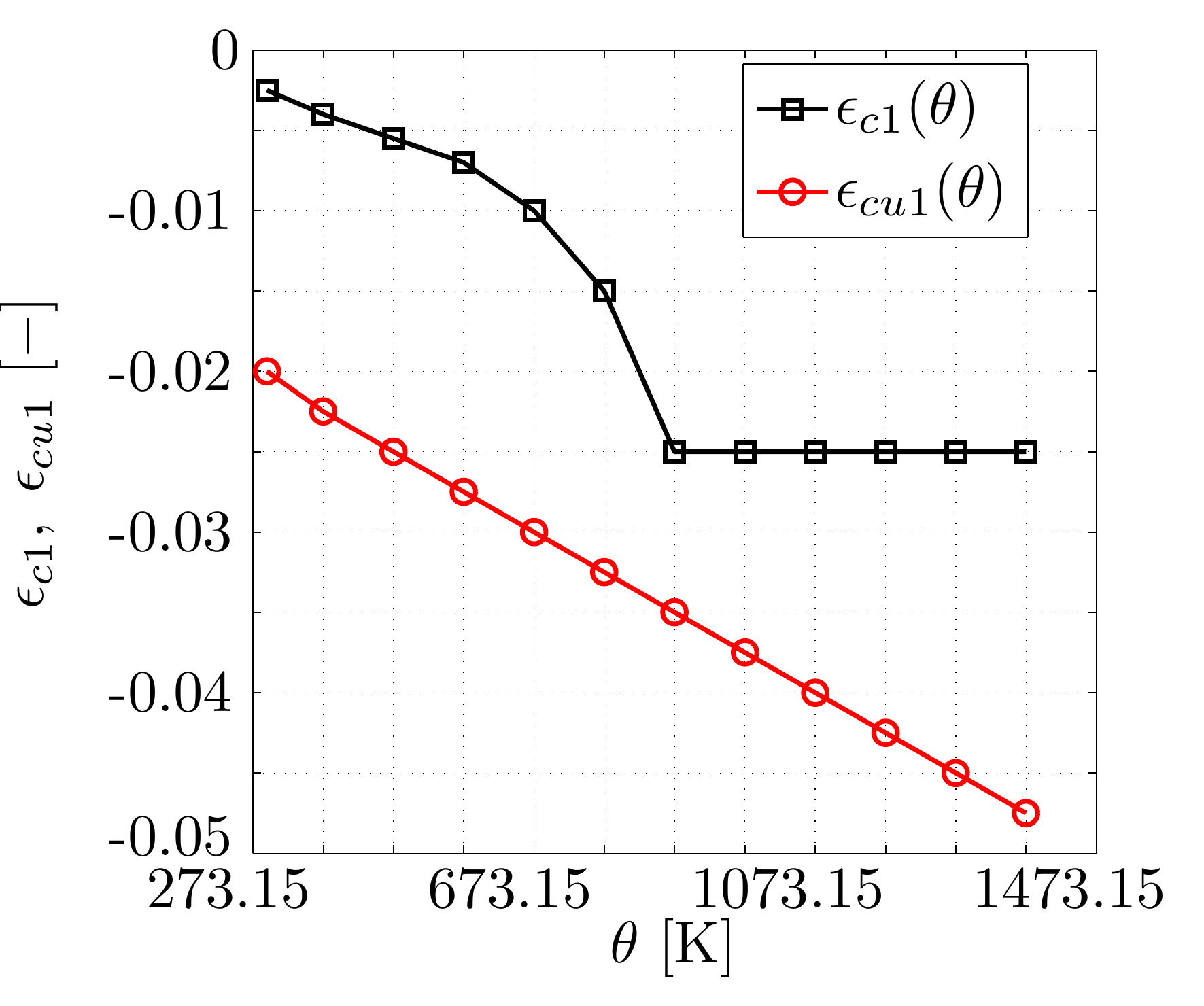}}
  \caption{Temperature evolution of the mechanical strains of concrete according to \citet{Eurocode2-1-2}}
  \label{Fig_Theory_Eurocode_Strain}
\end{minipage}
\end{figure}

The temperature dependent parameters of the constitutive law \eqref{concrete_constitutive_law}, $f_c(\theta)$, $\epsilon_{c1}(\theta)$, and $\epsilon_{cu1}(\theta)$, for the normal strength concrete with the siliceous and calcareous aggregates, NSC-S and NSC-C, respectively, as well as for three classes of high strength concrete (HSC-1, HSC-2, and HSC-3, see \citet[Section~6.1]{Eurocode2-1-2}) are stated in
\citet[Table~3.1, Table~6.1N]{Eurocode2-1-2} in the form of tabulated data that are illustrated in Figures~{\ref{Fig_Theory_Eurocode_Strength}} and {\ref{Fig_Theory_Eurocode_Strain}}.

The constitutive law of concrete in tension need not be defined in our approach. It is sufficient to describe the tensile strength, $f_t~\rm{[Pa]}$, that can be assumed as \citep[eq.~(34)]{DwaiKod2009}
\begin{equation}\label{tensile_strength}
f_t( \theta ) = f_{t,ref} \times \left\{
 \begin{array}{lll}
 1
 & {\rm{for}} & \theta \leq 373.15~{\rm{K}},
 \\
 (873.15 - \theta)/500
 & {\rm{for}} & 373.15~{\rm{K}} < \theta \leq 823.15~{\rm{K}},
 \\
 (1473.15 - \theta)/6500
 & {\rm{for}} & 823.15~{\rm{K}} < \theta \leq 1473.15~{\rm{K}},
 \\
 0
 & {\rm{for}} & \theta > 1473.15~{\rm{K}},
 \end{array} \right.
\end{equation}
where $f_{t,ref}~{\rm{[Pa]}}$ is the reference tensile strength of concrete at the room temperature.

If the reference tensile strength, $f_{t,ref}~{\rm{[Pa]}}$, is not known, it can be estimated from the reference compressive strength of concrete at the room temperature, $f_{c,ref}~{\rm{[MPa]}}$. In \citet[Table~3.1]{Eurocode2-1-1}, the tensile strength of high strength concrete is defined as
\begin{equation}\label{reference_tensile_strength}
f_{t,ref} = 2.12 \ln{\left(  1 + \frac{f_{c,ref}}{10}  \right) }   \quad {\rm{[MPa]}}.
\end{equation}

The Young's modulus of concrete, $E_c~{\rm{[Pa]}}$, can be easily derived from \eqref{concrete_constitutive_law} in the form \citep[see e.g.][eq.~(3.14)]{Fellinger2004}
\begin{equation}\label{concrete_modulus}
E_{c}(\theta) = \frac {3f_c(\theta)}{2\epsilon_{c1}(\theta)}.
\end{equation}

The Poisson's ratio of concrete, $\nu~{\rm{[-]}}$, is usually taken as temperature independent (see e.g. \citep[p.~46]{Alnajim2004}) since the data about its temperature dependency are ambiguous \citep[Section~2.2.1.2]{Naus2010}.
Here, we follow the assumption (which leads to conservative results) that the Poisson's ratio increases with increasing temperature. Based on the data experimentally determined by \citet[pp.~132--133]{Mindeguia2009} for high strength concrete ($f_{c,ref} \approx 60~\rm{MPa}$), we assume that
\begin{equation}\label{poisson}
\nu( \theta ) = \left\{
 \begin{array}{lll}
 0.2
 & {\rm{for}} & \theta \leq 293.15~{\rm{K}},
 \\
 0.2 + 0.5(\theta-293.15)/580
 & {\rm{for}} & 293.15~{\rm{K}} < \theta \leq 873.15~{\rm{K}},
 \\
 0.7
 & {\rm{for}} & \theta > 873.15~{\rm{K}}.
 \end{array} \right.
\end{equation}

\begin{figure}[h]
\centering
\begin{minipage}[t]{.47\textwidth}
  \centering
  {\includegraphics[angle=0,width=6.4cm]{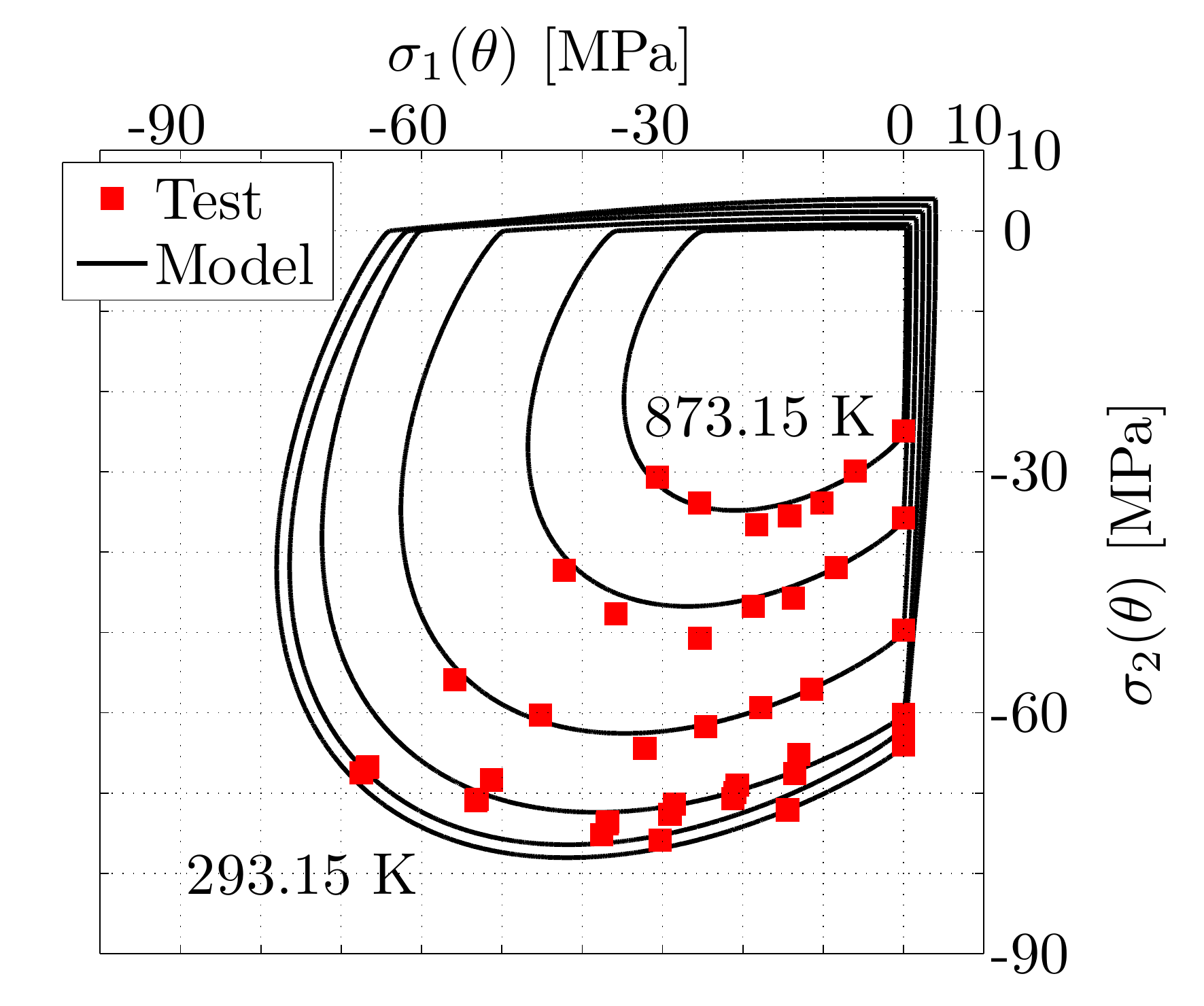}}
  \caption{Comparison of failure function \eqref{failure criterion} used in our model with the experimental data measured by \citet[Table~2]{HeSong2008}}
  \label{Fig_Theory_Failure_Function_Validation}
\end{minipage}%
\hspace{.01\textwidth}
\begin{minipage}[t]{.47\textwidth}
  \centering
  {\includegraphics[angle=0,width=6.4cm]{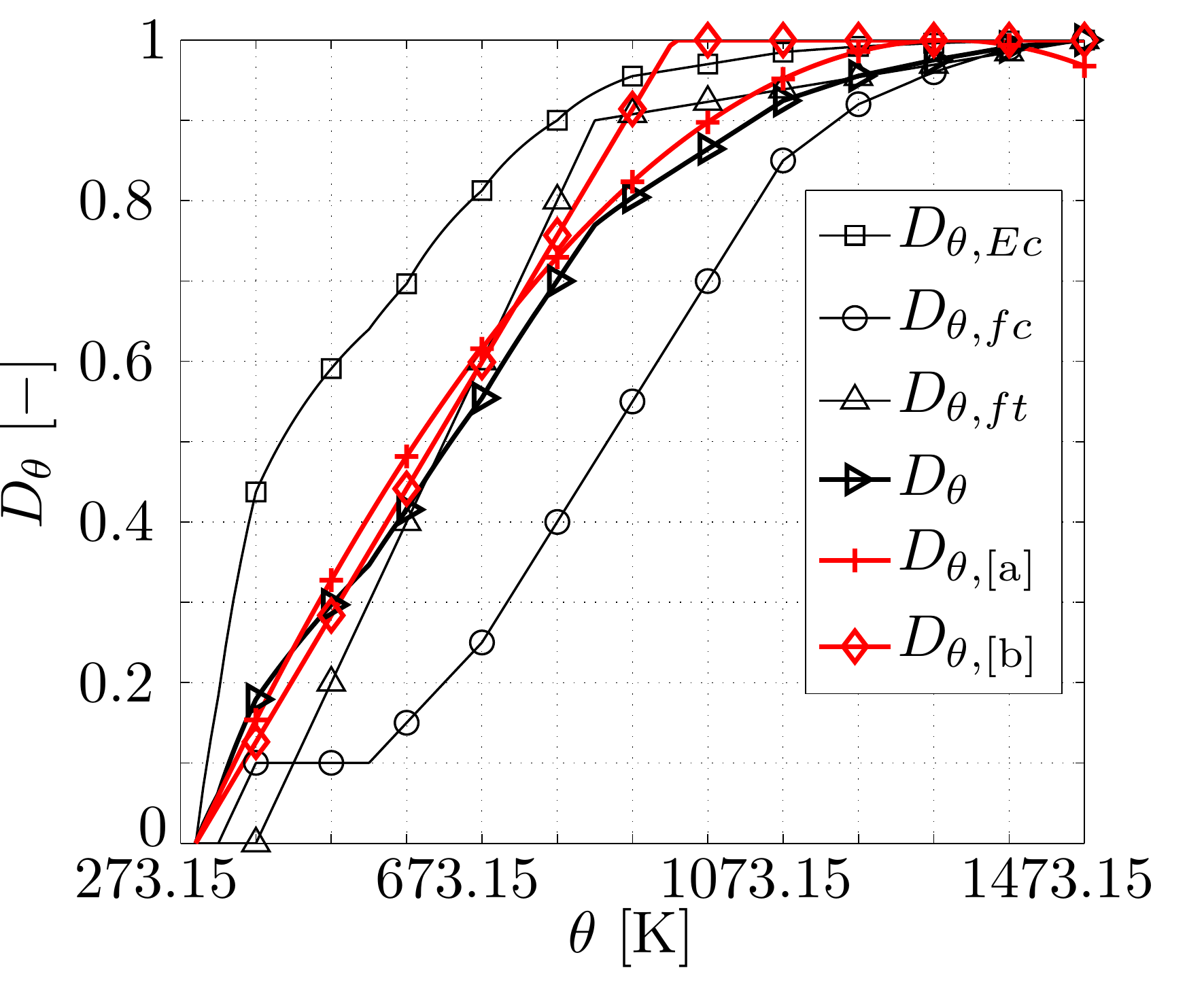}}
  \caption{Thermal damage assumed in our model ($D_{\theta}$) and its comparison with \citet[eq.~(30)]{Davie2010}, $D_{\theta,[\rm a]}$, and \citet[eq.~(52)]{Gawin2003}, $D_{\theta,[\rm  b]}$}
  \label{Fig_Theory_Thermal_Damage}
\end{minipage}
\end{figure}

The eccentricity, $e_F~{\rm{[-]}}$, used in the failure function of concrete (Section~\ref{Section_Failure_Function}) generally depends on the type of concrete and probably also on temperature. Here, we assume constant (temperature independent) value, $e_F = 0.505$, derived from the data measured by \citet[Table~2]{HeSong2008} for high strength concrete ($f_{c,ref} \approx 60~\rm{MPa}$) under biaxial compression tests at high temperatures, see Figure~\ref{Fig_Theory_Failure_Function_Validation}, where $f_c(\theta)$ is taken from \citet[Table~2]{HeSong2008}, $f_t(\theta)$ is assumed according to \eqref{tensile_strength}, with $f_{t,ref} = 4~{\rm{MPa}}$, and $F(\theta)$ is determined by \eqref{failure criterion}, with $\sigma_3 = 0$.

The porosity of concrete, $\phi~\rm{[-]}$, may be expressed as \citep[eq.~(41)]{Gawin1999}
\begin{equation}\label{GawinPorosity}
\phi(\theta) = \phi_{ref} + A_{\phi}(\theta - \theta_{ref}),
\end{equation}
where $\phi_{ref}~{\rm{[-]}}$ is the reference porosity (at the reference temperature, $\theta_{ref}~\rm{[K]}$), and $A_{\phi}~{\rm{[K^{-1}]}}$ is a concrete-type-depend constant (see \citep[pp.~46--47]{Gawin1999}).

The thermal conductivity of concrete, $\lambda_c~\rm{[W\,m^{-1}\,K^{-1}]}$, is given by \citep[eqs.~(46--47)]{Gawin1999}
\begin{equation}
\lambda_c(P,\theta) = \lambda_d(\theta) \left( 1 + \frac{4\,\phi(\theta)\,\rho_w(\theta)\,S_w(P,\theta)}{(1-\phi(\theta))\rho_s}   \right),
\end{equation}
with

\begin{equation} \label{ConductivityDry}
\lambda_d(\theta) = \lambda_{d,ref} \left[ 1 + A_{\lambda} (\theta - \theta_{ref}) \right],
\end{equation}
where $\lambda_{d,ref}~{\rm{[W\,m^{-1}\,K^{-1}]}}$ is the reference thermal conductivity of a dry concrete (at the reference temperature, $\theta_{ref}~{\rm{[K]}}$), and $A_{\lambda}~\rm{[K^{-1}]}$ is an experimentally determined coefficient.

For the intrinsic permeability of concrete, $K~\rm{[m^2]}$, we adopt the Bary function recommended by \citet[eq.~(13)]{Davie2012}
\begin{equation}\label{permeabiliy}
K(P,\theta) = K_{ref} \times 10^{4 D(P,\theta)},
\end{equation}
where $K_{ref}~{\rm{[m^2]}}$ is the reference permeability at the room temperature, and $D~{\rm{[-]}}$ is the multiplicative damage parameter that can be defined as (\citep[eq.~(70)]{Davie2010}; \citep[eq.~(54)]{Gawin2003})
\begin{equation}\label{damage}
D(P,\theta) = D_{m}(P,\theta) + D_{\theta}(\theta) - D_{m}(P,\theta) D_{\theta}(\theta),
\end{equation}
where $D_{m}~{\rm{[-]}}$ is the mechanical damage parameter, which is assumed to be equal to the failure parameter \eqref{failure criterion_simplified} in our approach, i.e.
\begin{equation}\label{damage_mechanical}
D_{m}(P,\theta) = F(P,\theta),
\end{equation}
and $D_{\theta}~{\rm{[-]}}$ is the thermal damage parameter that we define as (cf. \citep[eq.~(52)]{Gawin2003})
\begin{equation}\label{damage_therma}
D_{\theta}(\theta)
=
1 - \frac{1}{3} \left( D_{\theta,Ec} + D_{\theta,fc} + D_{\theta,ft} \right)
=
1 - \frac{1}{3} \left( \frac{E_c(\theta)}{E_{c,ref}} + \frac{f_{c}(\theta)}{f_{c,ref}} + \frac{f_{t}(\theta)}{f_{t,ref}} \right).
\end{equation}

It should be noted that usually, the thermal damage parameter is based only on the degradation of the Young's modulus (see \citep[eqs.~(29--30)]{Davie2010}; \citep[eq.~(52)]{Gawin2003}). This is however not possible to assume in our approach since we employ the constitutive model proposed by \citet{Eurocode2-1-2}, see \eqref{concrete_constitutive_law}, where the stress is expressed as a function of the total mechanical strain, $\epsilon_m$, instead of the instantaneous stress-related strain, $\epsilon_{\sigma}$, and hence the resulting softening of the material with increasing temperature is higher than in reality (see also \citep{Anderberg2008,GernayFransen2010}). In order to eliminate this effect, we propose to include not only the temperature dependent degradation of $E_c$ but also the reduction of the other mechanical properties ($f_c$, $f_t$), which leads to good agreement of the resulting thermal damage parameter with the data stated in literature, see Figure~\ref{Fig_Theory_Thermal_Damage}.

The gas relative permeability and the liquid water relative permeability, $K_{rg}~{\rm{[-]}}$ and $K_{rw}~{\rm{[-]}}$, respectively, can be expressed by the formulas experimentally determined by \citet[eqs.~(7--9)]{ChungCon2005}
\begin{equation}
K_{rg}(P,\theta) = 10^{S_w(P,\theta)\,\psi(\theta)} - 10^{\psi(\theta)}\,S_w(P,\theta),
\end{equation}
\begin{equation}
K_{rw}(P,\theta) = 10^{(1-S_w(P,\theta))\,\psi(\theta)} - 10^{\psi(\theta)}\,(1-S_w(P,\theta)),
\end{equation}
with
\begin{equation}
\psi(\theta) = 0.05 - 22.5\,\phi(\theta).
\end{equation}

The gas dynamic viscosity, $\mu_{g}~{\rm{[Pa\,s]}}$, is approximated as \citep[eq.~(50)]{Gawin1999}
\begin{equation}
\mu_{g}(P,\theta) = \mu_{gv}(\theta) + [\mu_{ga}(\theta) - \mu_{gv}(\theta)] \left( \frac{P_a}{P + P_a} \right) ^ {0.608},
\end{equation}
with
\begin{equation}
\mu_{gv}(\theta) = \mu_{gv,ref} + \alpha_v (\theta - \theta_{ref})
\end{equation}
and
\begin{equation}
\mu_{ga}(\theta) = \mu_{ga,ref} + \alpha_a (\theta - \theta_{ref}) + \beta_a (\theta - \theta_{ref})^2,
\end{equation}
where $\mu_{gv,ref} =  8.85 \times 10^{-6}~\rm{Pa\,s}$, $\alpha_v = 3.53 \times 10^{-8}~\rm{Pa\,s\,K^{-1}}$,
      $\mu_{ga,ref} = 17.17 \times 10^{-6}~\rm{Pa\,s}$, $\alpha_a = 4.73 \times 10^{-8}~\rm{Pa\,s\,K^{-1}}$, $\beta_a = 2.22 \times 10^{-11}~\rm{Pa\,s\,K^{-2}}$,
      and $\theta_{ref} = 273.15~\rm{K}$.

The liquid water dynamic viscosity, $\mu_{w}~{\rm{[Pa\,s]}}$, can be calculated from \citep[eq.~(51)]{Gawin1999}
\begin{equation}
\mu_{w}(\theta) = 0.6612 (\theta - 229) ^{-1.532}.
\end{equation}

The volume fraction of liquid (all free) water, $\eta_{w}~\rm{[-]}$, is governed by the sorption isotherm function proposed in \citep{BaCheTh1981,BaTh1978,BaTh1979}.
After some modification related to the saturated region, $P/P_s \geq 1.0$ (see \citep[eq.~(73)]{Davie2010}), we can write
\begin{equation}\label{sorptionisotherms}
\begin{array}{l}

{\rm{if}}~\theta \leq \theta_{cr}:

\\

\quad \eta_{w}(P,\theta) = \left\{ \begin{array}{lll}
\displaystyle \frac{c}{\rho_{w}(\theta)} \left( \frac {\phi(\theta_{ref}) \rho_{w}(\theta_{ref})} {c} \frac {P} {P_s(\theta)} \right) ^ {1/m(\theta)}
& \rm{for} & \displaystyle \frac {P} {P_s(\theta)} \leq 0.96,
\\
\displaystyle \sum\limits_{i=0}^{3}\xi_i(\theta) \left(\frac{P}{P_s(\theta)}-0.96\right)^i
& \rm{for} & \displaystyle 0.96 < \frac {P} {P_s(\theta)} < 1.00,
\\
\displaystyle \phi(\theta)
& \rm{for} & \displaystyle \frac {P} {P_s(\theta)} \geq 1.00,
\end{array} \right.

\\

{\rm{if}}~\theta > \theta_{cr}:

\\

\quad \eta_{w} = 0,

\end{array}
\end{equation}
with
\begin{equation}
m(\theta) = 1.04 - \frac {(\theta - 263.15)^2} {22.34(\theta_{ref}-263.15)^2+(\theta-263.15)^2},
\end{equation}
where $c~{\rm{[kg\,m^{-3}]}}$ is the mass of cement per unit volume of concrete, $\theta_{ref}~{\rm{[K]}}$ is the reference temperature ($\theta_{ref}=298.15~{\rm{K}}$), the term $[\phi(\theta_{ref}) \rho_{w}(\theta_{ref})]$ expresses the saturation water content at $298.15~{\rm{K}}$, and $\xi_i (\theta)$ are temperature dependent coefficients that ensure a smooth transition between the unsaturated and saturated region, thus the sorption isotherm function and its first derivative with respect to $(P/P_s(\theta))$ are continuous (cf. \citep{Davie2006,Davie2010,Majumdar1995,TenLiPur2001}). Therefore, we can write
\begin{eqnarray}
\xi_0(\theta) & = & \eta_{w,0.96}(\theta),
\\
\xi_1(\theta) & = & \eta'_{w,0.96}(\theta),
\\
\xi_2(\theta) & = & \frac{3(\eta_{w,1.00}(\theta)-\eta_{w,0.96}(\theta))}{0.04^2} - \frac{2\eta'_{w,0.96}(\theta)+\eta'_{w,1.00}(\theta)}{0.04},
\\
\xi_3(\theta) & = & \frac{2(\eta_{w,0.96}(\theta)-\eta_{w,1.00}(\theta))}{0.04^3} + \frac{\eta'_{w,0.96}(\theta)+\eta'_{w,1.00}(\theta)}{0.04^2},
\end{eqnarray}
where
\begin{equation}
\begin{array}{lcllcl}
\eta_{w,0.96}(\theta) & = &
\displaystyle \frac{c}{\rho_{w}(\theta)} \left( \frac {\phi(\theta_{ref}) \rho_{w}(\theta_{ref})} {c} 0.96 \right) ^ {1/m(\theta)},
\\
\eta_{w,1.00}(\theta) & = &
\displaystyle \phi(\theta),
\end{array}
\end{equation}
and
\begin{equation}
\eta'_{w,0.96}(\theta)  =
\displaystyle \frac{{\rm d}\eta_{w,0.96}(\theta)}{{\rm d}{\theta}}
\quad \textmd{ and } \quad
\eta'_{w,1.00}(\theta)  =
\displaystyle \frac{{\rm d}\eta_{w,1.00}(\theta)}{{\rm d}{\theta}}.
\end{equation}

The degree of saturation with liquid water, $S_w~\rm{[-]}$, is defined as
\begin{equation}\label{saturation}
S_w(P,\theta) = \frac{\eta_{w}(P,\theta)}{\phi(\theta)}.
\end{equation}

The mass of dehydrated water, $m_d~\rm{[kg\,m^{-3}]}$, is governed by equation \eqref{dehydration},
with (\citep[eq.~(8)]{Feraille-Fresnet:2003}; \citep[eq.~(C-22)]{PontEhrlacher2004}; \citep[eq.~(1.9)]{Alnajim2004})
\begin{eqnarray}\label{dehydration_eq}
m_{d,eq}(\theta) = & \displaystyle \frac{7.5}{100}m_{eq}^{378.15}
                     \left[ 1 - \exp {\left( - \frac{\theta - 378.15}{200}\right)} \right]
                     H(\theta - 378.15) \nonumber \\
                   & \displaystyle  + \frac{2}{100}m_{eq}^{378.15}
                     \left[ 1 - \exp {\left( - \frac{\theta - 673.15}{10}\right)} \right]
                     H(\theta - 673.15) \nonumber\\
                   & \displaystyle  + \frac{1.5}{100}m_{eq}^{378.15}
                     \left[ 1 - \exp {\left( - \frac{\theta - 813.15}{5}\right)} \right]
                     H(\theta - 813.15) ,
\end{eqnarray}
where $m_{eq}^{378.15}~\rm{[kg\,m^{-3}]}$ is the equilibrium mass at $378.15~\rm{K}$ and $H$ is the Heaviside function.

In \citep[p.~146]{PontEhrlacher2004}, the following values of material parameters that belong to equations \eqref{dehydration} and \eqref{dehydration_eq} are stated: $\tau = 10800~\rm{s}$, $m_{eq}^{378.15} = 210~\rm{kg\,m^{-3}}$, which are also adopted in our simulations, since these parameters are usually not measured within the fire tests.

The {density of solid skeleton}, $\rho_s~\rm{[kg\,m^{-3}]}$, is, together with the concrete porosity end the mass of water released into the pores by dehydration, governed by the solid mass conservation equation \eqref{balance_solid} (cf. \citep[pp.~47--48]{Gawin1999}). In our approach, the density of solid skeleton is assumed to be a constant value. As obvious from numerical experiments, this simplification has negligible effect on the obtained results.

The {density of liquid water}, $\rho_w~\rm{[kg\,m^{-3}]}$, can be expressed by the experimentally determined formula, originally proposed by \citet[eq.~(4.79)]{Furbish1997} and simplified by \citet[eq.~(35)]{Gawin2002} (neglecting the water pressere dependence of $\rho_w$), of the form
\begin{equation}
\label{WaterDensity}
\rho_w(\theta) = \sum\limits_{i=0}^{5} a_i(\theta - 273.15)^i \left( - 1 \times 10^7 \right)
+ \sum\limits_{i=0}^{5} b_i(\theta - 273.15)^i \quad \rm{for} \quad \theta \leq \theta_{cr},
\end{equation}
where $a_0 = 4.8863 \times 10^{-7}, a_1 = -1.6528 \times 10^{-9}, a_2 = 1.8621 \times 10^{-12}, a_3 = 2.4266 \times 10^{-13}, a_4 = -1.5996 \times 10^{-15}, a_5 = 3.3703 \times 10^{-18}, b_0 = 1.0213 \times 10^{3}, b_1 = -7.7377 \times 10^{-1}, b_2 = 8.7696 \times 10^{-3}, b_3 = -9.2118 \times 10^{-5}, b_4 = 3.3534 \times 10^{-7}, b_5 = -4.4034 \times 10^{-10}$.

The {{specific enthalpy of evaporation}}, $h_e~\rm{[J\,kg^{-1}]}$, can be expressed by the Watson formula (\citep[eq.~(49)]{Gawin1999}; \citep[eq.~(28)]{Gawin2011a}; \citep[eq.~(AI.27)]{Davie2006})
\begin{equation}
h_e(\theta) = \left\{ \begin{array}{lll}
2.672 \times 10^5 (\theta_{cr} - \theta)^{0.38} & \rm{for} & \theta \leq \theta_{cr},
\\
0 & \rm{for} & \theta > \theta_{cr}.
\end{array} \right.
\end{equation}

The {{specific enthalpy of dehydration}}, $h_d~\rm{[J\,kg^{-1}]}$, is considered to be a constant value \citep[eq.~(AI.26)]{Davie2006}
\begin{equation}
h_d = 2400 \times 10^3 ~ \rm{J\,kg^{-1}}.
\end{equation}

The {{specific heat capacity of dry air}}, $c_p^{a}~\rm{[J\,kg^{-1}\,K^{-1}]}$, is determined by \citet[eq.~(AI.28)]{Davie2006}
\begin{equation}
c_p^{a}(\theta) = a\,\theta^3 + b\,\theta^2 + c\,\theta + d,
\end{equation}
where $a = - 9.84936701814735 \times 10^{-8}$, $b = 3.56436257769861 \times 10^{-4}$, $c = - 1.21617923987757 \times 10^{-1}$, and  $d = 1.01250255216324 \times 10^{3}$.

The {{specific heat capacity of liquid water}}, $c_p^{w}~\rm{[J\,kg^{-1}\,K^{-1}]}$, is given by \citep[eq.~(AI.29)]{Davie2006}
\begin{equation}
c_p^{w}(\theta) = (2.4768\,\theta + 3368.2) + \left( \frac{a\,\theta}{513.15} \right) ^ b \quad \rm{for} \quad \theta \leq \theta_{cr},
\end{equation}
where $a = 1.08542631988638$ and $b = 31.4447657616636$.

The {{specific heat capacity of solid skeleton}}, $c_p^{s}~\rm{[J\,kg^{-1}\,K^{-1}]}$, can be calculated from \citep[eq.~(AI.30)]{Davie2006}
\begin{equation}
c_p^{s}(\theta) = 900 + 80 \frac{\theta - 273.15}{120} - 4 \left( \frac{\theta - 273.15}{120} \right) ^2.
\end{equation}

For the {{specific heat capacity of water vapour}}, $c_p^{v}~\rm{[J\,kg^{-1}\,K^{-1}]}$, the following equation is proposed by \citet[eq.~(AI.31)]{Davie2006}
\begin{equation}
c_p^{v}(\theta) = \left\{ \begin{array}{lll}
\displaystyle (7.1399\,\theta - 443) + \left( \frac{a\,\theta}{513.15} \right) ^b & \rm{for} & \theta \leq \theta_{cr},
\\
45821.01 & \rm{for} & \theta > \theta_{cr},
\end{array} \right.
\end{equation}
where $a = 1.13771502228162$ and $b = 29.4435287521143$.

It should be noted that above the critical temperature (if $\theta > \theta_{cr}$), there is no liquid water in concrete (see equation \eqref{sorptionisotherms}) and hence, the liquid water properties ($\rho_w$, $c_p^{w}$) need not be defined.

\section{Numerical results and experiments}
\label{Sec_Examples}
In order to validate the model presented in this paper, the results obtained by numerical simulations are compared with the experimentally determined data. Three experiments reported in literature are closely investigated:
\begin{itemize}
\item {the test of the hygro-thermal behaviour of high strength concrete ($f_{c,ref} \approx 90~\rm{MPa}$) prismatic specimen $300 \times 300 \times 120~\rm{mm^3}$ under unidirectional heating by the radiant heater of a temperature of $600~\rm{^{\circ}C}$ reported by \citet{Kalifa2000} and related publications \citet{Kalifa2001,Kalifa1998};}
\item {the test of the hygro-thermal behaviour of high strength concrete ($f_{c,ref} \approx 60~\rm{MPa}$) prismatic specimen $300 \times 300 \times 120~\rm{mm^3}$ under unidirectional heating by the radiant heater of a temperature of $600~\rm{^{\circ}C}$ reported by \citet{Mindeguia2009} and related publications \citet{Mindeguia2009a,Mindeguia2010};}
\item {the test of the spalling behaviour of high strength concrete ($f_{c,ref} \approx 60~\rm{MPa}$) prismatic specimen $700 \times 600 \times 150~\rm{mm^3}$ under unidirectional heating by the ISO~834 fire reported by \citet{Mindeguia2009} and related publications \citet{Mindeguia2009a,Mindeguia2013}.}
\end{itemize}

Within the first two of the above tests, the pore pressure and the temperature propagation through the specimen as well as its mass loss were recorded. This type of tests is therefore denoted as the "PTM test" by \citet{Mindeguia2009}, which is also adopted in this paper. It should be noted that no spalling was observed during the both PTM experiments and hence, these tests are employed primarily for the validation of the hygro-thermal behaviour simulation while the spalling prediction can be validated only on the qualitative level (whether it occurs or not). For that reason, the third test is also analysed, since it enables to validate our model also with respect to the quantitative prediction of spalling (i.e. the prediction of the amount of the potential spalling).

{
It should be noted that the usage of the model can provide only an approximate picture of the experiments described above. The inaccuracies may arise mainly from the fact that:
\begin{itemize}
\item within the mechanical part of the model, a simplified constitutive law of concrete given by \citet{Eurocode2-1-2} is utilized. In this model, the transient strain is included implicitly and the creep strain is neglected (see Section~\ref{Section_Thermo_mechanical_stress});
\item in the simulation, the specimen is assumed to be fully mechanically restrained in the plane perpendicular to its heated surface. This assumption is however disputable in this cases. It is not obvious, whether the ceramic blocks placed on the lateral sides of the small specimens within the PTM tests (see Fig.~\ref{Fig_Example1_Test_Setup}) or the unheated parts of the large specimens within the spalling test (see \citep[Figure~170]{Mindeguia2009}) provide a sufficient level of restraint as assumed in the model (probably not);
\item the model is not able to capture the "size effect" influencing the spalling behaviour of concrete specimens - i.e. that for small specimens, the spalling is less likely to occur in comparison with large specimens, as observed by recent experimental works \citep{Jansson2013,Mindeguia2009a}.
\end{itemize}
}

\subsection{Simulation of the PTM test 1}
\label{Sec_Example1}
The Kalifa's experiments \citep{Kalifa2001,Kalifa2000,Kalifa1998} have been accepted as a benchmark problem by many authors focused on modelling of concrete
subjected to high temperatures
(cf.~\citep[Section~III-3]{Alnajim2004};
    \citep[Section~3.2]{Davie2010};
    \citep[Section~5]{DwaiKod2009};
    \citep[Section~6.1]{Gawin2012};
    \citep[Section~6.2]{Gawin2002};
    \citep[Sections~4.3, 5.4]{Kukla2010};
    \citep[Section~6.4]{Pesavento2000};
    \citep[Section~3.1]{Witek2007}).

In \citep{Kalifa1998}, we can find the high temperature thermal and hygral properties measured by \citet{Kalifa1998} for 4 types of concrete named as M30, M75C, M75SC and M100C. The M30 and M100 concretes were used for the subsequent PTM tests reported by \citet{Kalifa2000}. Within this experiment, the specimens were subjected to an unidirectional heating up to $450~\rm{^{\circ}C}$, $600~\rm{^{\circ}C}$, and $800~\rm{^{\circ}C}$ using a radiant heater. Moreover, similar experiment -- in this case for the specimens made of C110 concrete (practically the same as the M100 concrete, only with a little bit higher amount of cement and superplasticizer \citep{Kalifa2001}) with polypropylene fibres of the content of (0--3) $\rm{kg\,m^{-3}}$, were made by \citet{Kalifa2001} in order to assess the effect of fibres on the high-temperature behaviour of high strength concrete.

For the validation of our model, the results measured by \citet{Kalifa2000} for the M100 concrete heated up to $600~\rm{^{\circ}C}$ are employed. This case is chosen for the following reasons: our model is primarily designed to study the behaviour of high strength concrete (such as M100); material properties of the M100 concrete are described in detail in \citep{Kalifa1998} (more comprehensive description than for the C110 concrete in \citep{Kalifa2001}); and the results obtained for the other heating conditions (up to $450~\rm{^{\circ}C}$ or $800~\rm{^{\circ}C}$) are not sufficiently reported in \citep{Kalifa2000}.

\subsubsection{Experiment description}
\label{Sec_Example1_Experiment}
The tests performed by Kalifa et al. are described in detail in \citep{Kalifa2000}. Therefore, only a brief description of the experiment is stated below.

As mentioned above, two types of concrete (M30 and M100) and three different heating conditions ($450~\rm{^{\circ}C}$, $600~\rm{^{\circ}C}$, and $800~\rm{^{\circ}C}$) were used by \citet{Kalifa2000}. Hereafter, only the case of M100 -- $600~\rm{^{\circ}C}$ is assumed. Material (thermal and hygral) properties of the M100 concrete have been reported in \citep{Kalifa1998} and are discussed in Section~\ref{Sec_Example1_Modelling}.

\begin{figure}[h]
  \centering
  {\includegraphics[angle=0,width=7.5cm]{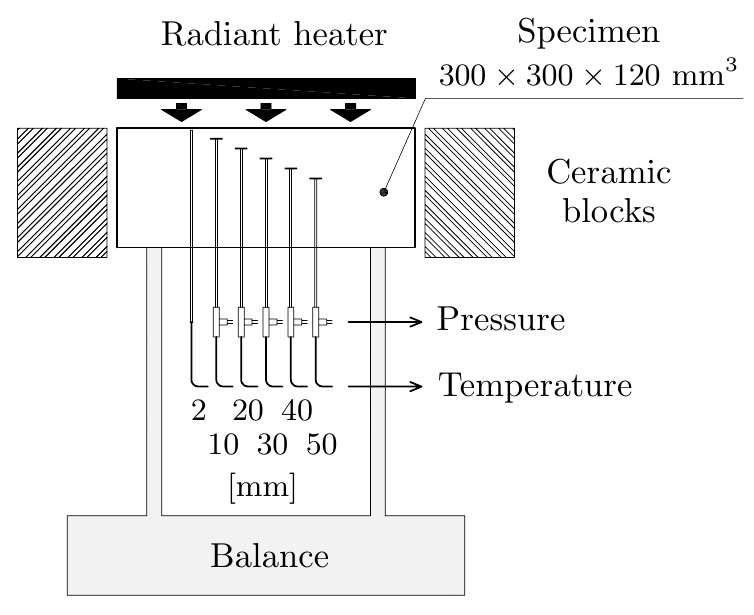}}
  \caption{Scheme of the test set-up (according to \citet[Figure~2]{Kalifa2000})}
  \label{Fig_Example1_Test_Setup}
\end{figure}

As shown in Figure~\ref{Fig_Example1_Test_Setup}, the $300 \times 300 \ 120~\rm{mm^3}$ test specimen was subjected to an unidirectional heating (in the 120-mm direction) by a radiant heater of the temperature of $600~\rm{^{\circ}C}$ placed near its surface for a heating period of 6 hours.

Within the test, the temperature, the pore pressure, and the mass loss of the specimen were recorded. The temperature together with the pore pressure were measured using the combined pressure-temperature gauges (see \citep[Figure~3]{Kalifa2001}) placed at the distance of (10, 20, 30, 40, and 50) mm from the heated surface. The heated surface temperature was measured by a thermocouple placed at the depth of 2~mm. Moreover, as obvious
from the graphs listed in \citep{Kalifa2000}, the temperature on the unexposed side was also recorded (probably by an additional thermocouple or by a~surface thermometer). The mass loss was monitored by a balance on which the specimen was placed during the test (see Figure~\ref{Fig_Example1_Test_Setup}).

\subsubsection{Modelling}
\label{Sec_Example1_Modelling}
In this section, the material properties of concrete, the initial and boundary conditions as well as the discretization employed for the numerical modelling are discussed. The thermal and hygral properties of concrete are assumed according to the relationships given in Section~\ref{Sec_MaterialProperties}, with the parameters determined from the data measured by \citet{Kalifa1998}. All the parameters are summarized in Table~\ref{Tab_Example1_Parameters}.

\begin{table}[h]
\centering
\caption{Material properties and parameters used in our simulation of the PTM test 1}
\small
\begin{tabular}{l l l l}
\hline
Parameter      & Value                   & Unit          & Reference \\
\hline\hline
\multirow{2}{*}{Type of concrete} & \multirow{2}{*}{HSC2-C} & \multirow{2}{*}{$-$} & \citet{Eurocode2-1-2};\\
               &                         &               & \citet{Kalifa1998} \\
\hline
$f_{c,ref}$    & $91.8$                  & $\rm{MPa}$    & \citet[Table~1]{Kalifa1998} \\
\hline
$f_{t,ref}$    & $4.9$                   & $\rm{MPa}$    & \citet{Eurocode2-1-1}, see \eqref{reference_tensile_strength} \\
\hline
$c$  & $414.8$ & $\rm{kg\,m^{-3}}$       & \citet[Table~1]{Kalifa1998} \\
\hline
$\theta_{ref}$ & $293.15$                & $\rm{K}$      & determined from \\
$\phi_{ref}$   & $0.0897$                & $-$           & \citet[Table~2]{Kalifa1998}, \\
$A_{\phi}$     & $2.4457 \times 10^{-5}$ & $\rm{K^{-1}}$ & see Figure~\ref{Fig_Example1_Porosity} \\
\hline
\multirow{3}{*}{$\rho_s $} & \multirow{3}{*}{$2660$} & \multirow{3}{*}{$\rm{kg\,m^{-3}}$} & determined from \\
               &                         &               & \citet[Table~4]{Kalifa1998}, \\
               &                         &               & see Figure~\ref{Fig_Example1_Density} \\
\hline
$\theta_{ref}$ & $293.15$                & $\rm{K}$      & determined from \\
$\lambda_{d,ref}$ & $1.9759$             & $\rm{W\,m^{-1}\,K^{-1}}$ & \citet[Table~7]{Kalifa1998}, \\
$A_{\lambda}$  & $- 6.4215 \times 10^{-4}$ & $\rm{K^{-1}}$ & see Figure~\ref{Fig_Example1_Conductivity} \\
\hline
$K_{ref}$      & $1.3 \times 10^{-20}$   & $\rm{m^{2}}$  & $-$ \\
\hline
\end{tabular}
\label{Tab_Example1_Parameters}
\end{table}

The concrete of the test specimens (named as M100 concrete in \citep{Kalifa2000,Kalifa1998}) is assumed to be a high strength concrete, class 2 (see \citet[Section~6.1]{Eurocode2-1-2} and Section~\ref{Sec_MaterialProperties}), with calcareous aggregates, which can be denoted as "HSC2-C". This classification determines the free thermal strain of concrete and the temperature dependent reduction of its compressive strength according to \citet{Eurocode2-1-2}, as described in Section~\ref{Sec_MaterialProperties}.

The reference compressive strength of concrete at the room temperature is taken from \citet[Table~1]{Kalifa1998}. The reference tensile strength of concrete at the room temperature is not specified in \citep{Kalifa2000,Kalifa1998} and hence, it is calculated by \eqref{reference_tensile_strength}.

The mass of cement per unit volume of concrete is taken from \citet[Table~1]{Kalifa1998}.

\begin{figure}[h]
\centering
\begin{minipage}[t]{.47\textwidth}
  \centering
  {\includegraphics[angle=0,width=6.4cm]{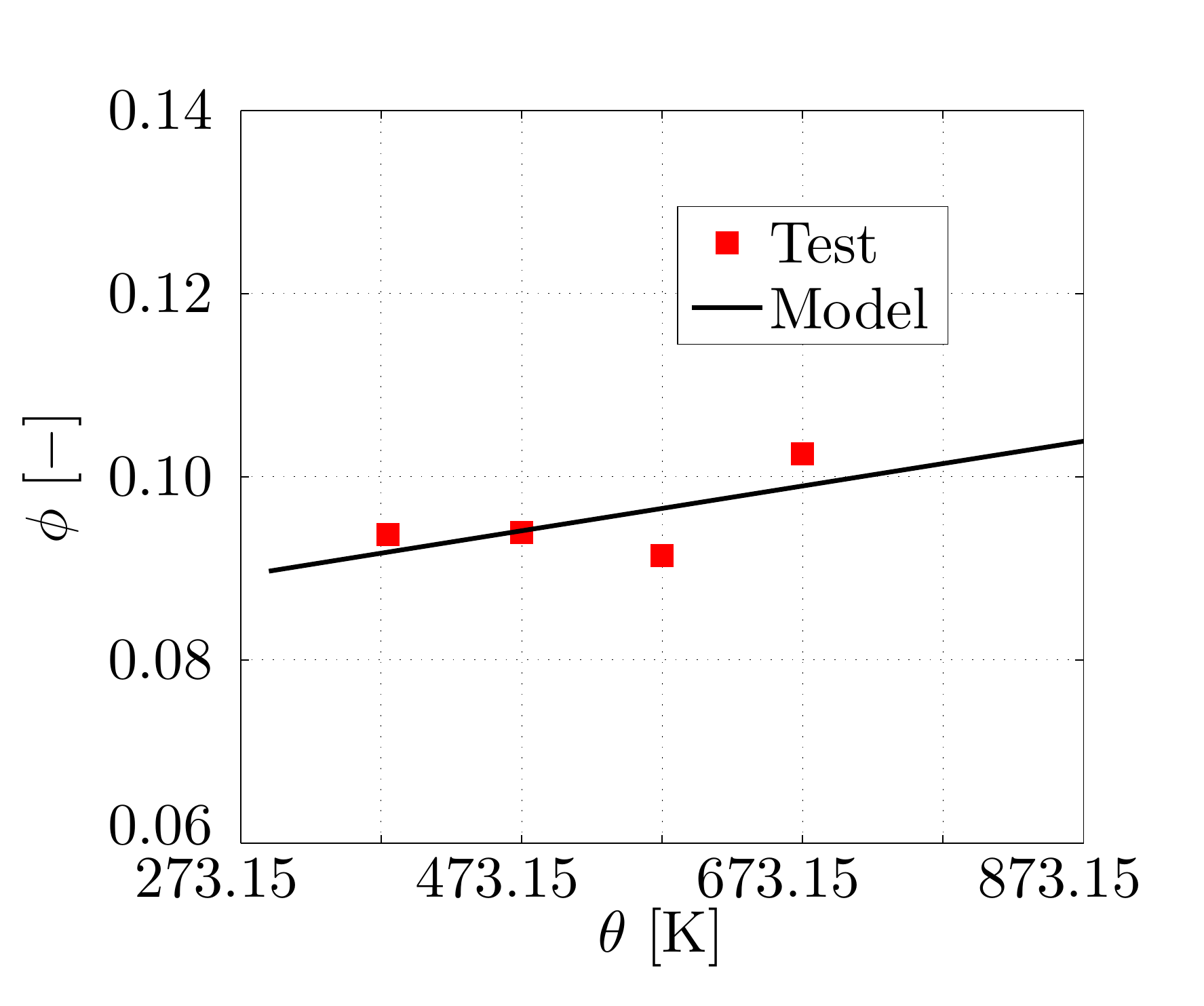}}
  \caption{Porosity of concrete measured by \citet{Kalifa1998} (points) and assumed in our model (line)}
  \label{Fig_Example1_Porosity}
\end{minipage}%
\hspace{.01\textwidth}
\begin{minipage}[t]{.47\textwidth}
  \centering
  {\includegraphics[angle=0,width=6.4cm]{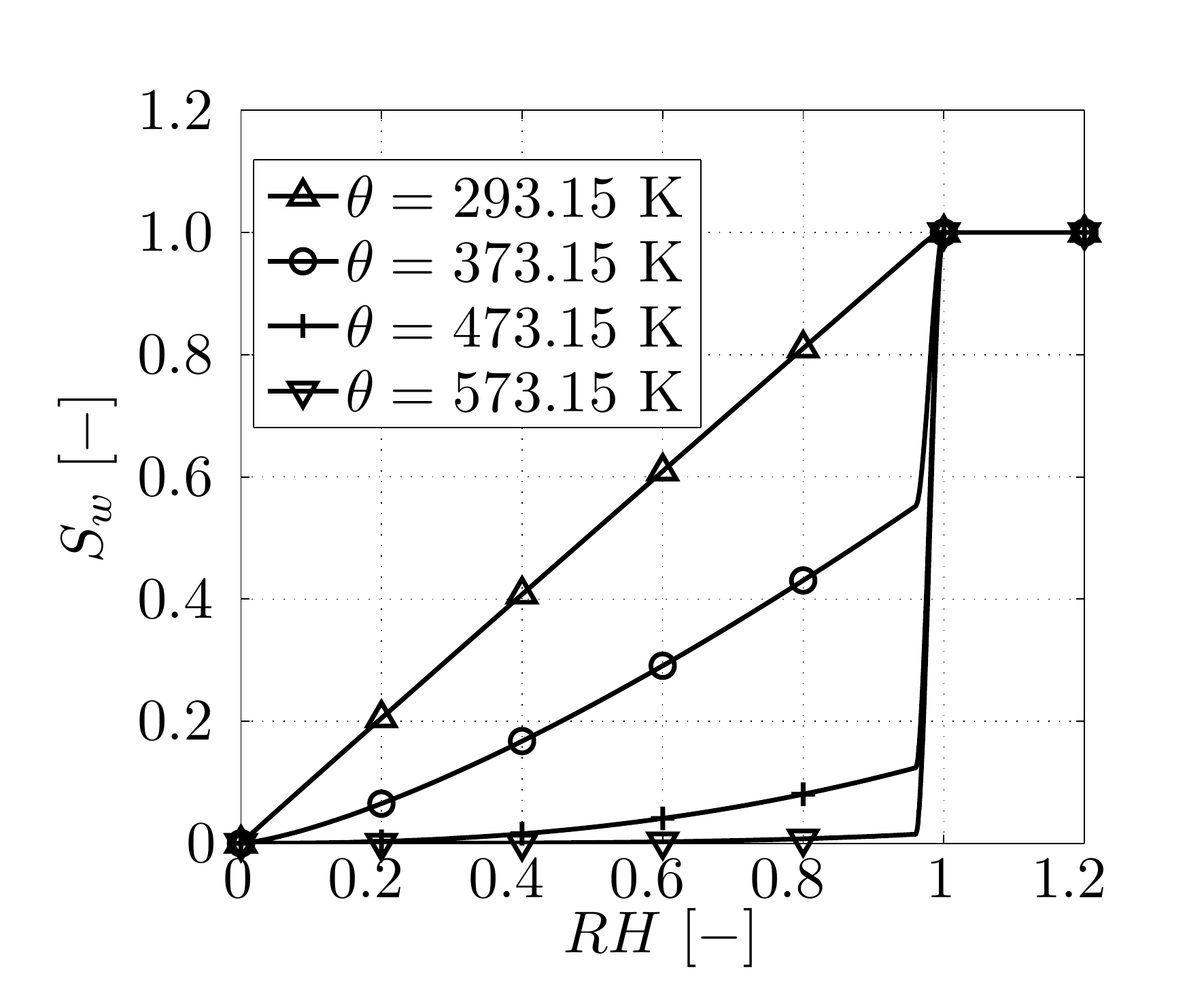}}
  \caption{Degree of saturation with liquid water assumed in our model}
  \label{Fig_Example1_Saturation}
\end{minipage}
\end{figure}

For the porosity of concrete, the parameters of equation \eqref{GawinPorosity} are determined by a linear regression of the data stated in \citep[Table~2]{Kalifa1998}, see Figure~\ref{Fig_Example1_Porosity}.

Assuming the above values of $c$ and $\phi(\theta)$, the resulting degree of saturation with liquid water \eqref{saturation}, based on the sorption isotherms \eqref{sorptionisotherms}, is illustrated in
Figure~\ref{Fig_Example1_Saturation}.

\begin{figure}[h]
\centering
\begin{minipage}[t]{.47\textwidth}
  \centering
  {\includegraphics[angle=0,width=6.4cm]{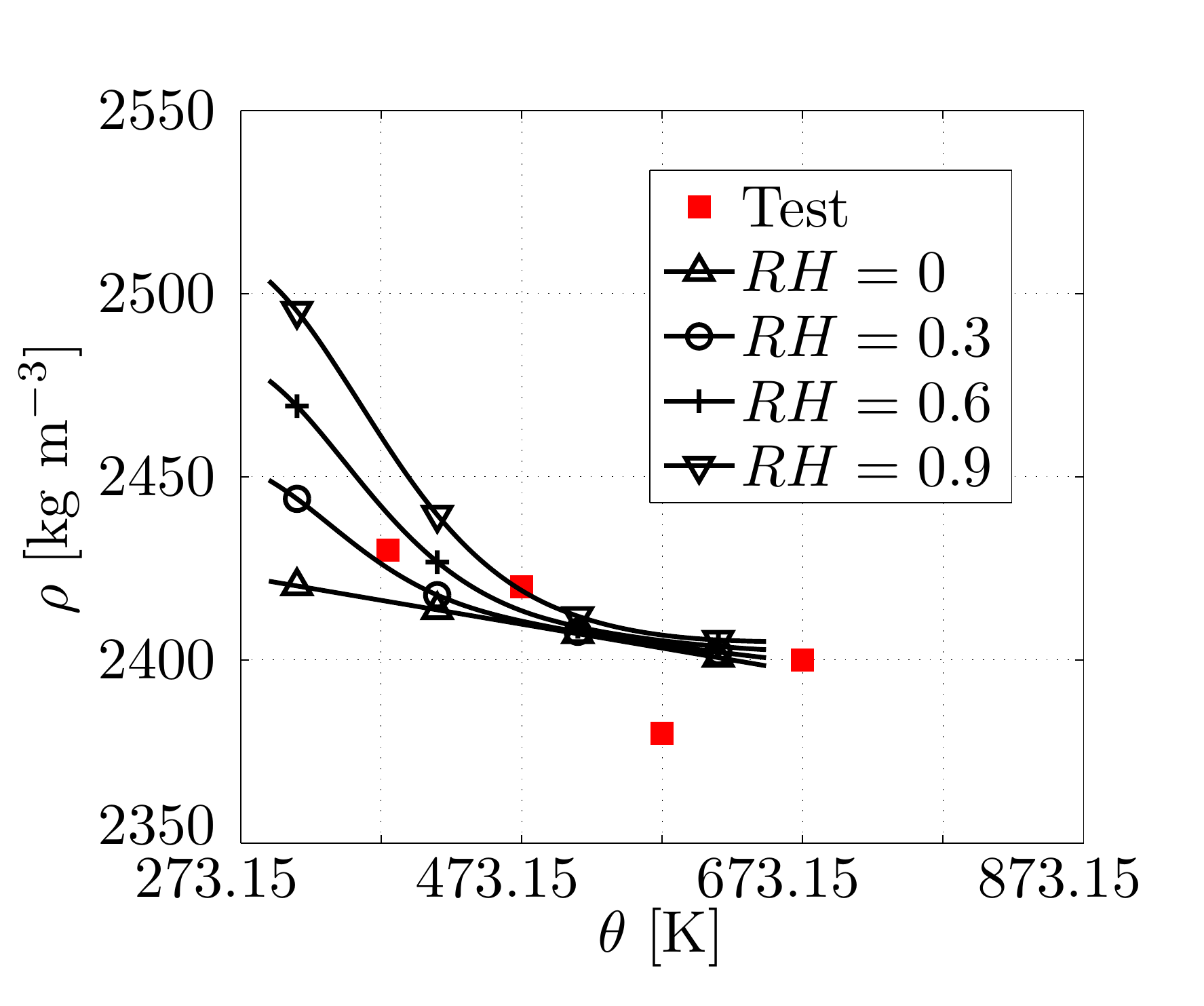}}
  \caption{Apparent density of concrete measured by \citet{Kalifa1998} at a dry state (points) and assumed in our model (lines)}
  \label{Fig_Example1_Density}
\end{minipage}%
\hspace{.01\textwidth}
\begin{minipage}[t]{.47\textwidth}
  \centering
  {\includegraphics[angle=0,width=6.4cm]{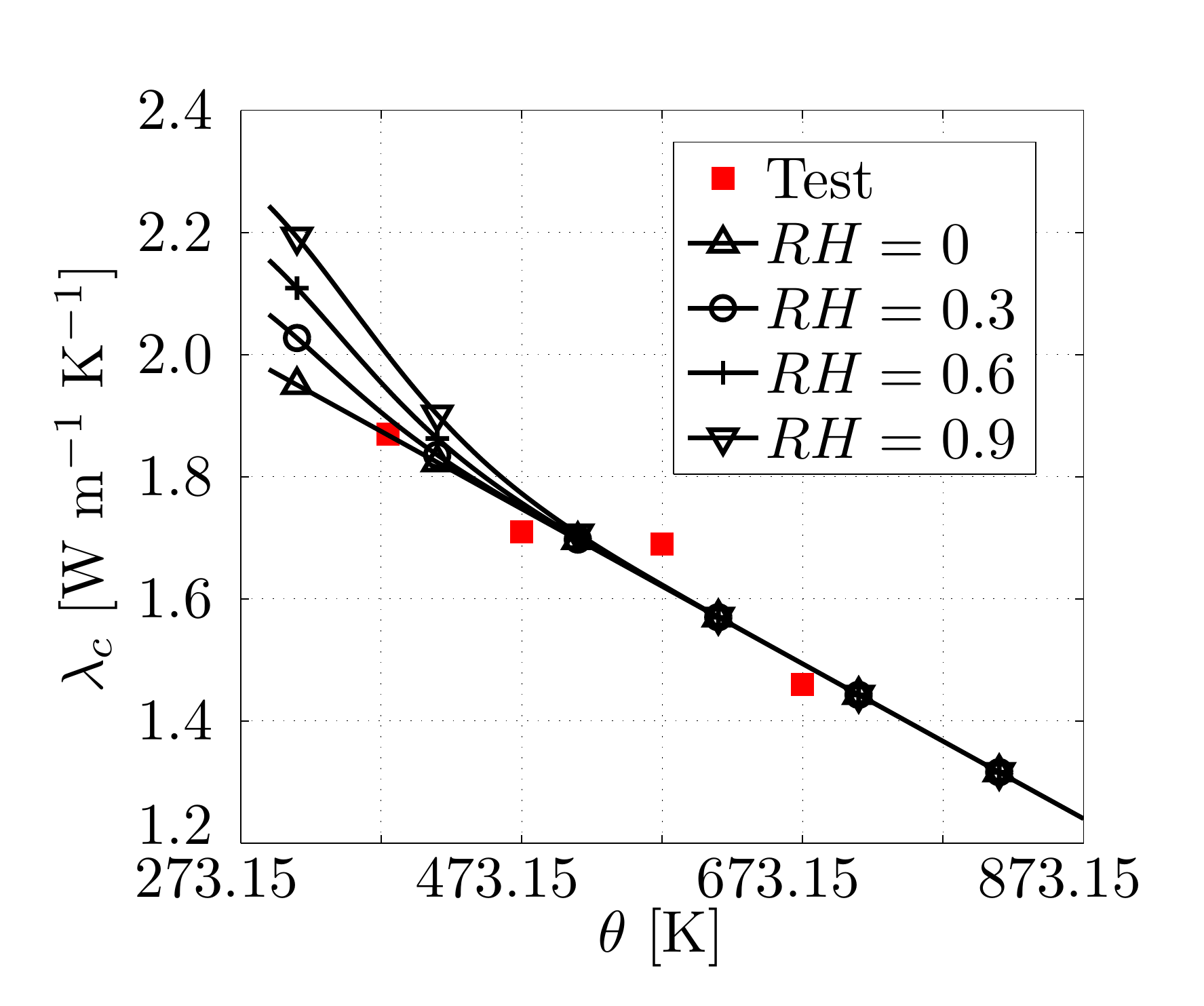}}
  \caption{Thermal conductivity of concrete measured by \citet{Kalifa1998} at a dry state (points) and assumed in our model (lines)}
  \label{Fig_Example1_Conductivity}
\end{minipage}
\end{figure}

As mentioned in Section~\ref{Sec_MaterialProperties}, the density of solid skeleton, $\rho_{s}$, is assumed to be a constant value in our approach. This value can be estimated from the data measured by \citet[Table~4]{Kalifa1998} for the apparent density. The corresponding apparent density assumed in our model, that can be expressed as (cf. \citep[eq.~(26)]{Gawin2003})
\begin{equation}\label{density}
\rho
=
\rho_{w}\,\eta_{w} + \rho_g\,\eta_{g} + \rho_{s}\,\eta_{s}
=
\rho_{w}(\phi\,S_{w}) + (\rho_{v}+\rho_{a})[\phi(1-S_{w})] + \rho_{s}(1-\phi),
\end{equation}
appears in Figure~\ref{Fig_Example1_Density}.

The temperature dependence of the thermal conductivity of concrete was measured by \citet[Table~7]{Kalifa1998} at a dry state. The parameters of equation \eqref{ConductivityDry} are determined by a linear regression of the measured data, see Figure~\ref{Fig_Example1_Conductivity} ($RH = 0$). The resulting thermal conductivity of concrete assumed in our model is shown in Figure~\ref{Fig_Example1_Conductivity}.

The intrinsic permeability of concrete was tested by \citet[Table~6]{Kalifa1998} on concrete samples dried at $105~^\circ{\rm C}$ and then heated up to several temperature levels. However, the reference permeability of undamaged concrete at the room temperature is not reported in \citep{Kalifa1998} and hence it is determined by a trial-and-error method in our simulation (cf. \citep[p.~277]{Witek2007}).

\begin{figure}[h]
  \centering
  {\includegraphics[angle=0,width=8.7cm]{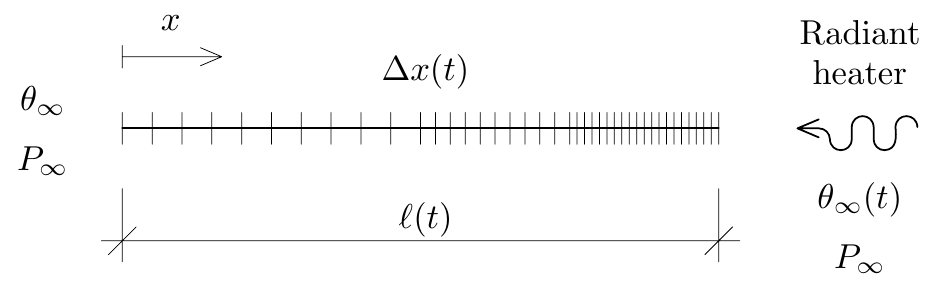}}
  \caption{Scheme of the analysed problem}
  \label{Fig_Example1_Scheme}
\end{figure}

Scheme of the analysed problem is displayed in Figure~\ref{Fig_Example1_Scheme}. The spatial discretization is performed with the use of linear 1-D elements. In total, 120 elements are employed -- 30 elements in the interval $x \in (0,\ell/2)$), 30 elements for $x \in (\ell/2,3\ell/4)$ and 60 elements in the interval $x \in (3\ell/4,\ell)$. For the time discretization, the time step is set to $\Delta t = 1~\rm{s}$. The characteristic time of spalling is assumed as $\gamma = 10~\rm{s}$.

The initial conditions are assumed as $P_0 = 1.9039 \times 10^3~\rm{Pa}$, $\theta_0 = 293.15~\rm{K}$, $\ell_0 = 0.12~\rm{m}$. For these values, the initial saturation with liquid water $S_{w,0} = 0.77$, which is the value reported by \citet[Table~2]{Kalifa2000}.

\begin{figure}[h]
  \centering
  {\includegraphics[angle=0,width=6.4cm]{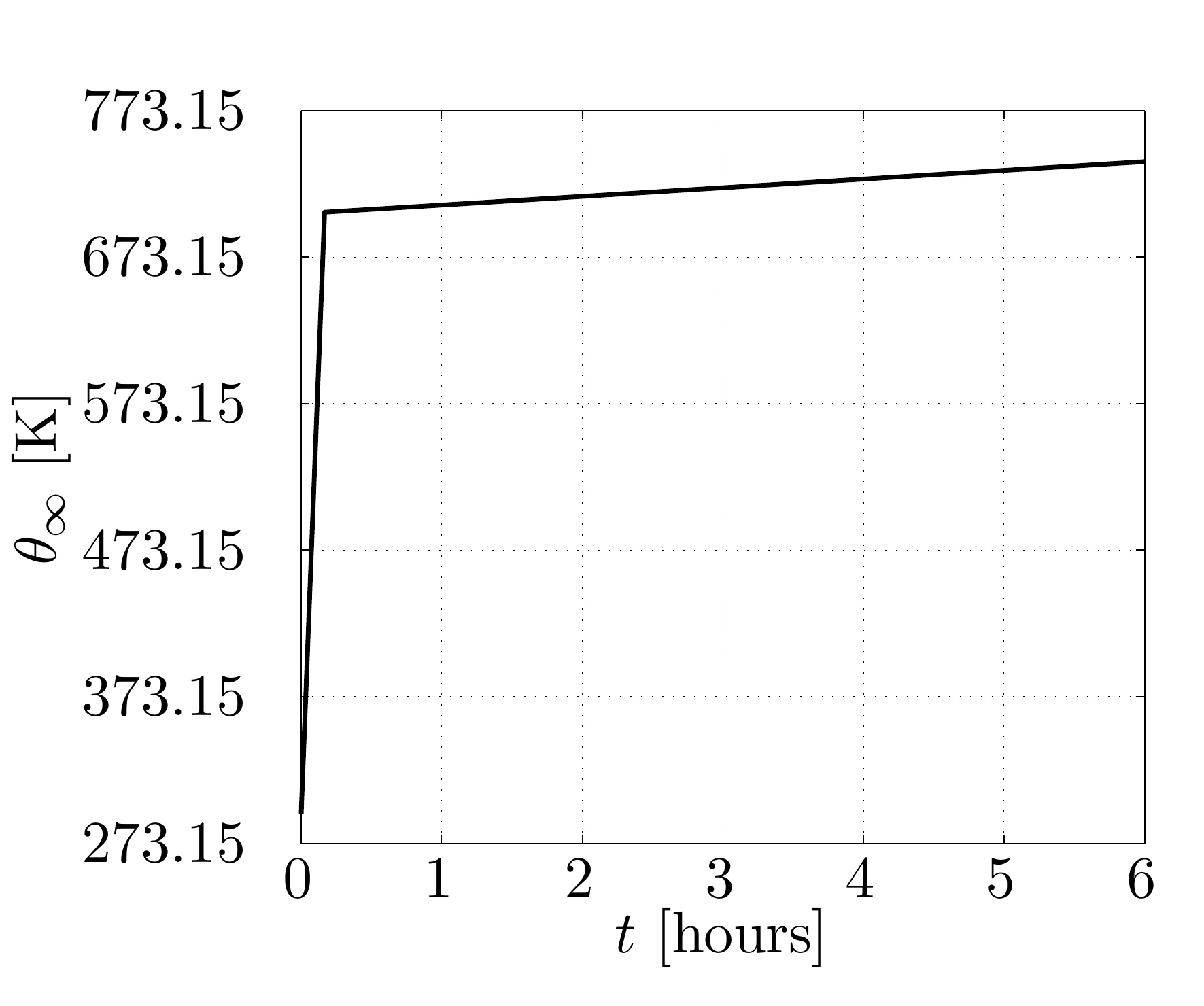}}
  \caption{Air temperature on the heated side of the specimen used for the simulation of the PTM test 1}
  \label{Fig_Example1_Heating}
\end{figure}

As stated above, a radiant heater of a temperature of $600~\rm{^{\circ}C}$ was employed for the experiment. However, as mentioned by e.g. \citet[pp. 554--556]{Gawin2002}, the temperature of the ambient air on the heated side, which needs to be defined for the boundary conditions assumed in the model (see Section~\ref{BoundaryConditions}), is lower than the temperature of the heater. For our simulation, the ambient temperature on the heated side, $\theta_{\infty}~\rm{[K]}$, has been determined by a trial-and-error method as (see Figure~\ref{Fig_Example1_Heating})
\begin{equation}\label{Heating1}
\theta_{\infty}(t) = \left\{ \begin{array}{lll}
\displaystyle 293.15 + t \frac{410}{300} & \rm{for} & t \leq 300~\rm{s},
\\
\displaystyle 703.15 + (t-300) \frac{35}{21300} & \rm{for} & t > 300~\rm{s},
\end{array} \right.
\end{equation}
where $t~\rm{[s]}$ is the time of heating.

The boundary conditions are summarized in Table~\ref{Tab_Example1_Boundary_Conditions}.

\begin{table}[h]
\centering
\caption{Boundary conditions parameters for simulation of PTM test 1}
\small
\begin{tabular}{c c c c}
\hline
\multirow{2}{*}{Variable} &
\multicolumn{2}{c}{Value} &
\multirow{2}{*}{Unit} \\ \cline{2-3}
& Unexposed side & Exposed side & \\
\hline\hline
$P_{\infty}$      & $P_0$      & $P_0$   & $\rm{Pa}$\\
$\theta_{\infty}$ & $\theta_0$ & $\theta_{\infty}(t)$, see \eqref{Heating1} & $\rm{K}$ \\
$\alpha_c$        & $4$        & $20$    & $\rm{W\,m^{-2}\,K^{-1}}$ \\
$e\sigma_{SB}$    & $0.7 \times 5.67 \times 10^{-8}$  & $0.7 \times 5.67 \times 10^{-8}$ & $\rm{W\,m^{-2}\,K^{-4}}$ \\
$\beta_c$         & $0.009$    & $0.019$ & $\rm{m\,s^{-1}}$ \\
\hline
\end{tabular}
\label{Tab_Example1_Boundary_Conditions}
\end{table}

\subsubsection{Discussion} \label{Sec_Example1_Discussion}
\begin{figure}[h]
  \centering
  {\includegraphics[angle=0,width=11.6cm]{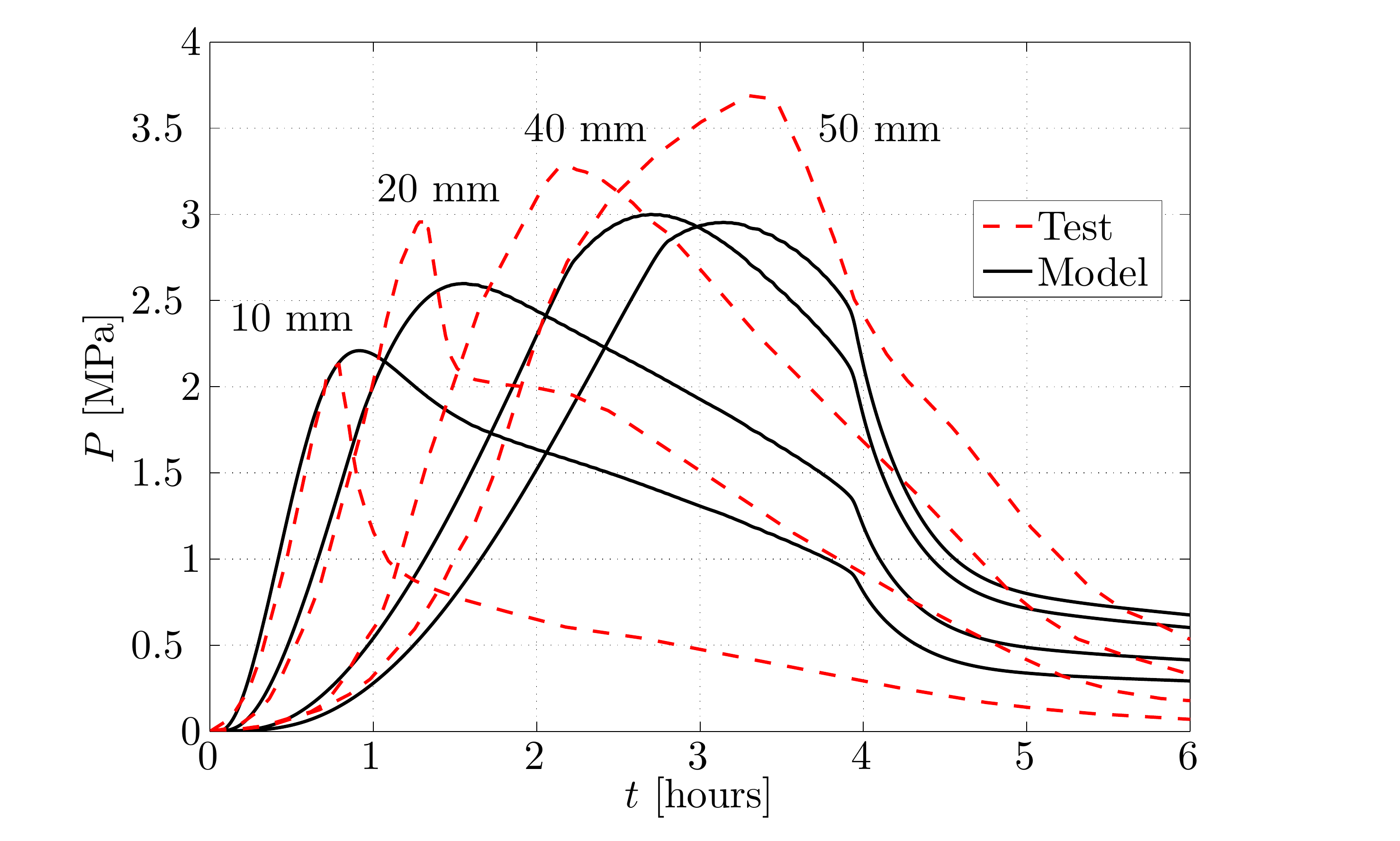}}
  \caption{Pore pressure evolution measured by \citet[Figure~4]{Kalifa2000} (dashed lines) and determined by the present model (solid lines)}
  \label{Fig_Example1_Comparison_Press}
\end{figure}

\begin{figure}[h]
  \centering
  {\includegraphics[angle=0,width=11.6cm]{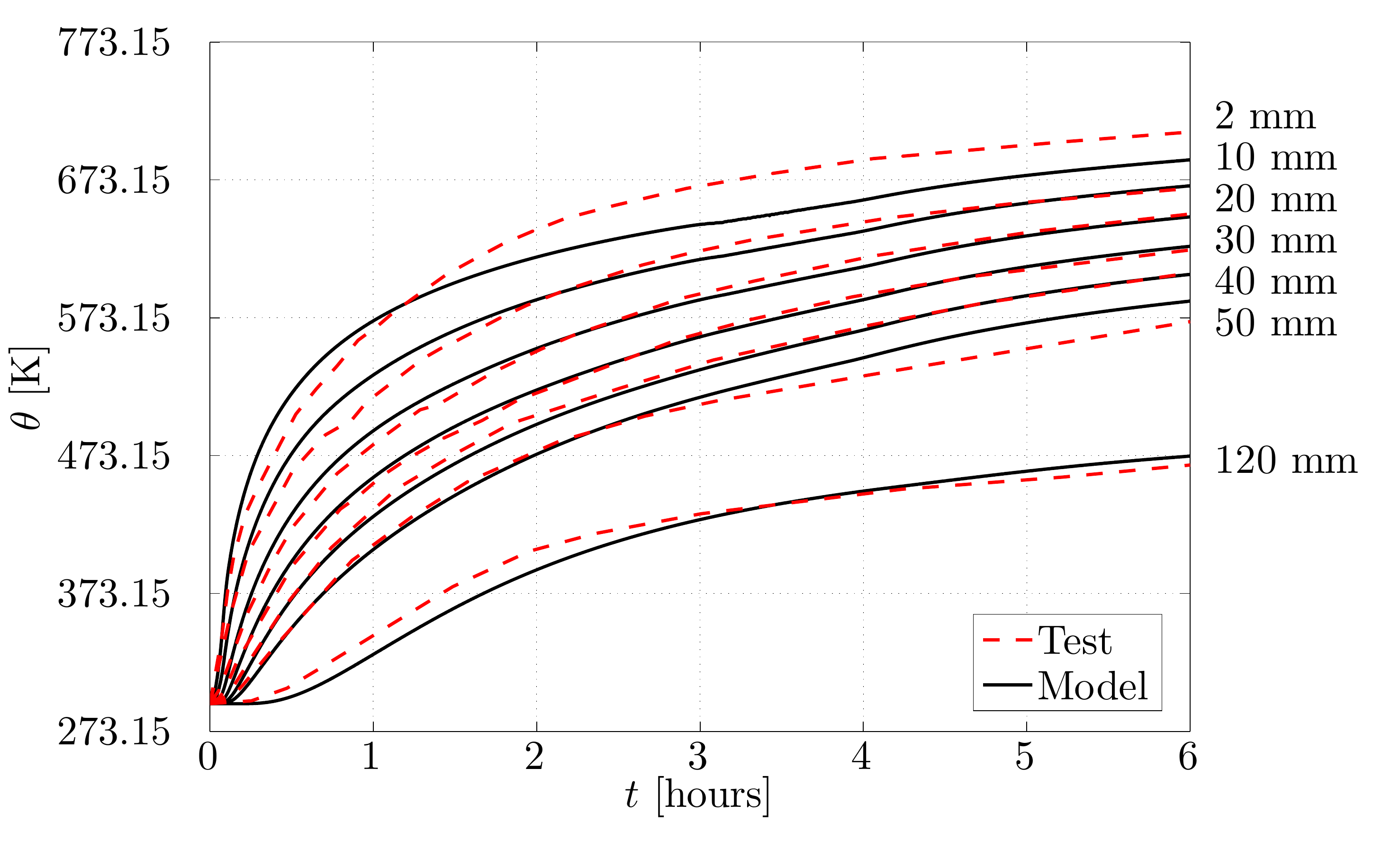}}
  \caption{Temperature evolution measured by \citet[Figure~4]{Kalifa2000} (dashed lines) and determined by the present model (solid lines)}
  \label{Fig_Example1_Comparison_Temp}
\end{figure}

\begin{figure}[h]
\centering
\begin{minipage}[t]{.47\textwidth}
  \centering
  {\includegraphics[angle=0,width=6.4cm]{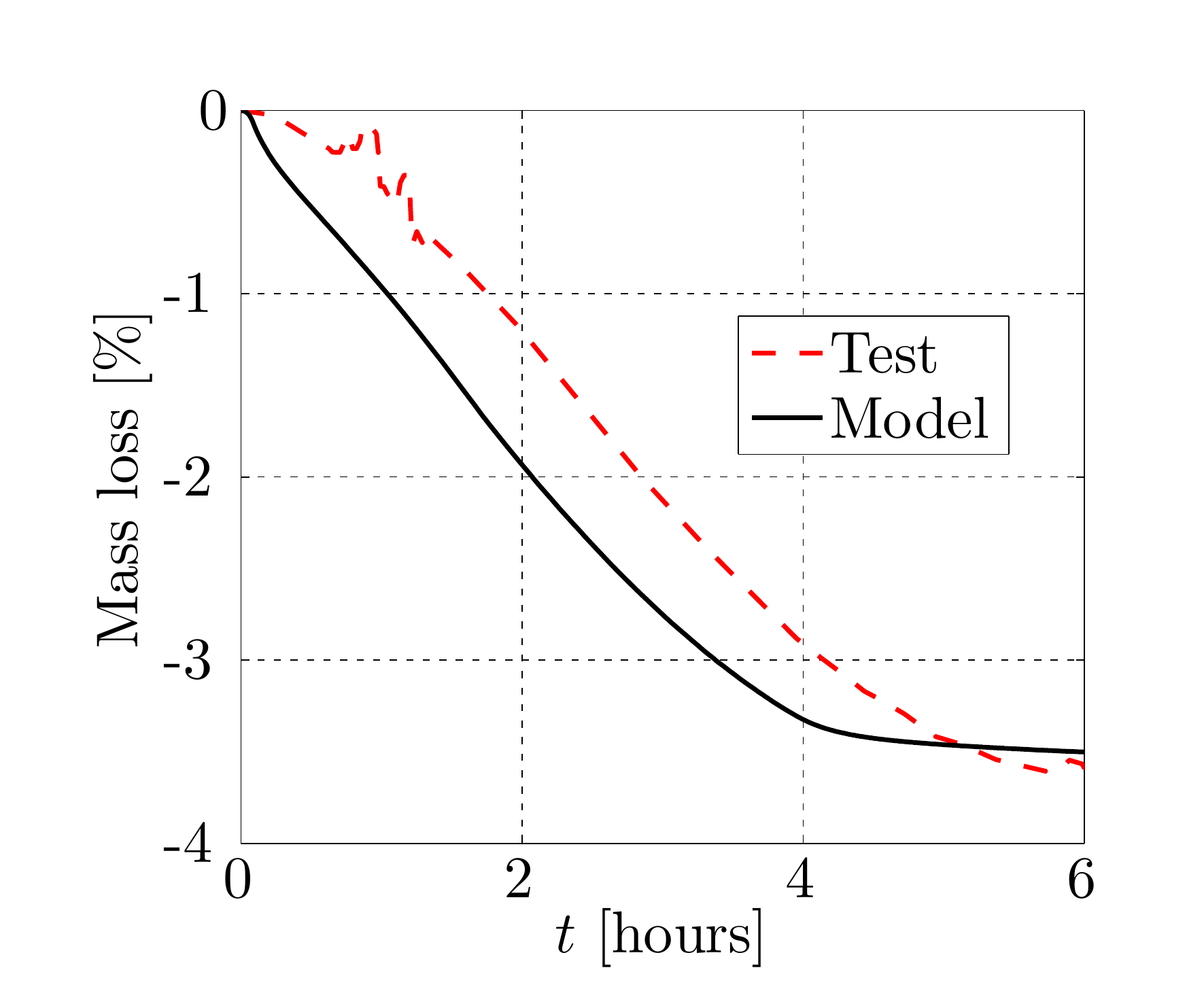}}
  \caption{Mass loss measured by \citet[Figure~7]{Kalifa2000}) (dashed line) and determined by the present model (solid line)}
  \label{Fig_Example1_Comparison_Mass_Loss}
\end{minipage}%
\hspace{.01\textwidth}
\begin{minipage}[t]{.47\textwidth}
  \centering
  {\includegraphics[angle=0,width=6.4cm]{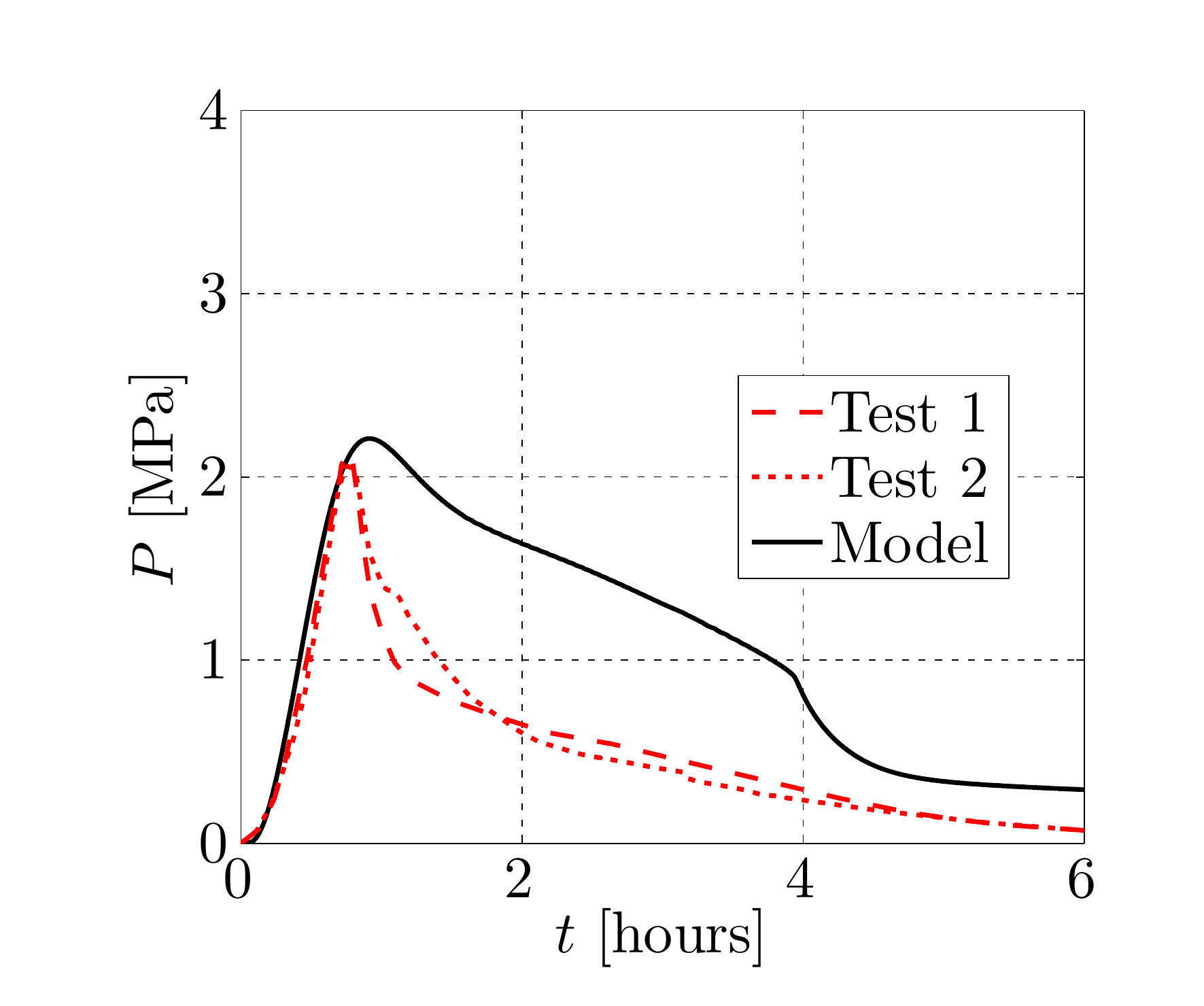}}
  \caption{Pore pressure development measured by \citet[Figure~6]{Kalifa2000} and determined by the present model at the depth of 10~mm from the heated surface}
  \label{Fig_Example1_Comparison_Press10}
\end{minipage}
\begin{minipage}[t]{.47\textwidth}
  \centering
  {\includegraphics[angle=0,width=6.4cm]{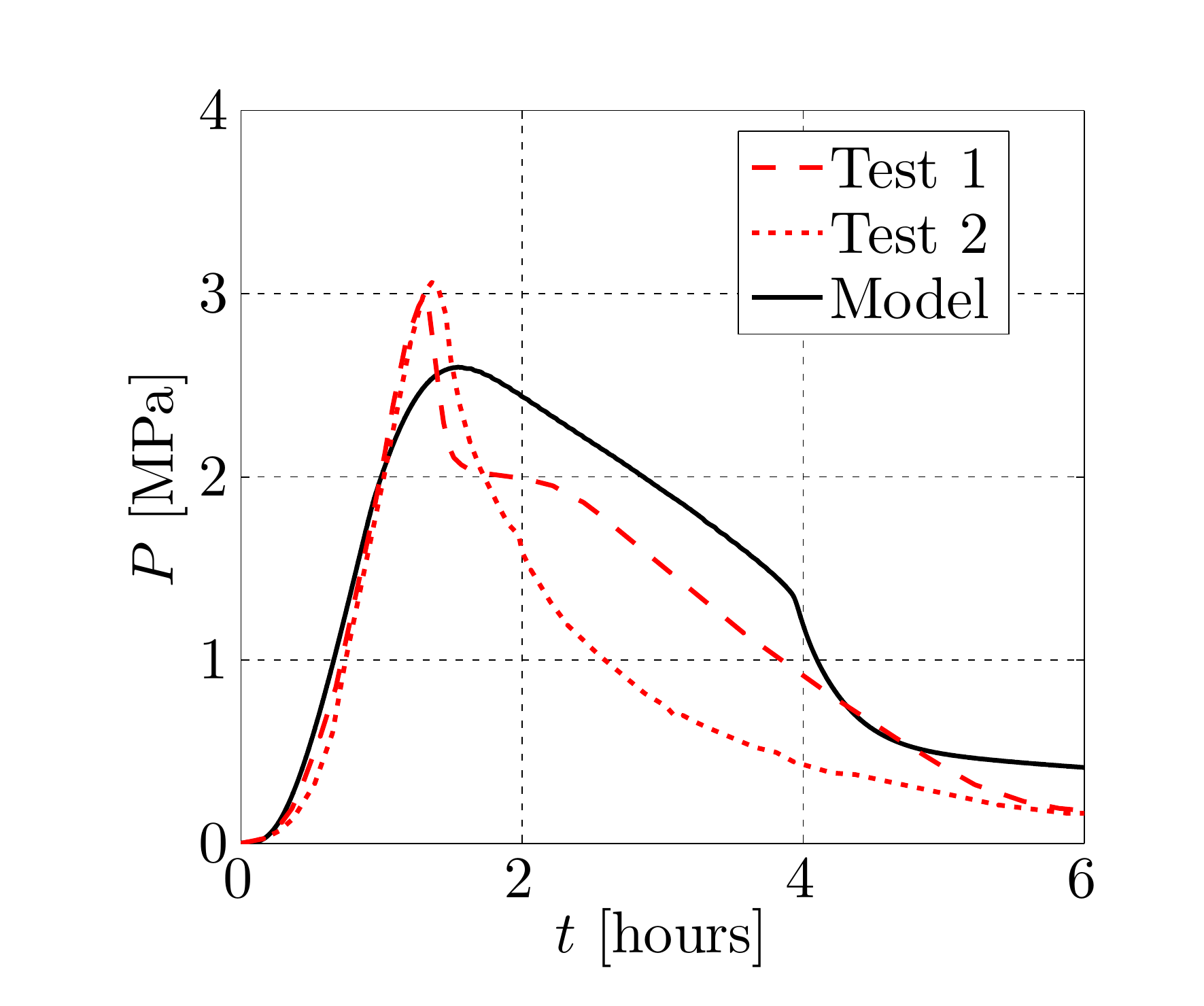}}
  \caption{Pore pressure development measured by \citet[Figure~6]{Kalifa2000} and determined by the present model at the depth of 20~mm from the heated surface}
  \label{Fig_Example1_Comparison_Press20}
\end{minipage}%
\hspace{.01\textwidth}
\begin{minipage}[t]{.47\textwidth}
  \centering
  {\includegraphics[angle=0,width=6.4cm]{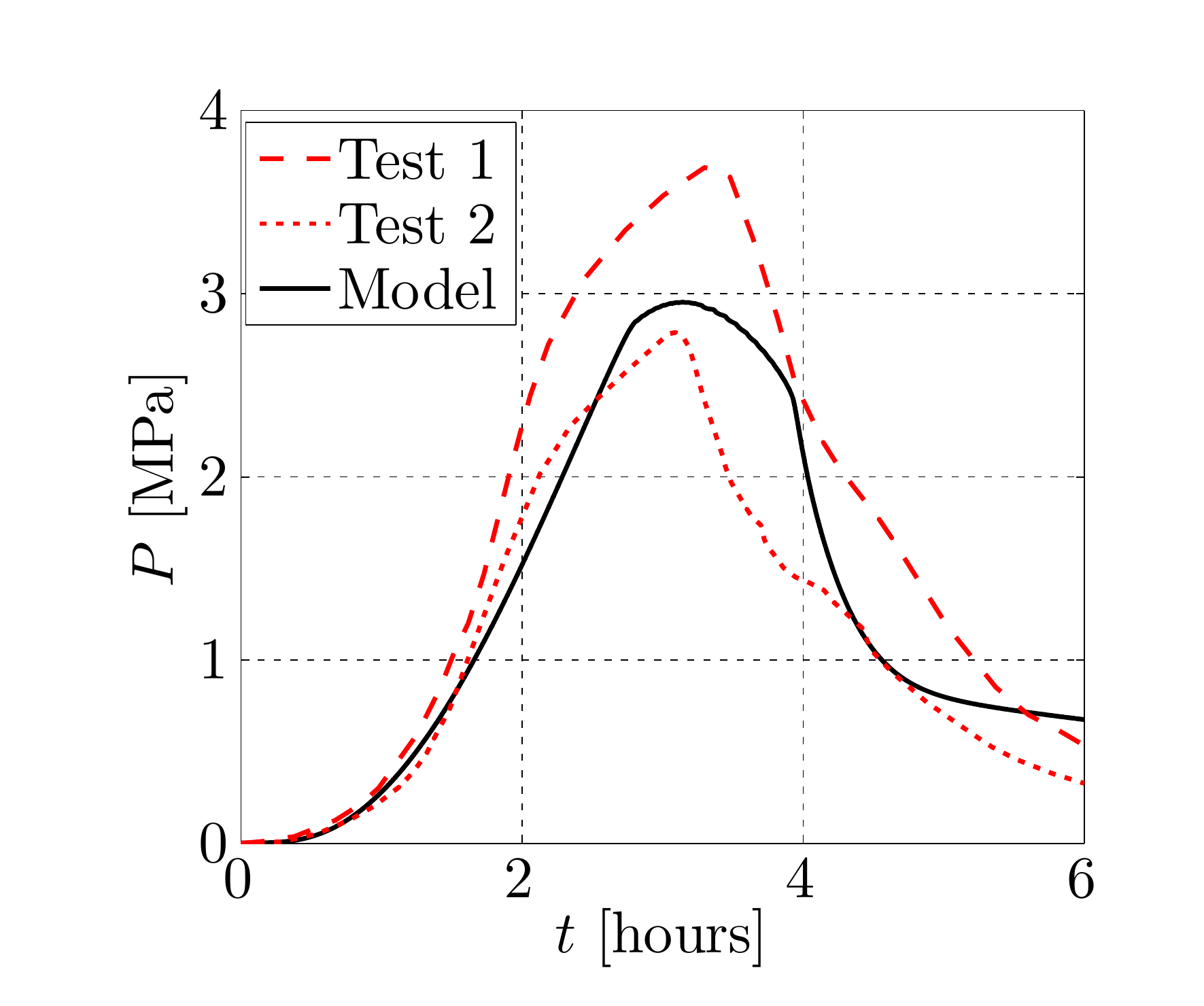}}
  \caption{Pore pressure development measured by \citet[Figure~6]{Kalifa2000} and determined by the present model at the depth of 50~mm from the heated surface}
  \label{Fig_Example1_Comparison_Press50}
\end{minipage}
\end{figure}

The resulting pore pressure, temperature and mass loss evolutions determined by our simulation are compared with the experimentally measured data in Figures~\ref{Fig_Example1_Comparison_Press}--\ref{Fig_Example1_Comparison_Mass_Loss} (note that the distances in millimeters stated in the figures are, in accordance with \citet{Kalifa2000}, measured from the heated surface).

The pore pressure evolution in the depth of $30~\rm{mm}$ from the heated surface is not displayed in Figure~\ref{Fig_Example1_Comparison_Press} since it was not correctly measured by \citet{Kalifa2000}, probably due to the damage of the gauge (see \citep[p.~1923]{Kalifa2000}). On the other hand, the pore pressure evolution in the depths of (10, 20, and 50) mm can be verified more precisely because the investigated PTM test was duplicated and the data from the both measurements are stated in \citep[Figure~6]{Kalifa2000}, see Figures~\ref{Fig_Example1_Comparison_Press10}--\ref{Fig_Example1_Comparison_Press50}.

From Figures~\ref{Fig_Example1_Comparison_Press}--\ref{Fig_Example1_Comparison_Press50}, it is obvious that in this case,
{despite of the simplifications mentioned in the introductory part of this section,}
the present model predicts the hygro-thermal behaviour of the heated concrete specimen on the sufficient level of accuracy.
{It should be however mentioned that some of the potential inaccuracies arising from the simplifications embedded in the model are probably eliminated by the appropriate setting of the reference permeability of undamaged concrete at the room temperature (fitted by a trial-and-error method)}.

\begin{figure}[h]
\centering
\begin{minipage}[t]{.47\textwidth}
  \centering
  {\includegraphics[angle=0,width=6.4cm]{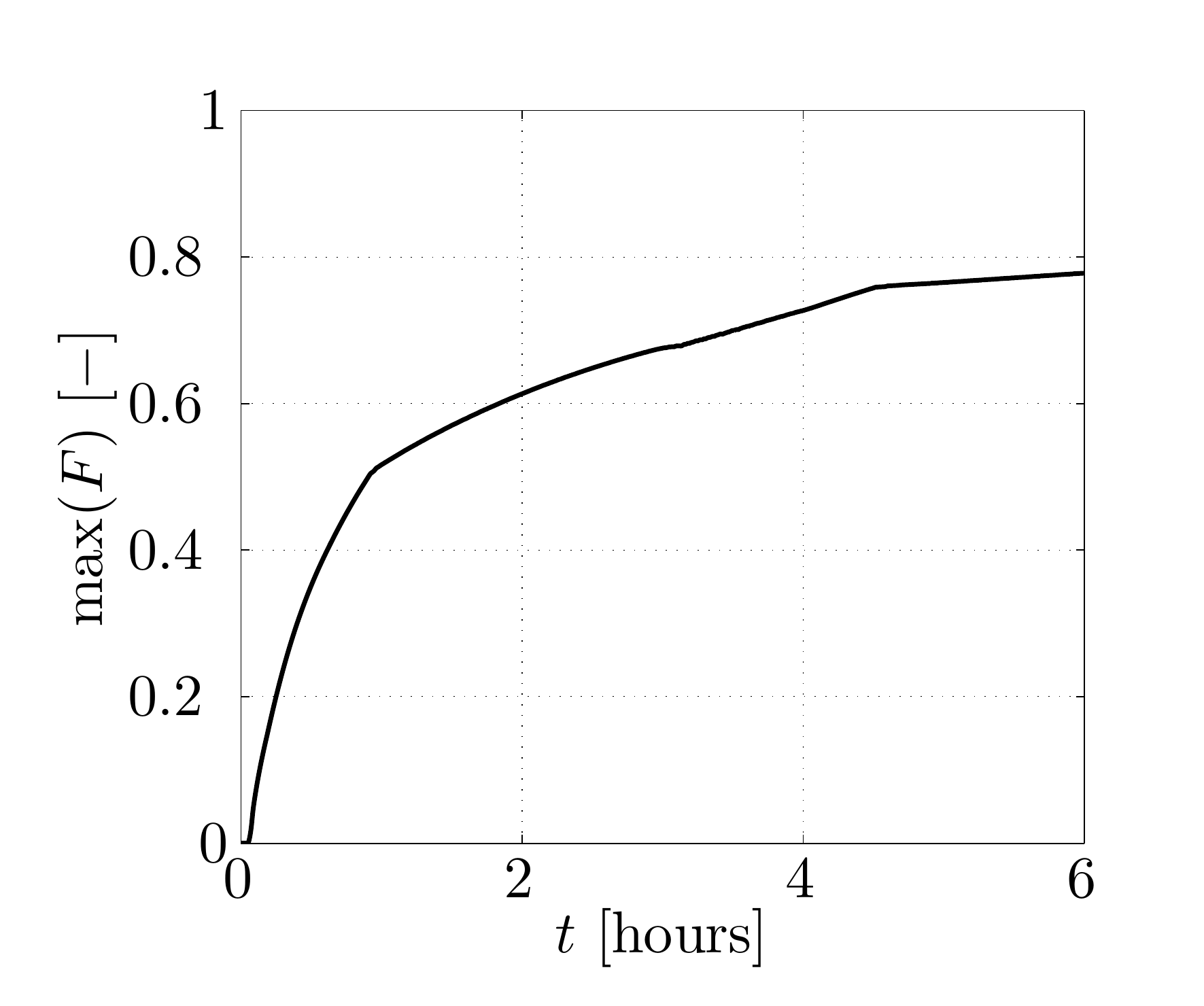}}
  \caption{Evolution of the failure parameter}
  \label{Fig_Example1_Failure}
\end{minipage}%
\hspace{.01\textwidth}
\begin{minipage}[t]{.47\textwidth}
  \centering
  {\includegraphics[angle=0,width=6.4cm]{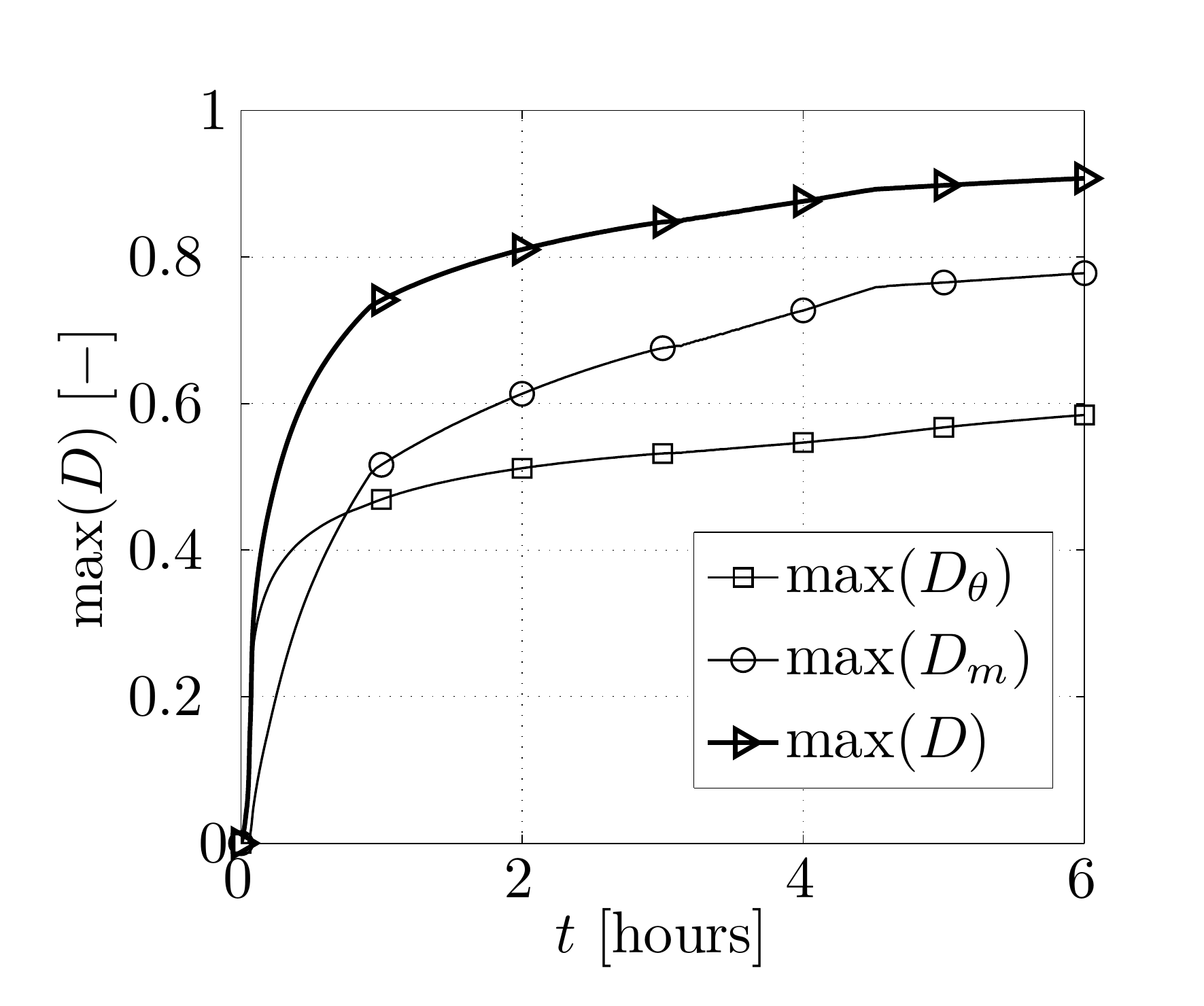}}
  \caption{Evolution of the damage parameter}
  \label{Fig_Example1_Damage}
\end{minipage}
\end{figure}

The spalling behaviour is also simulated correctly since, as mentioned above, during the test, no spalling was observed by \citet{Kalifa2000}, which is in accordance with our simulation, see Figure~\ref{Fig_Example1_Failure} -- the maximal value of failure function \eqref{failure criterion_simplified} achieved within the specimen during the test period does not exceeds the value of 1. The evolution of the maximal values of the damage parameter and its components within the specimen during the test period determined by our model is shown in Figure~\ref{Fig_Example1_Damage}.

\begin{figure}[h]
\centering
\begin{tabular}{lr}
  \includegraphics[angle=0,width=6.4cm]{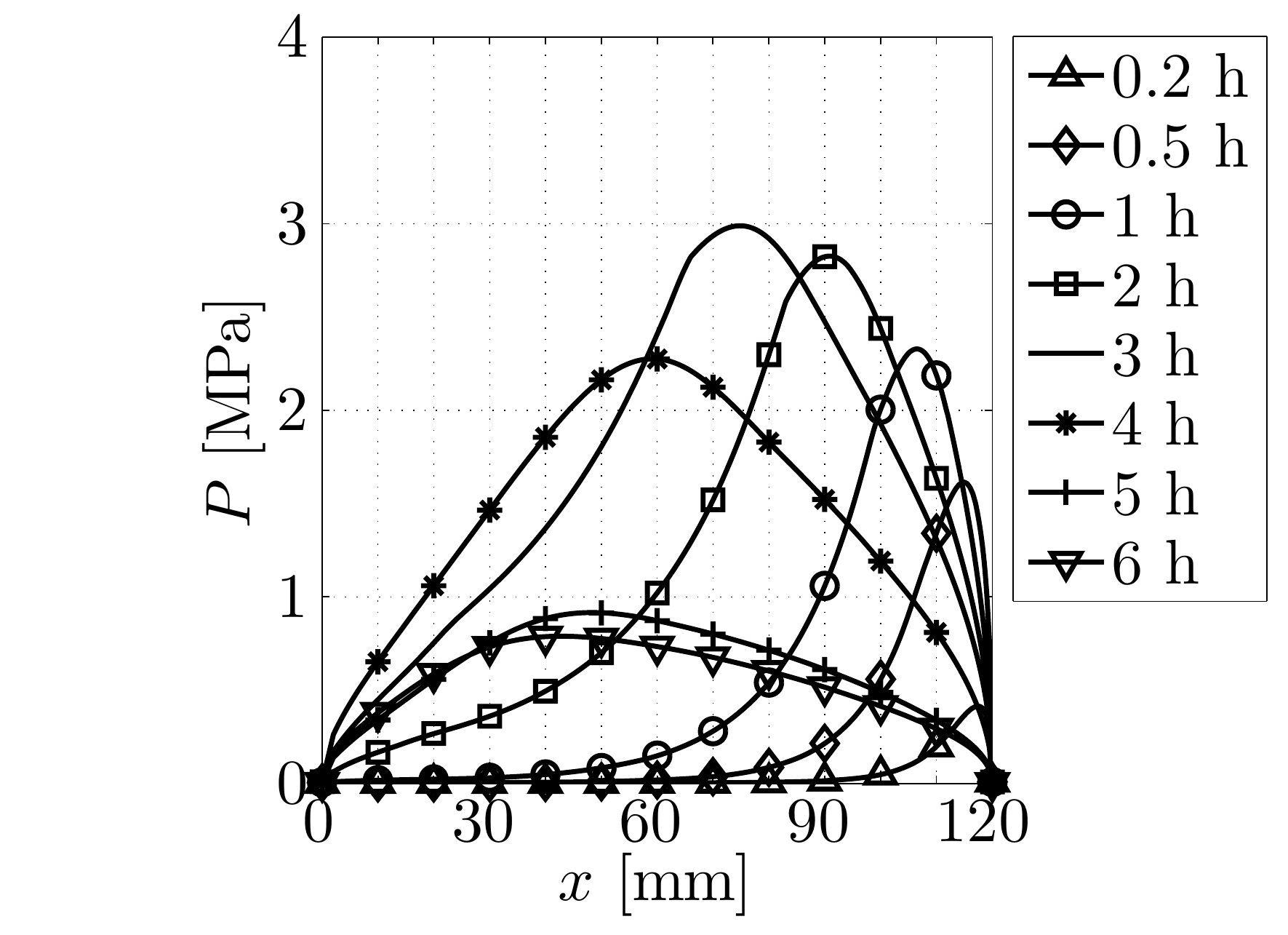}
  &
  \includegraphics[angle=0,width=5.9cm]{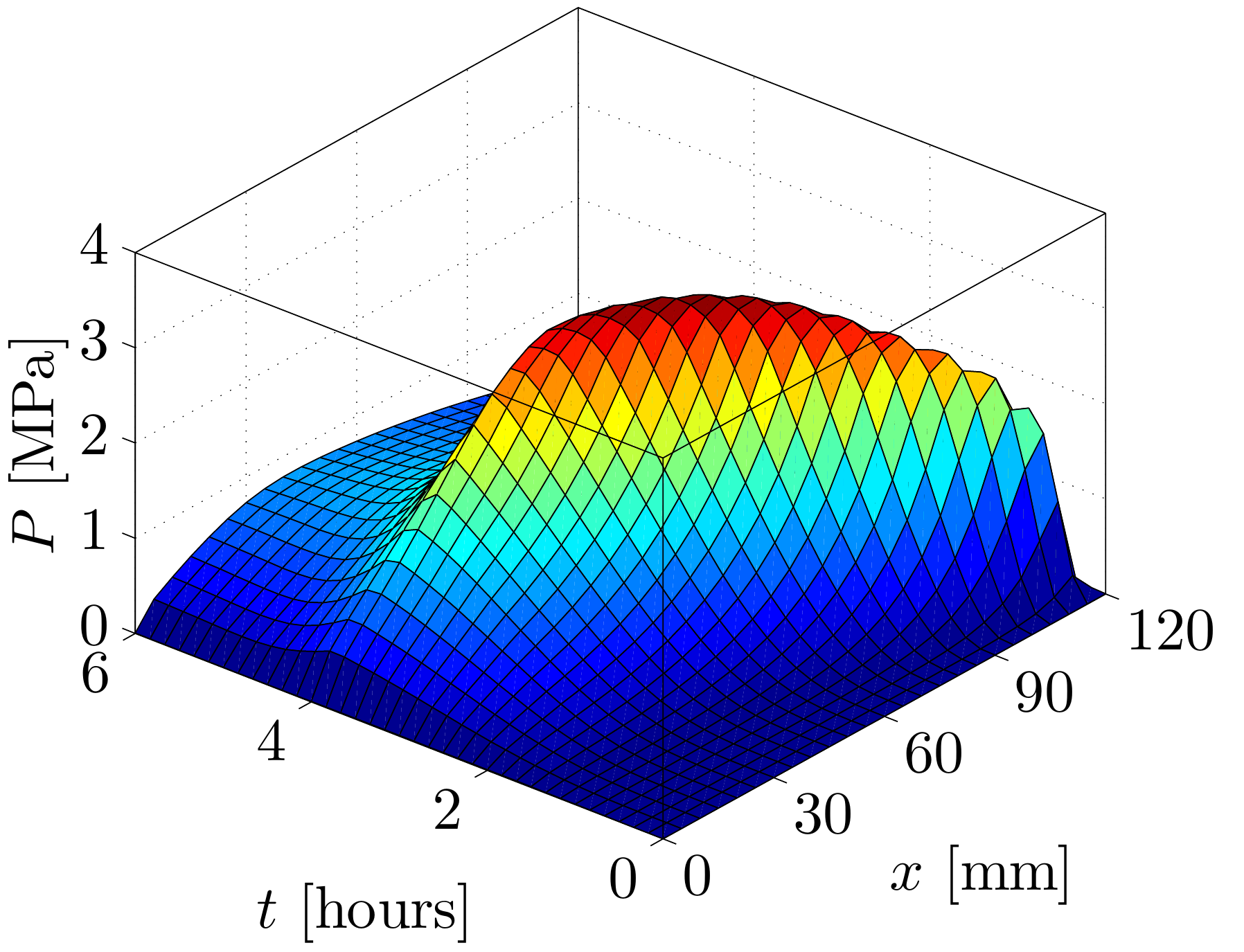}
  \\
  \includegraphics[angle=0,width=6.4cm]{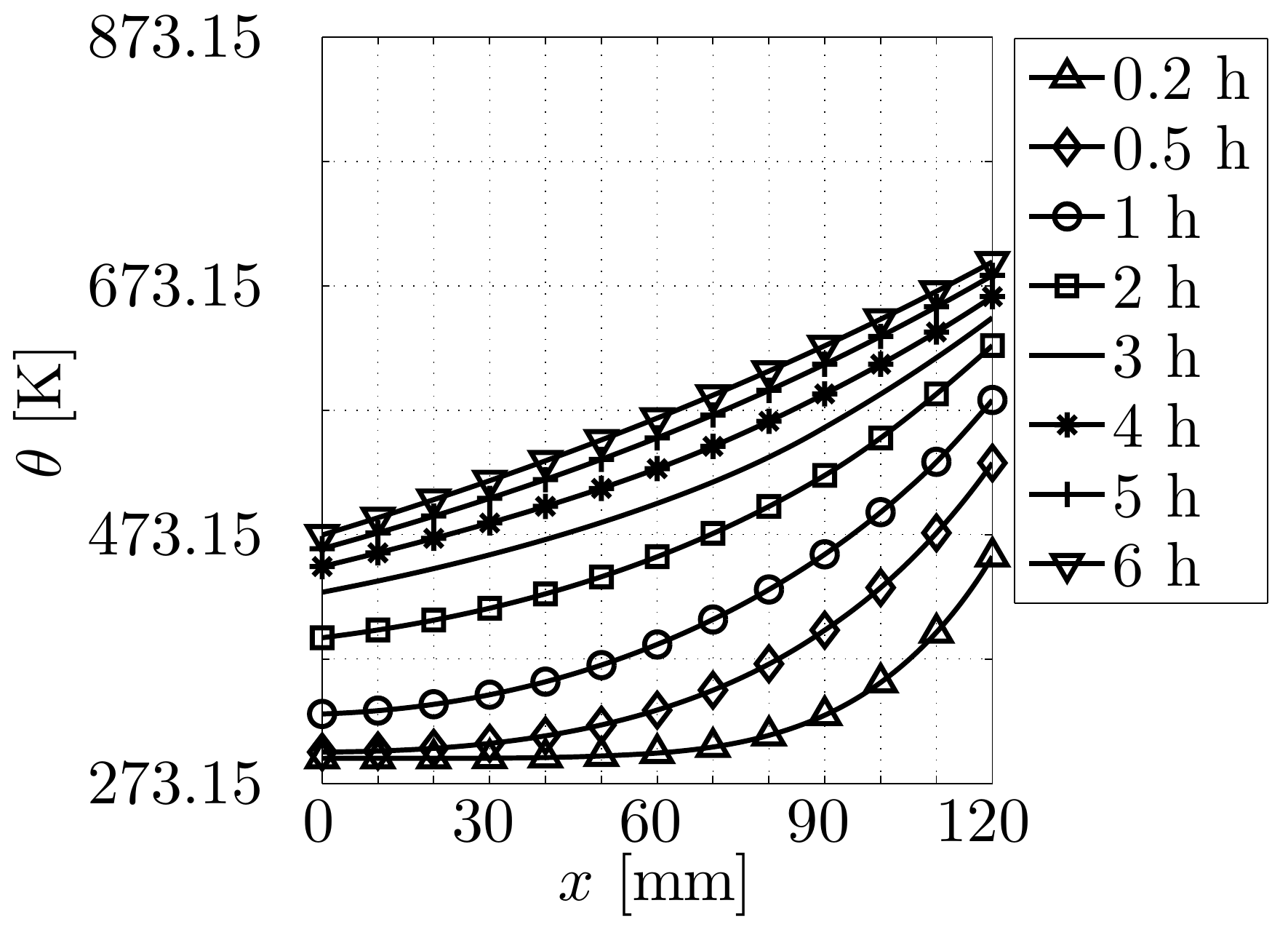}
  &
  \includegraphics[angle=0,width=6.4cm]{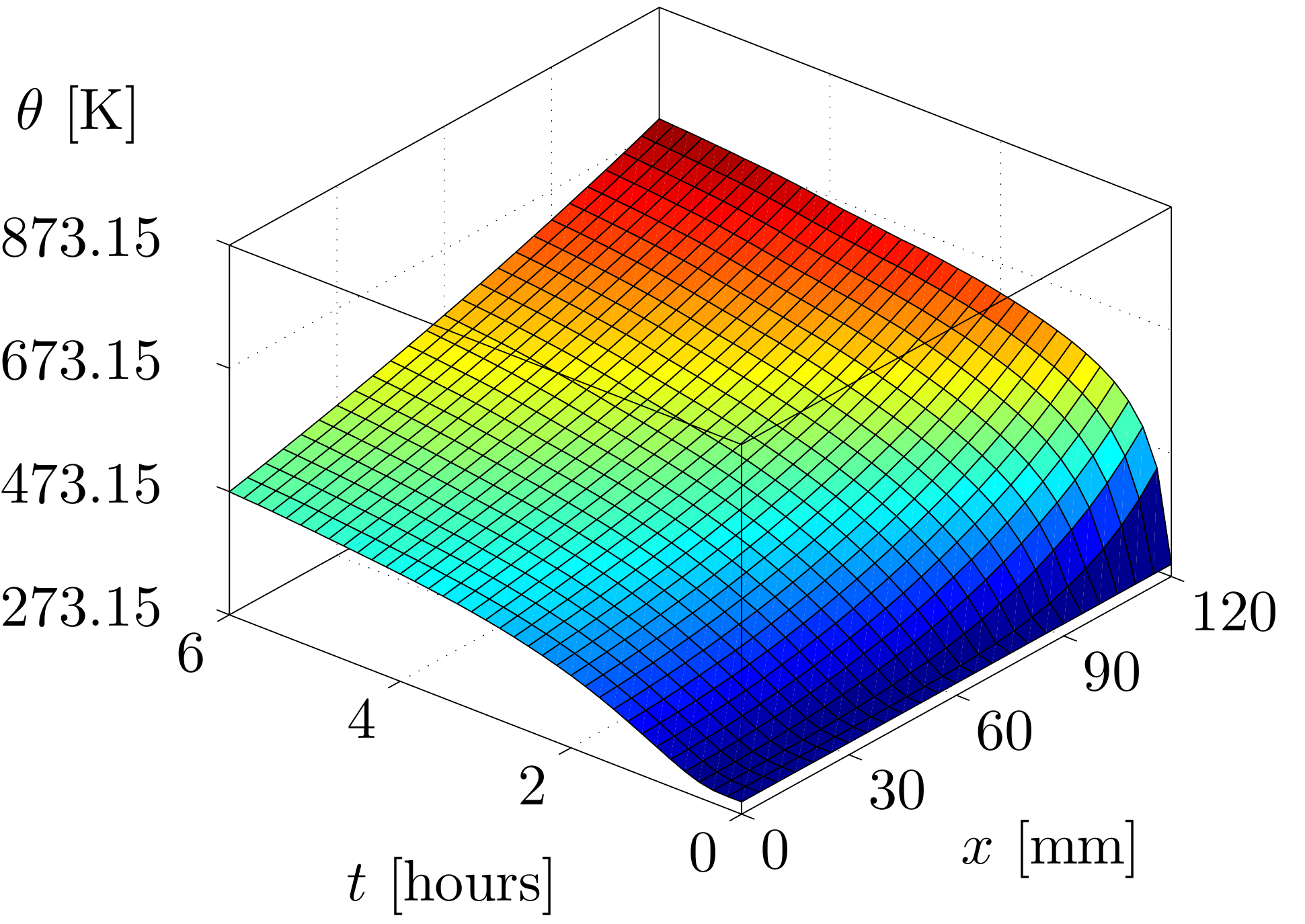}
\end{tabular}
\caption{Distribution of $P$ and $\theta$ for the analysed PTM test 1}
\label{Fig_Example1_Distribution_P_T}
\end{figure}

\begin{figure}[h]
\centering
\begin{tabular}{lr}
  \includegraphics[angle=0,width=6.4cm]{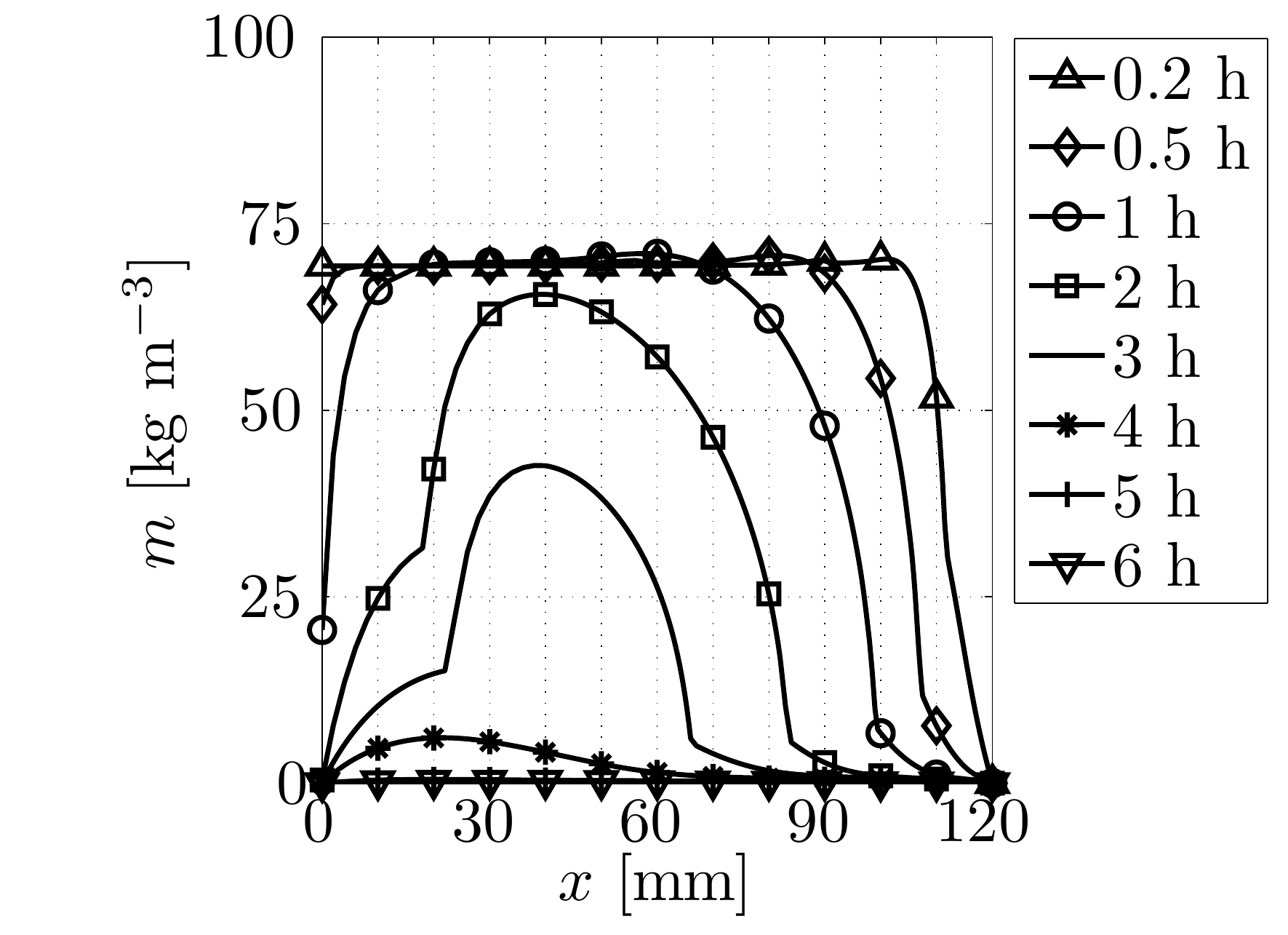}
  &
  \includegraphics[angle=0,width=6.25cm]{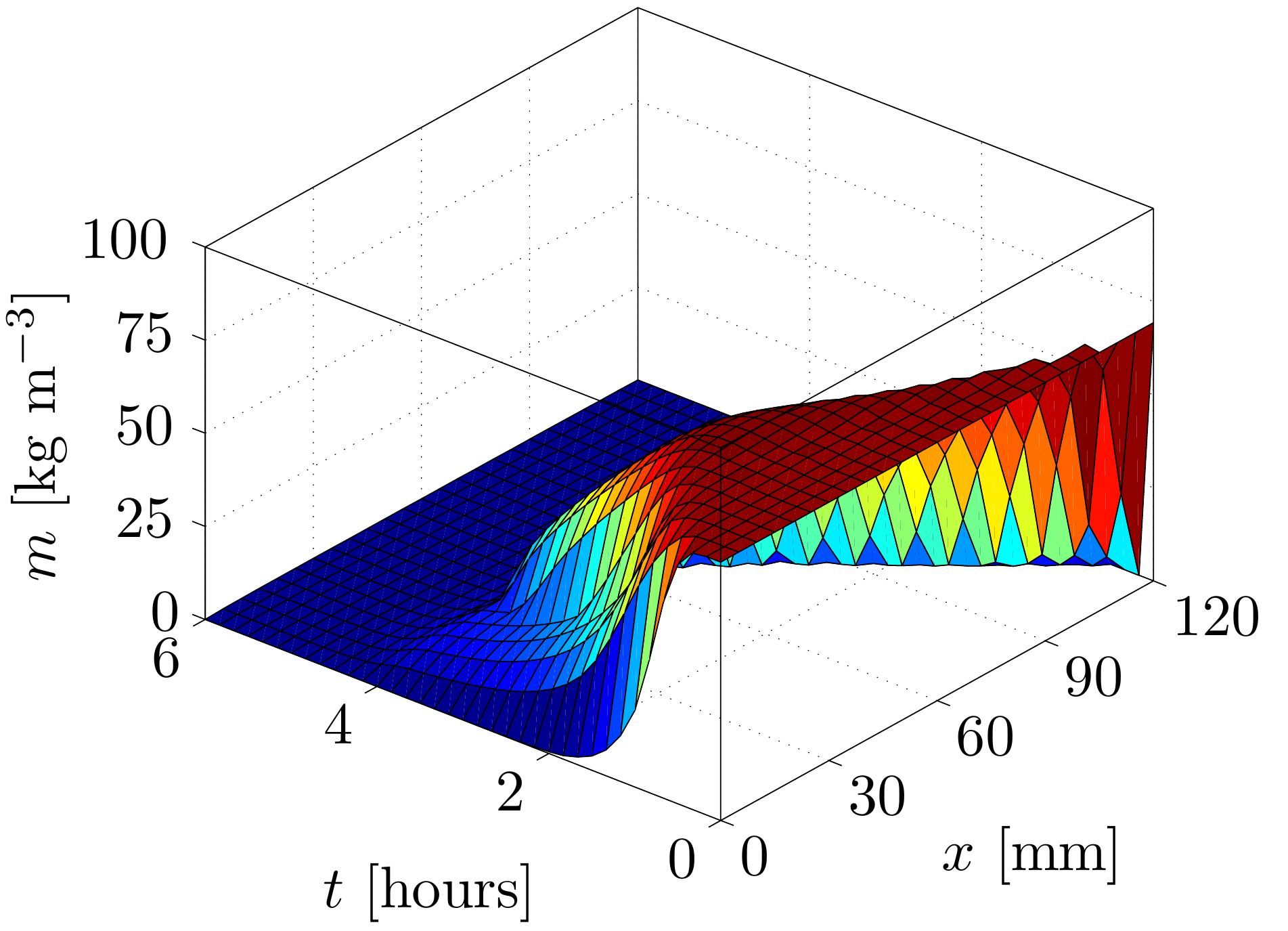}
  \\
  \includegraphics[angle=0,width=6.4cm]{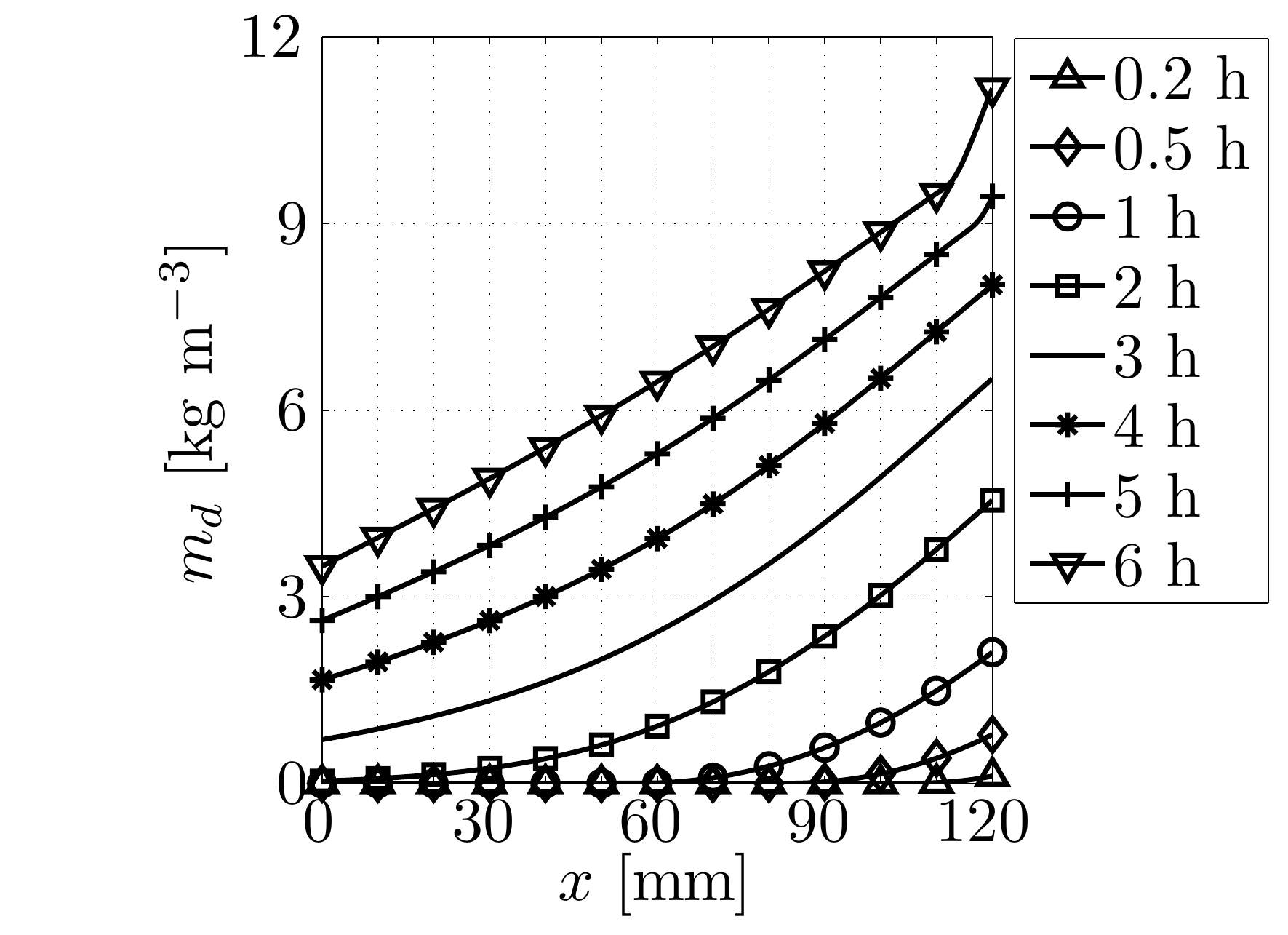}
  &
  \includegraphics[angle=0,width=6.1cm]{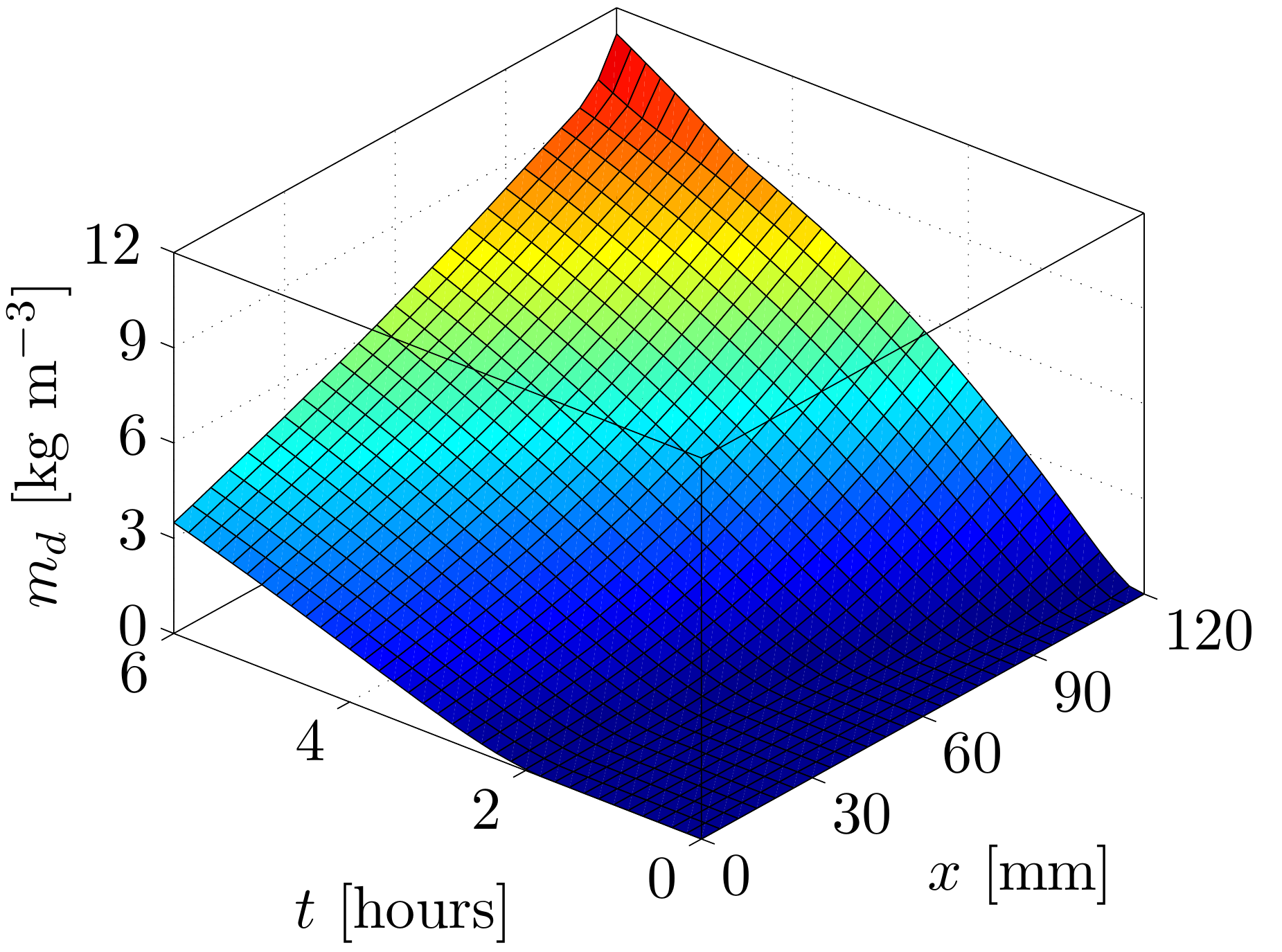}
\end{tabular}
\caption{Distribution of $m$ and $m_d$ for the analysed PTM test 1}
\label{Fig_Example1_Distribution_m_md}
\end{figure}

For illustration, the spatial and time distributions of the primary unknowns of the present model (except the thickness of the specimen, which is constant in this case) for the Kalifa's PTM test 1 \citep{Kalifa2000} are
shown in Figures~\ref{Fig_Example1_Distribution_P_T} and \ref{Fig_Example1_Distribution_m_md}.

\clearpage

\subsection{Simulation of the PTM test 2}
\label{Sec_Example2}

\subsubsection{Experiment description}
\label{Sec_Example2_Experiment}
The PTM test procedure proposed by \citet{Kalifa2000} was adopted by Mindeguia as a part of an extended experimental program reported in \citep{Mindeguia2009,Mindeguia2009a,Mindeguia2013,Mindeguia2010}.
Within this program, the material properties investigation, hygro-thermal behaviour measurements (the PTM tests) and the spalling experiments were performed (a detailed summary of the experiments included in the program appears in \citep[Section~1]{Mindeguia2009}). Hereafter, one of the PTM tests performed by \citet{Mindeguia2009} is closely investigated -- the PTM test of a high strength concrete (denoted as "B60" in \citep{Mindeguia2009}) prismatic specimens (conditioned in "Air\_1" according to \citet{Mindeguia2009}) heated by a radiant heater of the temperature of $600~\rm{^{\circ}C}$ (the type of heating denoted as "{\emph{mod\'{e}r\'{e}}}" in \citep{Mindeguia2009}) placed near its surface for a heating period of 5 hours. The test set-up is the same as described for the Kalifa's experiment (Section~\ref{Sec_Example1_Experiment}). Note that (i) within the investigated PTM test 2, no spalling was observed by \citet{Mindeguia2009}, and (ii) the test was duplicated and the data from the both measurements of $P$ are stated in \citep{Mindeguia2009}, see Section~\ref{Sec_Example2_Discussion}.

\subsubsection{Modelling}
\label{Sec_Example2_Modelling}
The thermal and hygral properties of concrete are assumed according to the relationships given in Section~\ref{Sec_MaterialProperties}, with the parameters determined from the data measured by \citet{Mindeguia2009}. All the parameters are summarized in Table~\ref{Tab_Example2_Parameters}.

\begin{table}[h]
\centering
\caption{Material properties and parameters used in our simulation of the PTM test 2}
\small
\begin{tabular}{l l l l}
\hline
Parameter         & Value    & Unit       & Reference \\
\hline\hline
\multirow{2}{*}{Type of concrete} & \multirow{2}{*}{HSC1-C} & \multirow{2}{*}{$-$} & \citet{Eurocode2-1-2};\\
                  &          &            & \citet{Mindeguia2009} \\
\hline
$f_{c,ref}$       & $61.0$   & $\rm{MPa}$ & \multirow{2}{*}{\citet[Table~24]{Mindeguia2009}} \\
$f_{t,ref}$       & $3.76$   & $\rm{MPa}$ & \\
\hline
$c$  & $550$      & $\rm{kg\,m^{-3}}$     & \citet[Table~14]{Mindeguia2009} \\
\hline
$\theta_{ref}$    & $293.15$ & $\rm{K}$   & determined from \\
$\phi_{ref}$      & $0.1027$ & $-$        & \citet[Table~25]{Mindeguia2009}, \\
$A_{\phi}$        & $1.0624 \times 10^{-4}$ & $\rm{K^{-1}}$ & see Figure~\ref{Fig_Example2_Porosity} \\
\hline
\multirow{3}{*}{$\rho_s $} & \multirow{3}{*}{$2660$} & \multirow{3}{*}{$\rm{kg\,m^{-3}}$} & determined from \\
                  &          &            & \citet[Table~26]{Mindeguia2009}, \\
                  &          &            & see Figure~\ref{Fig_Example2_Density} \\
\hline
$\theta_{ref}$    & $293.15$ & $\rm{K}$   & determined from \\
$\lambda_{d,ref}$ & $2.0153$ & $\rm{W\,m^{-1}\,K^{-1}}$ & \citet[Table~28]{Mindeguia2009}, \\
$A_{\lambda}$     & $- 9.8533 \times 10^{-4}$  & $\rm{K^{-1}}$ & see Figure~\ref{Fig_Example2_Conductivity} \\
\hline
$K_{ref}$         & $4.0 \times 10^{-20}$ & $\rm{m^{2}}$ & $-$ \\
\hline
\end{tabular}
\label{Tab_Example2_Parameters}
\end{table}

The concrete of the test specimens (denoted as C60 concrete in \citep{Mindeguia2009}) is assumed to be a high strength concrete, class 1 (see \citet[Section~6.1]{Eurocode2-1-2} and Section~\ref{Sec_MaterialProperties}), with calcareous aggregates, which can be denoted as "HSC1-C". This classification determines the free thermal strain of concrete and the temperature dependent reduction of its compressive strength according to \citet{Eurocode2-1-2}, as described in Section~\ref{Sec_MaterialProperties}.

The reference compressive and tensile strengths of concrete at the room temperature are taken from \citet[Table~24]{Mindeguia2009}.

The mass of cement per unit volume of concrete is taken from \citet[Table~14]{Mindeguia2009}.

\begin{figure}[h]
\centering
\begin{minipage}[t]{.47\textwidth}
  \centering
  {\includegraphics[angle=0,width=6.4cm]{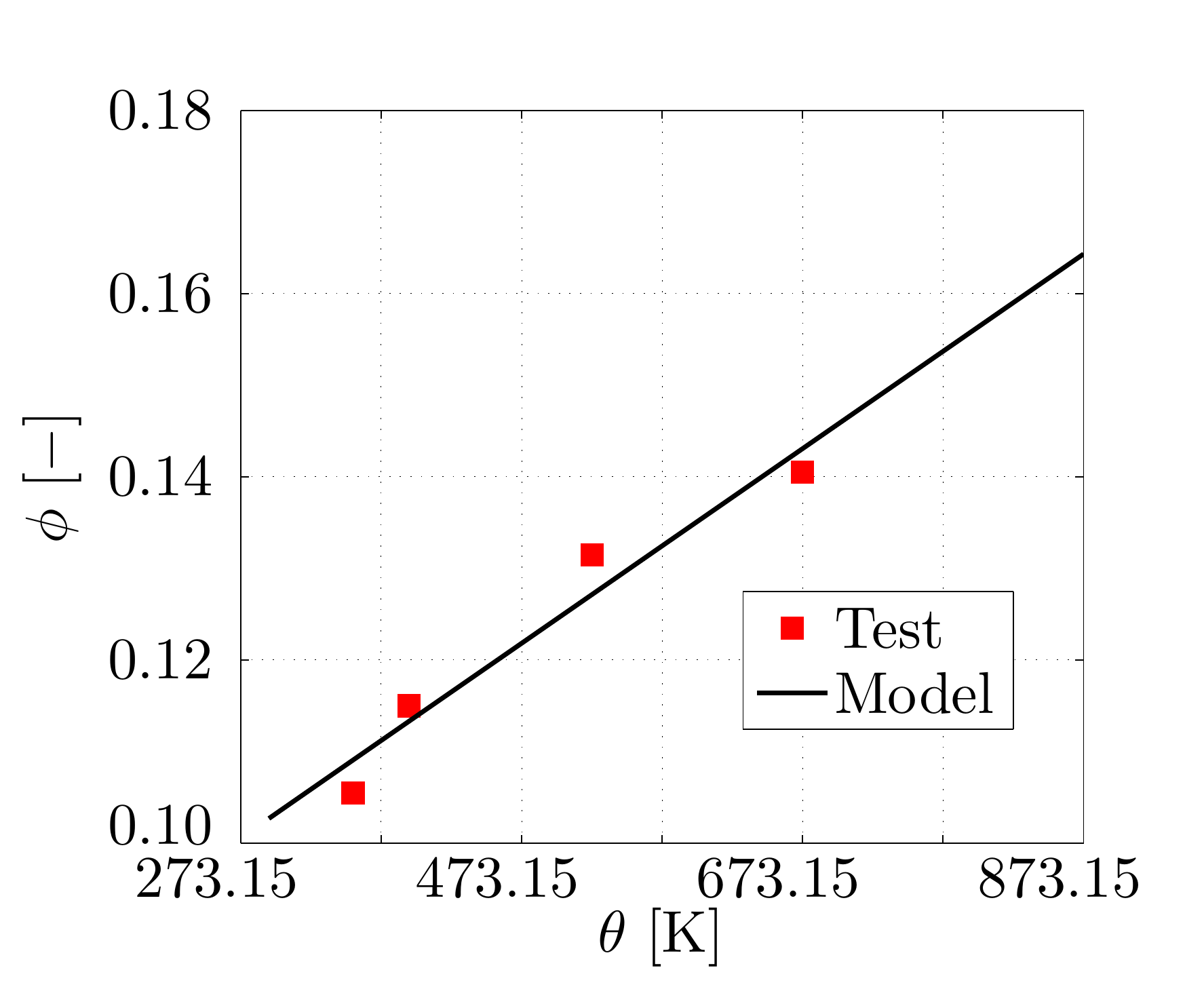}}
  \caption{Porosity of concrete measured by \citet{Mindeguia2009} (points) and assumed in our model (line)}
  \label{Fig_Example2_Porosity}
\end{minipage}%
\hspace{.01\textwidth}
\begin{minipage}[t]{.47\textwidth}
  \centering
  {\includegraphics[angle=0,width=6.4cm]{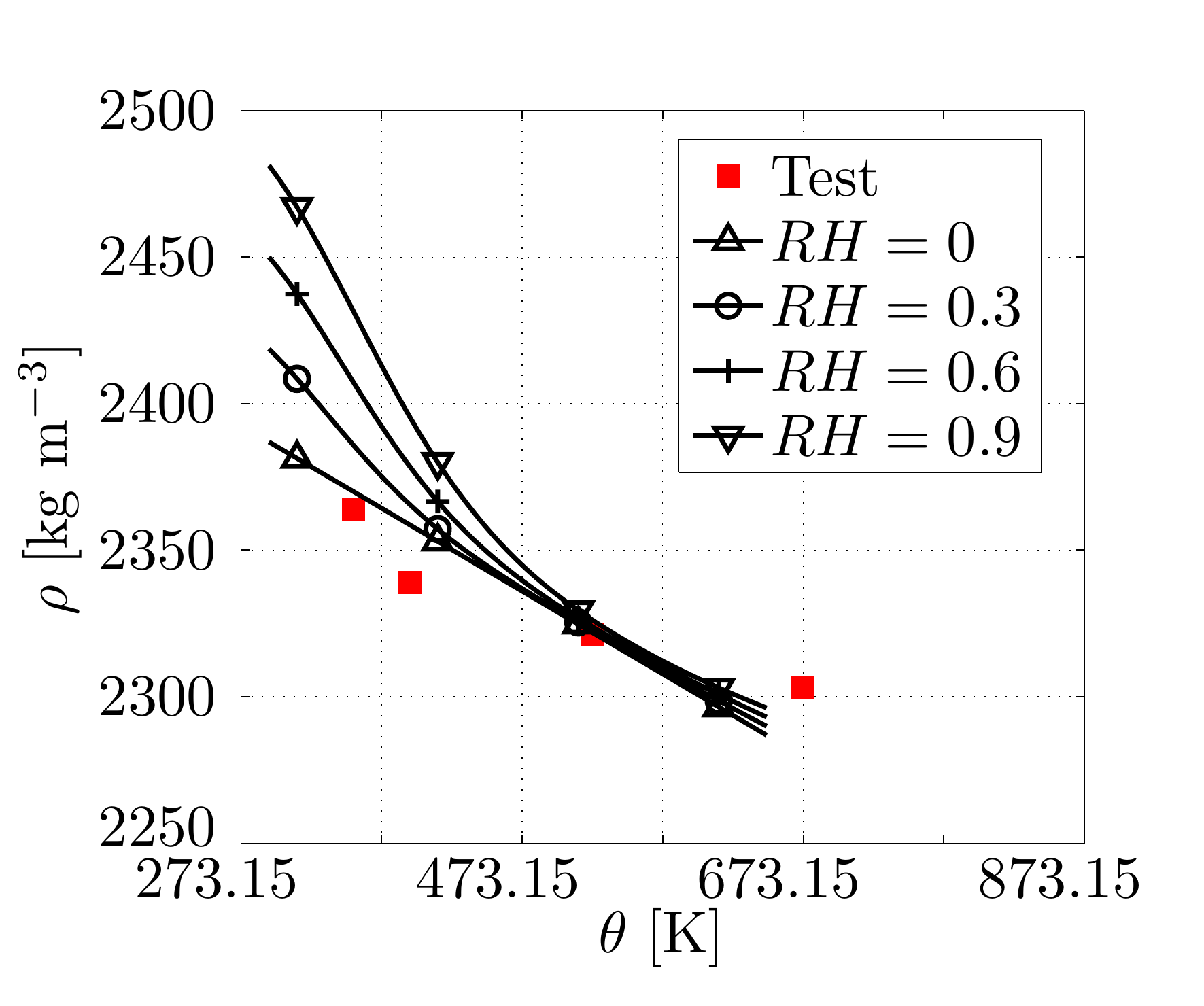}}
  \caption{Apparent density of concrete measured by \citet{Mindeguia2009} (points) and assumed in our model (lines)}
  \label{Fig_Example2_Density}
\end{minipage}
\end{figure}

For the porosity of concrete, the parameters of equation \eqref{GawinPorosity} are determined by a linear regression of the data stated in \citep[Table~24]{Mindeguia2009}, see Figure~\ref{Fig_Example2_Porosity}.

The density of solid skeleton, $\rho_{s}$, is estimated from the data measured by \citet[Table~26]{Mindeguia2009}. The corresponding apparent density assumed in our model appears in Figure~\ref{Fig_Example2_Density}.

\begin{figure}[h]
\centering
\begin{minipage}[t]{.47\textwidth}
  \centering
  {\includegraphics[angle=0,width=6.4cm]{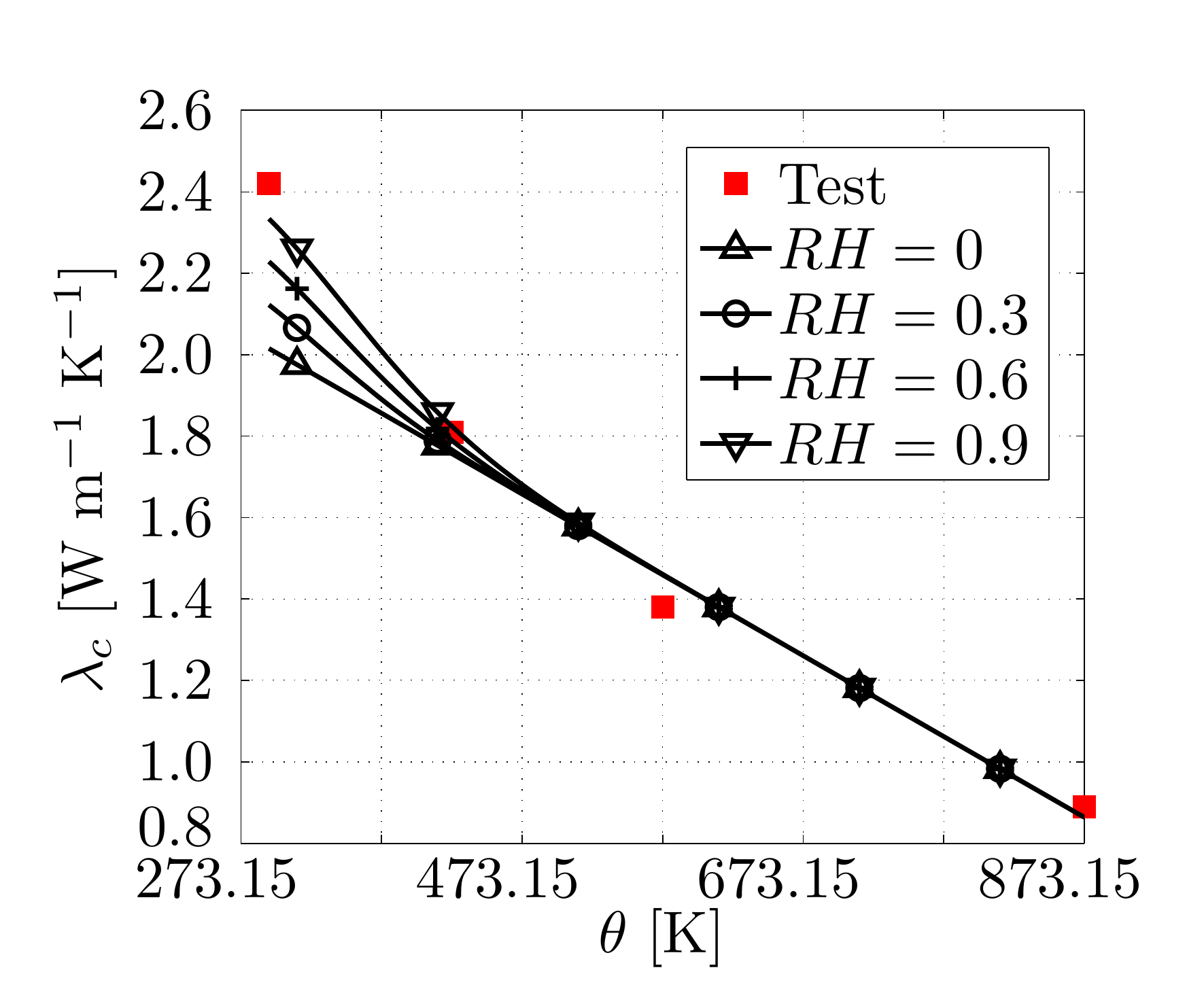}}
  \caption{Thermal conductivity of concrete measured by \citet{Mindeguia2009} (points) and assumed in our model (lines)}
  \label{Fig_Example2_Conductivity}
\end{minipage}%
\hspace{.01\textwidth}
\begin{minipage}[t]{.47\textwidth}
  \centering
  \includegraphics[angle=0,width=6.4cm]{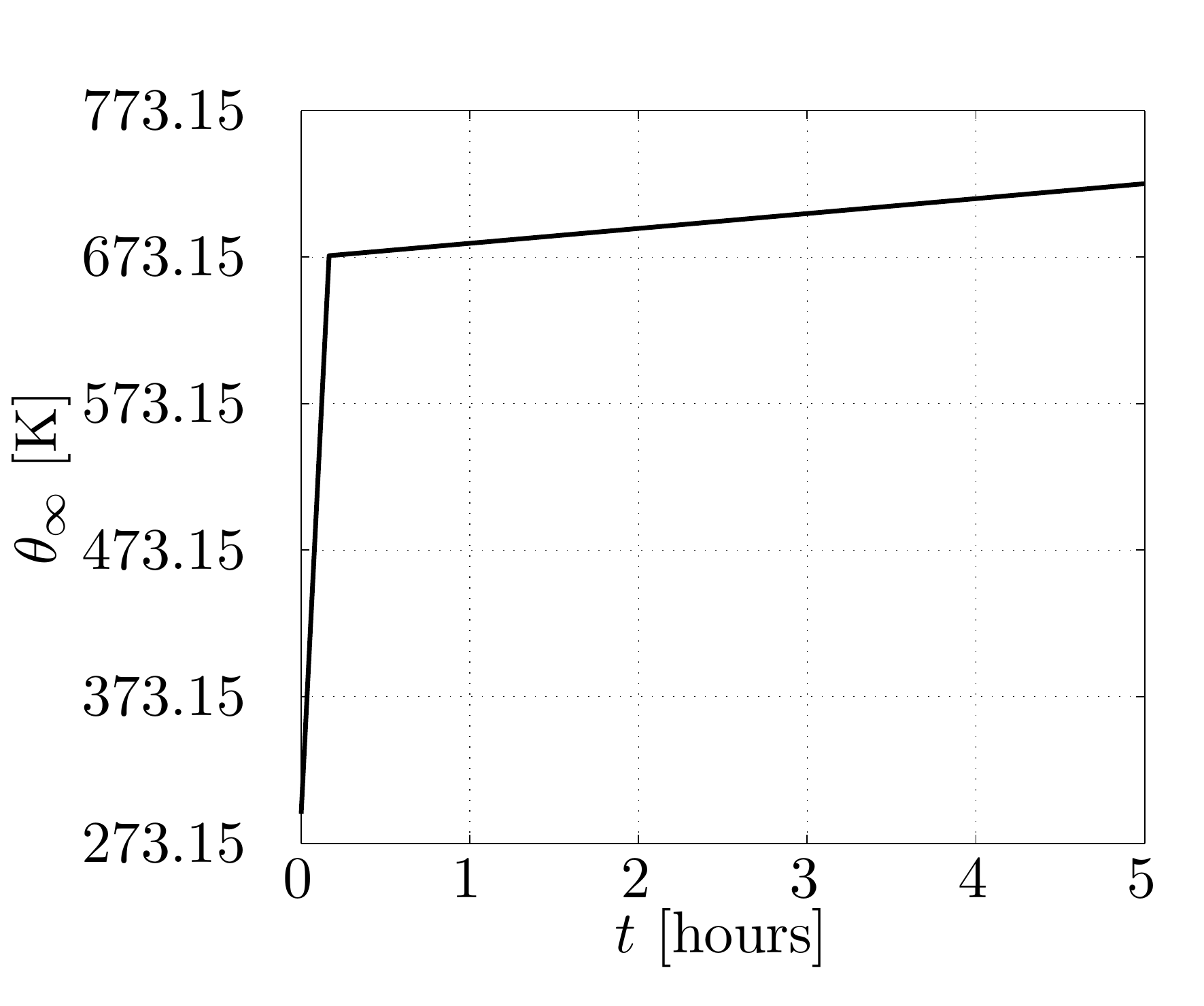}
  \caption{Air temperature on the heated side of the specimen used for the simulation of the PTM test 2}
  \label{Fig_Example2_Heating}
\end{minipage}
\end{figure}

The parameters of equation \eqref{ConductivityDry} for the thermal conductivity of dry concrete are determined by a linear regression of the data stated in \citep[Table~28]{Mindeguia2009} (excluding the value for the temperature of $20~^\circ{\rm C}$), see Figure~\ref{Fig_Example2_Conductivity} ($RH = 0$). The resulting thermal conductivity of concrete assumed in our model is shown in Figure~\ref{Fig_Example2_Conductivity}.

The intrinsic permeability of concrete was tested by \citet{Mindeguia2009} on concrete samples dried at $80~^\circ{\rm C}$ and then heated up to several temperature levels.
However, the reference permeability of undamaged concrete at the room temperature is not reported in \citep{Mindeguia2009} and hence it is determined by a trial-and-error method in our simulation (cf. \citep[p.~277]{Witek2007}).

The scheme of the analysed problem, the spatial discretization, the time discretization, and the characteristic time of spalling are identical as in the previous example (see Section~\ref{Sec_Example1_Modelling}).

The initial conditions are assumed as $P_0 = 1.9194 \times 10^3~\rm{Pa}$, $\theta_0 = 293.15~\rm{K}$, $\ell_0 = 0.12~\rm{m}$. For these values, the initial saturation with liquid water $S_{w,0} = 0.78$, which is the value reported by \citet[Table~18]{Mindeguia2009}.

The ambient temperature on the heated side, $\theta_{\infty}~\rm{[K]}$, has been determined by a trial-and-error method as (see Figure~\ref{Fig_Example2_Heating})
\begin{equation}\label{Heating2}
\theta_{\infty}(t) = \left\{ \begin{array}{lll}
\displaystyle 293.15 + t \frac{380}{300} & \rm{for} & t \leq 300~\rm{s},
\\
\displaystyle 673.15 + (t-300) \frac{50}{17700} & \rm{for} & t > 300~\rm{s},
\end{array} \right.
\end{equation}
where $t~\rm{[s]}$ is the time of heating, and it is practically the same as for the Kalifa's PTM test 1, cf. Figures~\ref{Fig_Example1_Heating} and \ref{Fig_Example2_Heating}.

The boundary conditions are summarized in Table~\ref{Tab_Example2_Boundary_Conditions}.

\begin{table}[h]
\centering
\caption{Boundary conditions parameters for simulation of PTM test 2}
\small
\begin{tabular}{c c c c}
\hline
\multirow{2}{*}{Variable} &
\multicolumn{2}{c}{Value} &
\multirow{2}{*}{Unit} \\ \cline{2-3}
& Unexposed side & Exposed side & \\
\hline\hline
$P_{\infty}$      & $P_0$      & $P_0$   & $\rm{Pa}$\\
$\theta_{\infty}$ & $\theta_0$ & $\theta_{\infty}(t)$, see \eqref{Heating2} & $\rm{K}$ \\
$\alpha_c$        & $4$        & $20$    & $\rm{W\,m^{-2}\,K^{-1}}$ \\
$e\sigma$         & $0.7 \times 5.67 \times 10^{-8}$  & $0.7 \times 5.67 \times 10^{-8}$ & $\rm{W\,m^{-2}\,K^{-4}}$ \\
$\beta_c$         & $0.009$    & $0.019$ & $\rm{m\,s^{-1}}$ \\
\hline
\end{tabular}
\label{Tab_Example2_Boundary_Conditions}
\end{table}

\subsubsection{Discussion}
\label{Sec_Example2_Discussion}
\begin{figure}[h]
  \centering
  {\includegraphics[angle=0,width=11.6cm]{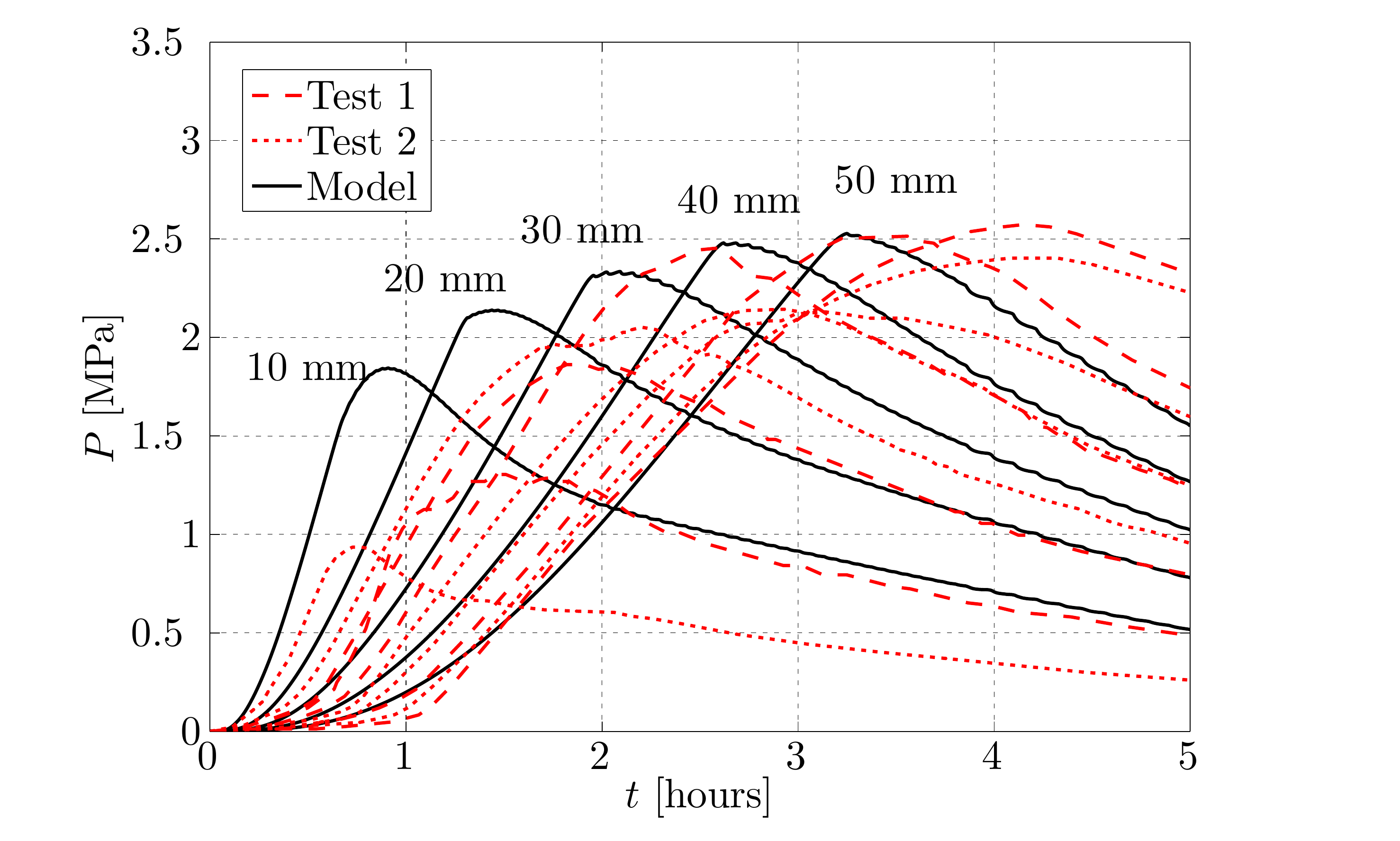}}
  \caption{Pore pressure evolution measured by \citet[Figure~154c]{Mindeguia2009} (dashed lines, dotted lines) and determined by the present model (solid lines)}
  \label{Fig_Example2_Comparison_Press}
\end{figure}

\begin{figure}[h]
  \centering
  {\includegraphics[angle=0,width=11.6cm]{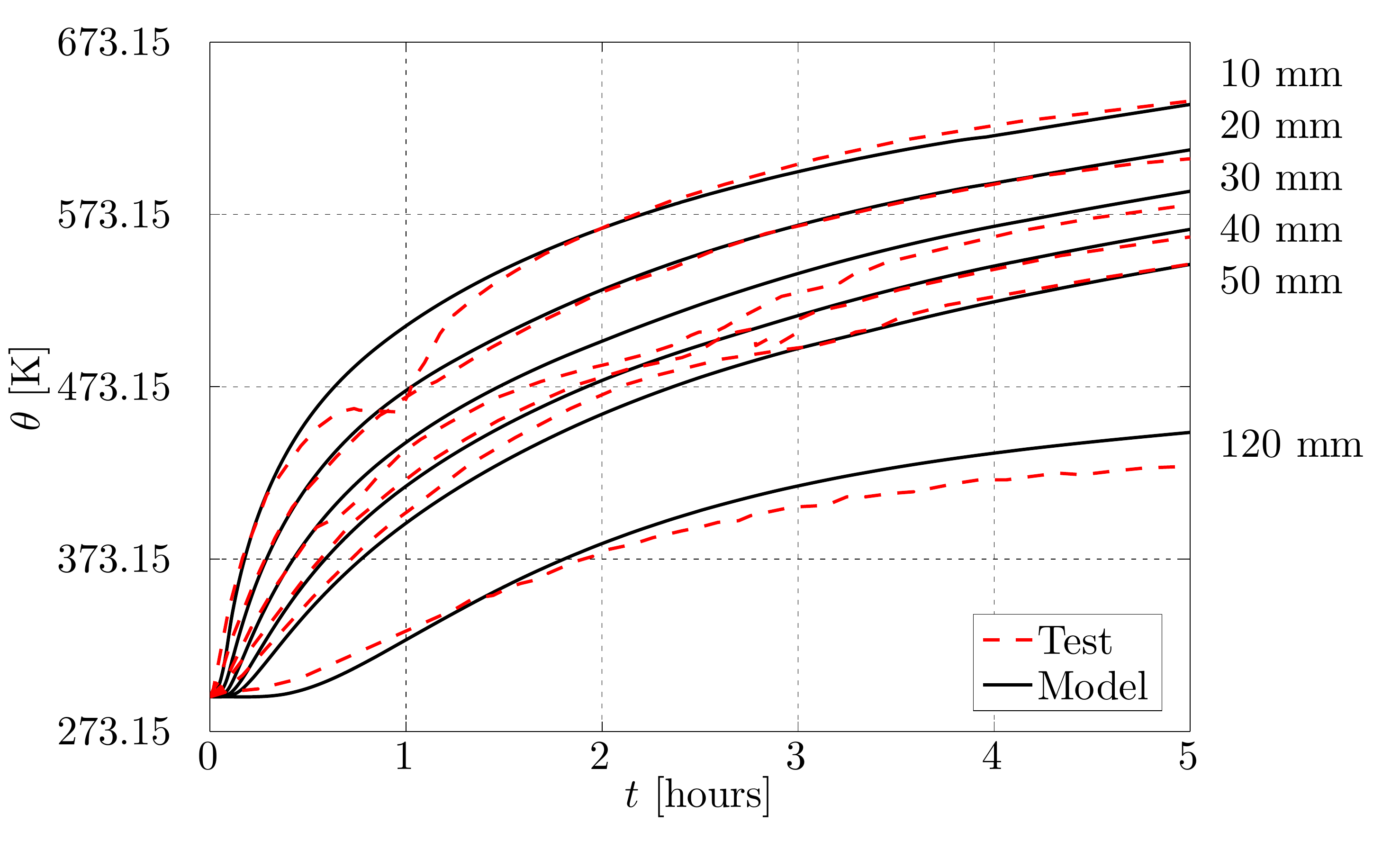}}
  \caption{Temperature evolution measured by \citet[Figure~157c]{Mindeguia2009} (dashed lines) and determined by the present model (solid lines)}
  \label{Fig_Example2_Comparison_Temp}
\end{figure}

\begin{figure}[h]
\centering
\begin{tabular}{cc}
  \\
  \includegraphics[angle=0,width=6.4cm]{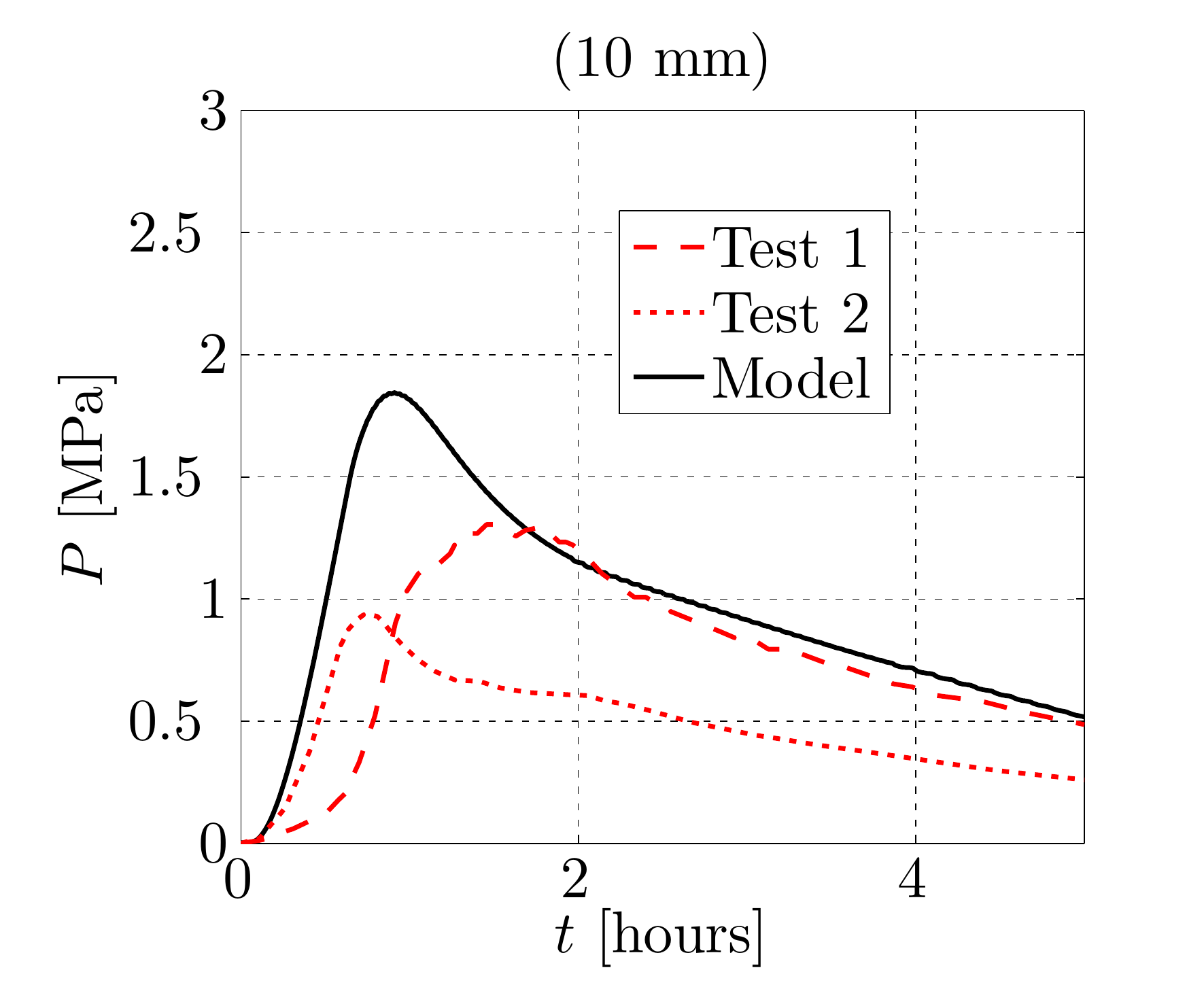} &
  \includegraphics[angle=0,width=6.4cm]{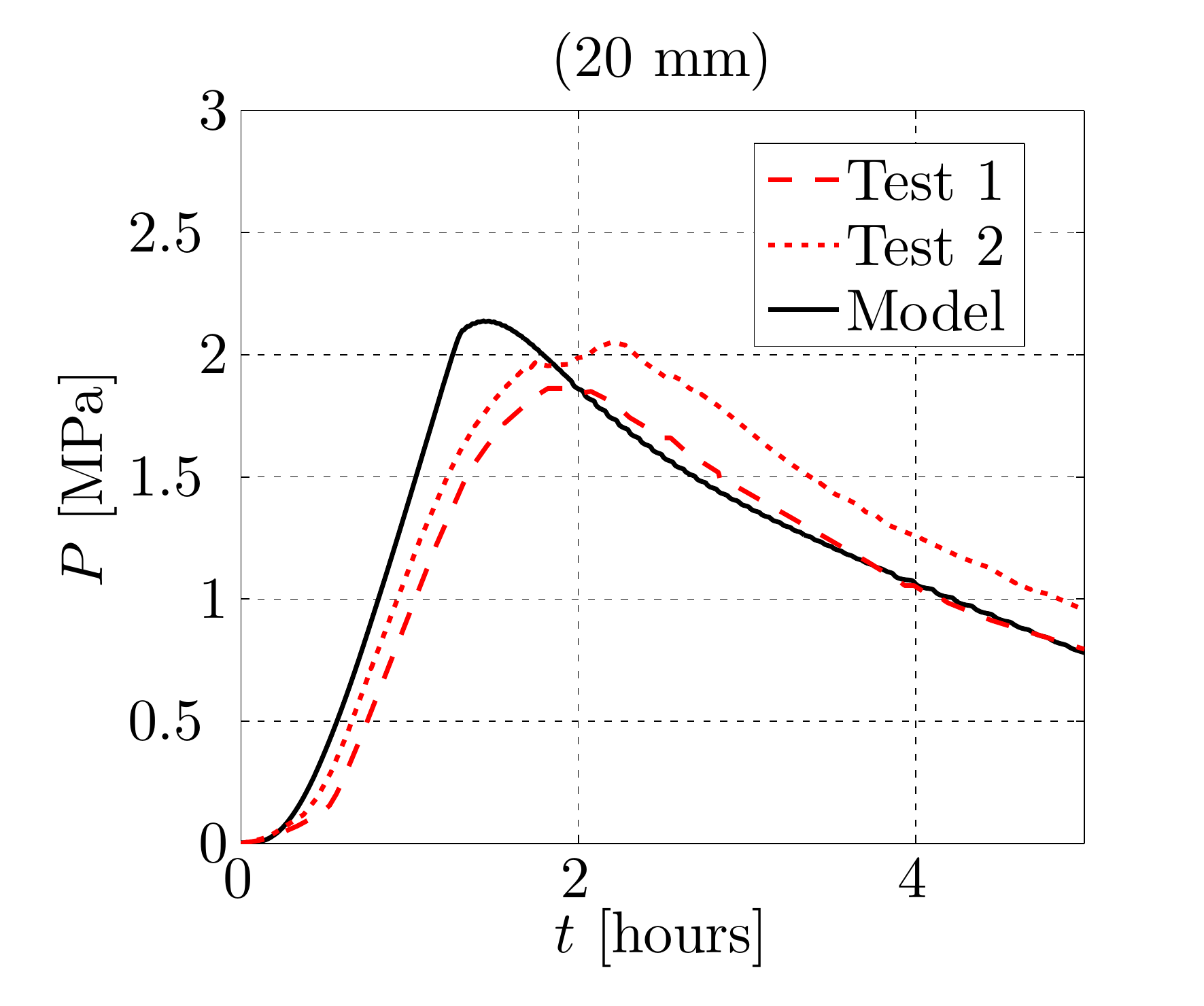}
  \\
  \includegraphics[angle=0,width=6.4cm]{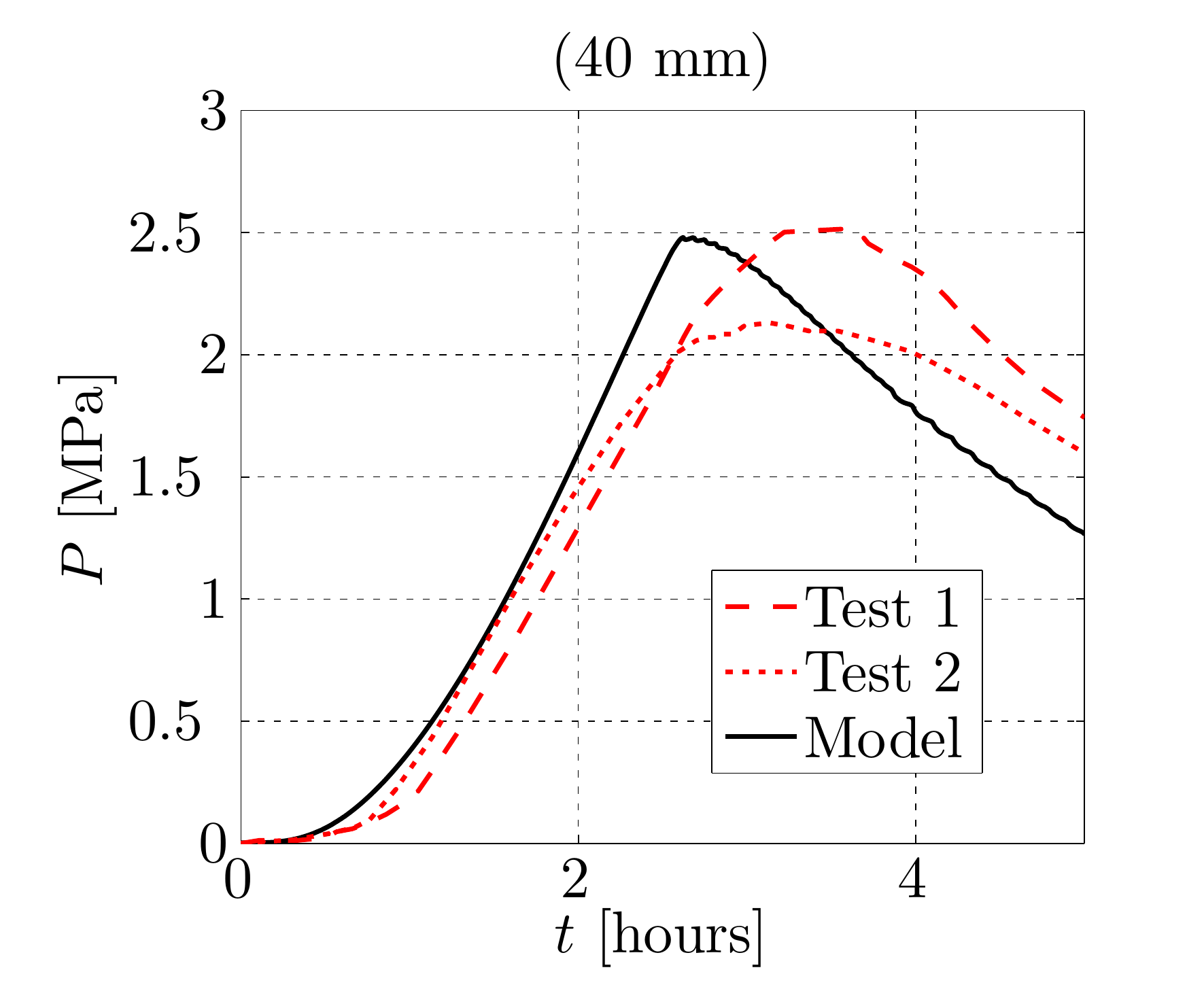} &
  \includegraphics[angle=0,width=6.4cm]{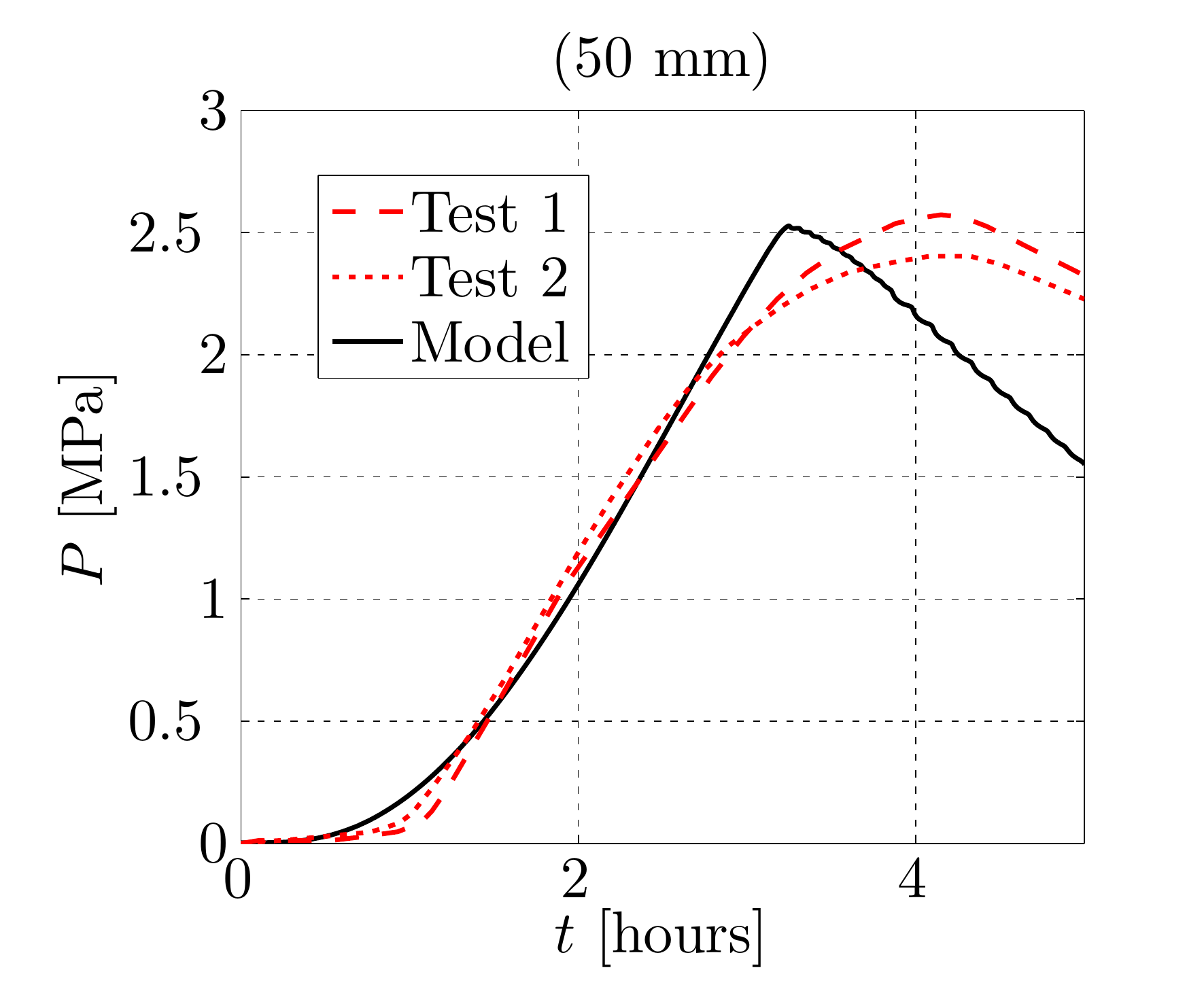}
\end{tabular}
\caption{Pore pressure evolution measured by \citet[Figure~157c]{Mindeguia2009} and determined by the present model at the depth of (10, 20, 40, 50)~mm from the heated surface}
\label{Fig_Example2_Comparison_Press_yy}
\end{figure}

The resulting pore pressure and temperature evolutions determined by our simulation are compared with the experimentally measured data in Figures~\ref{Fig_Example2_Comparison_Press} and \ref{Fig_Example2_Comparison_Temp} (note that the distances in millimeters stated in the figures are, in accordance with \citet{Mindeguia2009}, measured from the heated surface). The more detailed illustration of the pore pressure prediction and its comparison with the test data is shown in Figure~\ref{Fig_Example2_Comparison_Press_yy}.

From Figures~\ref{Fig_Example2_Comparison_Press}--\ref{Fig_Example2_Comparison_Press_yy}, it is obvious that also in this case, the present model provides an accurate prediction of the hygro-thermal behaviour of the heated concrete specimen.
{Similarly as in the previous case, the potential inaccuracies arising from the simplifications of the model (see the introductory part of this section) are probably excluded by the appropriate setting of the reference permeability of undamaged concrete at the room temperature (fitted by a trial-and-error method)}.

\begin{figure}[h]
\centering
\begin{minipage}[t]{.47\textwidth}
  \centering
  {\includegraphics[angle=0,width=6.4cm]{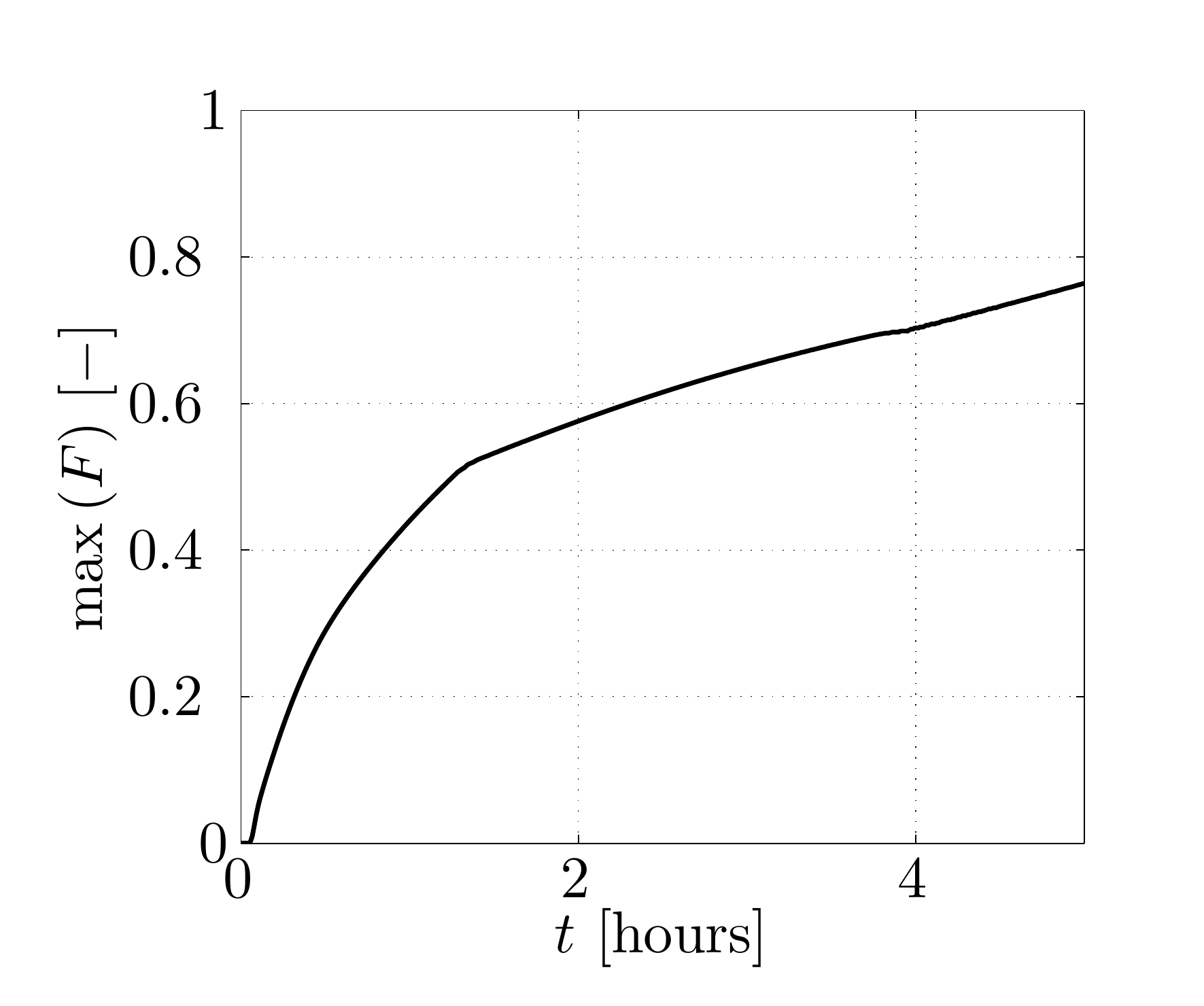}}
  \caption{Evolution of the failure parameter}
  \label{Fig_Example2_Failure}
\end{minipage}%
\hspace{.01\textwidth}
\begin{minipage}[t]{.47\textwidth}
  \centering
  {\includegraphics[angle=0,width=6.4cm]{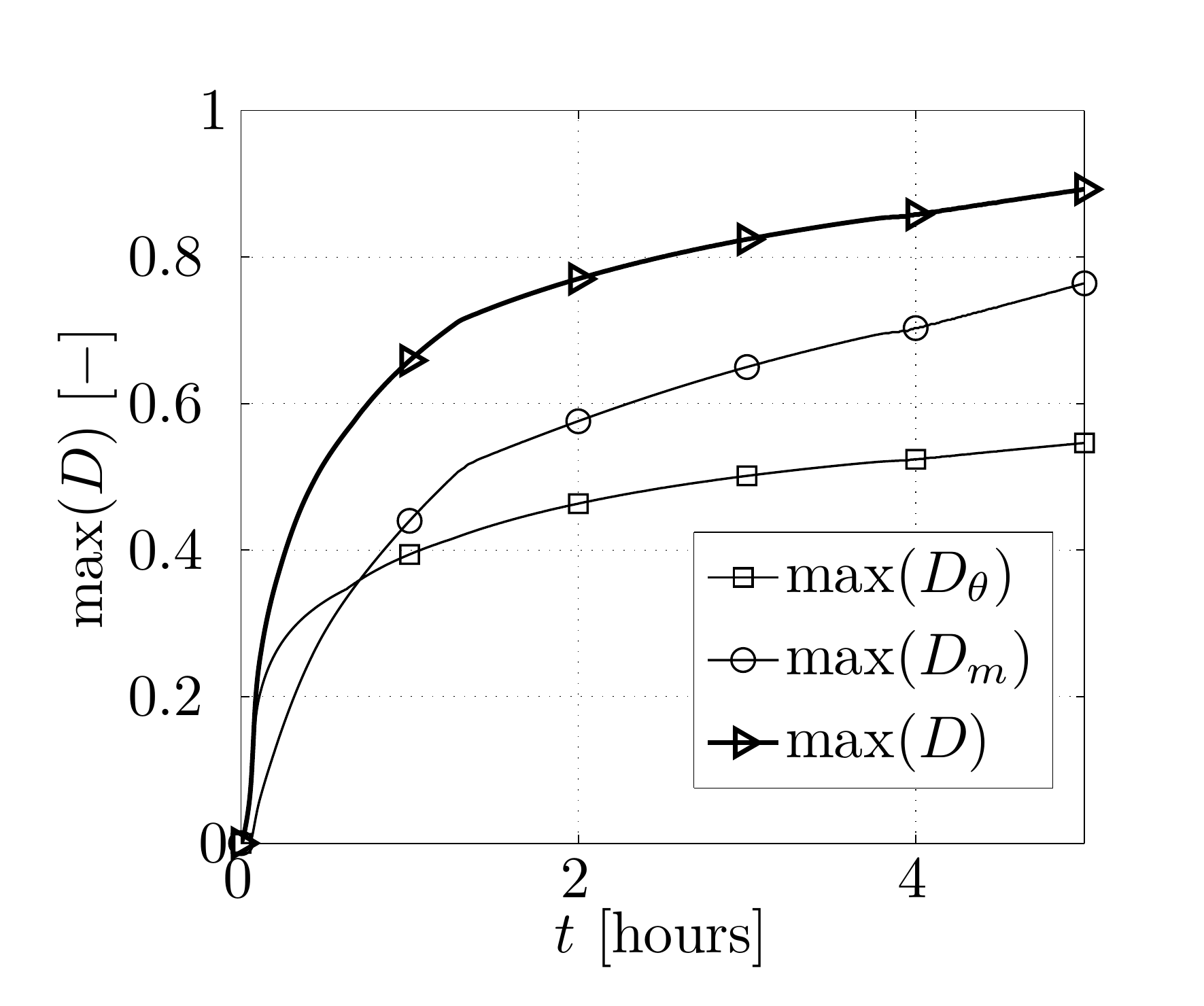}}
  \caption{Evolution of the damage parameter}
  \label{Fig_Example2_Damage}
\end{minipage}
\end{figure}

The spalling behaviour is also simulated correctly since, as mentioned above, during the test, no spalling was observed by \citet{Mindeguia2009}, which is in accordance with our simulation, see Figure~\ref{Fig_Example2_Failure} -- the maximal value of failure function \eqref{failure criterion_simplified} achieved within the specimen during the test period does not exceeds the value of 1. The evolution of the maximal values of the damage parameter and its components within the specimen during the test period determined by our model is shown in Figure~\ref{Fig_Example2_Damage}.

\subsection{Simulation of the spalling test}
\label{Sec_Example3}

\subsubsection{Experiment description}
\label{Sec_Example3_Experiment}
The spalling experiments were performed by \citet{Mindeguia2009} on the $700 \times 600 \times 150~\rm{mm^3}$ prismatic specimens made of various types of concrete and exposed to various fire scenarios. Hereafter, the spalling test of a high strength concrete (denoted as "B60" in \citep{Mindeguia2009}) sample (conditioned in "Air\_2" according to \citet{Mindeguia2009}) heated by the ISO~834 fire (in the 150-mm direction) for 60~minutes. This test was duplicated and the data from the both measurements of the spalling depths are stated in \citep{Mindeguia2009}. The specimen was placed on the top of the furnace and it was exposed to heating on the area of $420 \times 600~\rm{mm^2}$ (see \citep[Figure~170]{Mindeguia2009}). As mentioned by \citet{Mindeguia2009,Mindeguia2013}, during the test, the specimen was not subjected to any mechanical load.

Within the test, the temperature and the pore pressure propagations through the specimen were recorded. After the termination of heating, the spalling depths on the exposed surface were closely measured which enables to determine the maximal and average depths of spalling and depict the well-arranged "spalling maps" (see \citep{Mindeguia2009,Mindeguia2013}).

The detailed description of the spalling experiments performed by Mindeguia can be found in \citep{Mindeguia2009,Mindeguia2013}.

\subsubsection{Modelling}
\label{Sec_Example3_Modelling}
The thermal and hygral properties of the B60 concrete as well as its hygro-thermal behaviour are described in Section~\ref{Sec_Example2_Modelling}.

\begin{figure}[h]
  \centering
  {\includegraphics[angle=0,width=8.7cm]{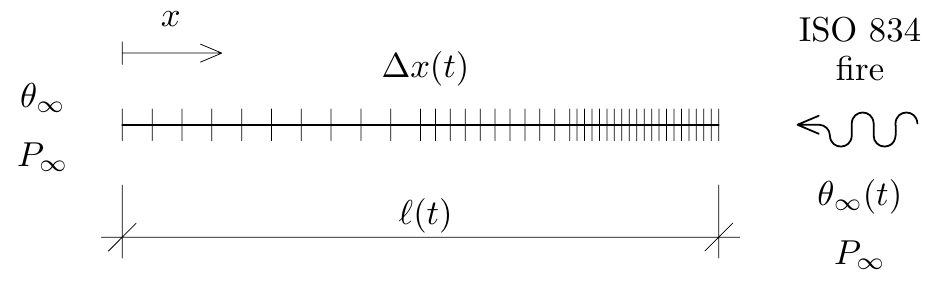}}
  \caption{Scheme of the analysed problem}
  \label{Fig_Example3_Scheme}
\end{figure}

Scheme of the analysed problem is displayed in Figure~\ref{Fig_Example3_Scheme}. The spatial discretization is performed with the use of linear 1-D elements. In total, 160 elements are employed -- 40 elements in the interval $x \in (0,\ell/2)$), 40 elements for $x \in (\ell/2,3\ell/4)$ and 80 elements in the interval $x \in (3\ell/4,\ell)$. For the time discretization as well for the characteristic time of spalling, three values are assumed: $\Delta t = 1~\rm{s}$, $\Delta t = 0.5~\rm{s}$, $\Delta t = 0.1~\rm{s}$, and $\gamma = 1~\rm{s}$, $\gamma = 10~\rm{s}$, and $\gamma = 100~\rm{s}$, respectively, which leads to 9 numerical experiments discussed in Section~\ref{Sec_Example3_Discussion}.

The initial conditions are assumed as $P_0 = 1.9194 \times 10^3~\rm{Pa}$, $\theta_0 = 293.15~\rm{K}$, $\ell_0 = 0.15~\rm{m}$. For these values, the initial saturation with liquid water $S_{w,0} = 0.78$, which is the average value of the two initial saturations reported by \citet[Table~20]{Mindeguia2009}.

The ambient temperature on the heated side is governed by the ISO~834 fire curve
\begin{equation}\label{Heating3}
\theta_{\infty}(t) = 293.15 + 345 \log(8t + 1),
\end{equation}
where $t~\rm{[minutes]}$ is the time of heating.

The boundary conditions are summarized in Table~\ref{Tab_Example3_Boundary_Conditions}.

\begin{table}[h]
\centering
\caption{Boundary conditions parameters for simulation of spalling test}
\small
\begin{tabular}{c c c c}
\hline
\multirow{2}{*}{Variable} &
\multicolumn{2}{c}{Value} &
\multirow{2}{*}{Unit} \\ \cline{2-3}
& Unexposed side & Exposed side & \\
\hline\hline
$P_{\infty}$       & $P_0$      & $P_0$   & $\rm{Pa}$\\
$\theta_{\infty}$  & $\theta_0$ & $\theta_{\infty}(t)$, see \eqref{Heating3} & $\rm{K}$ \\
$\alpha_c$         & $4$        & $25$    & $\rm{W\,m^{-2}\,K^{-1}}$ \\
$e\sigma$          & $0.7 \times 5.67 \times 10^{-8}$  & $0.7 \times 5.67 \times 10^{-8}$ & $\rm{W\,m^{-2}\,K^{-4}}$ \\
$\beta_c$          & $0.009$    & $0.019$ & $\rm{m\,s^{-1}}$ \\
\hline
\end{tabular}
\label{Tab_Example3_Boundary_Conditions}
\end{table}

\subsubsection{Discussion}
\label{Sec_Example3_Discussion}
\begin{figure}[h]
\centering
\begin{tabular}{cc}
  \multicolumn{2}{c}{\includegraphics[angle=0,width=10.65cm]{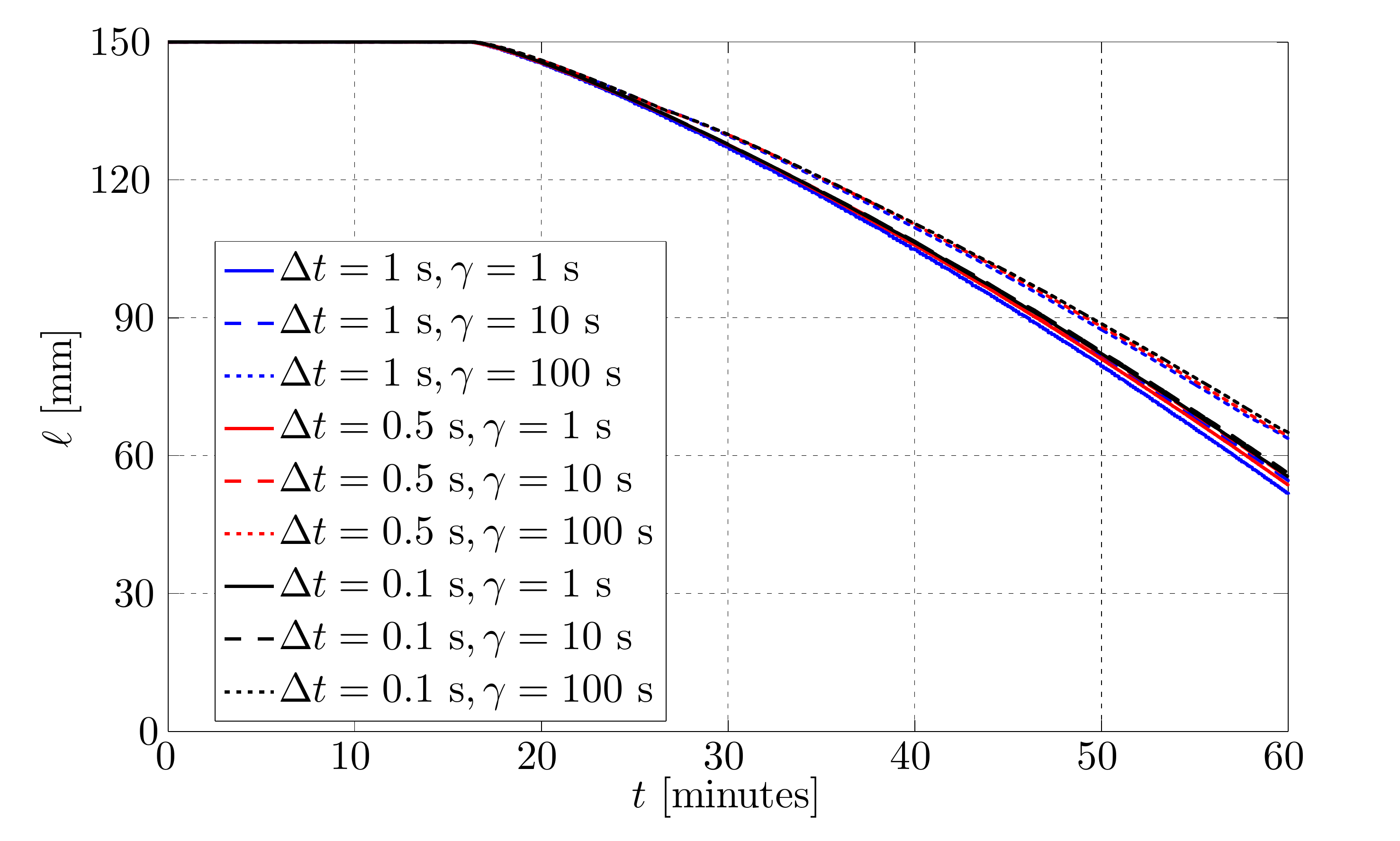}}
  \\
  \includegraphics[angle=0,width=6.4cm]{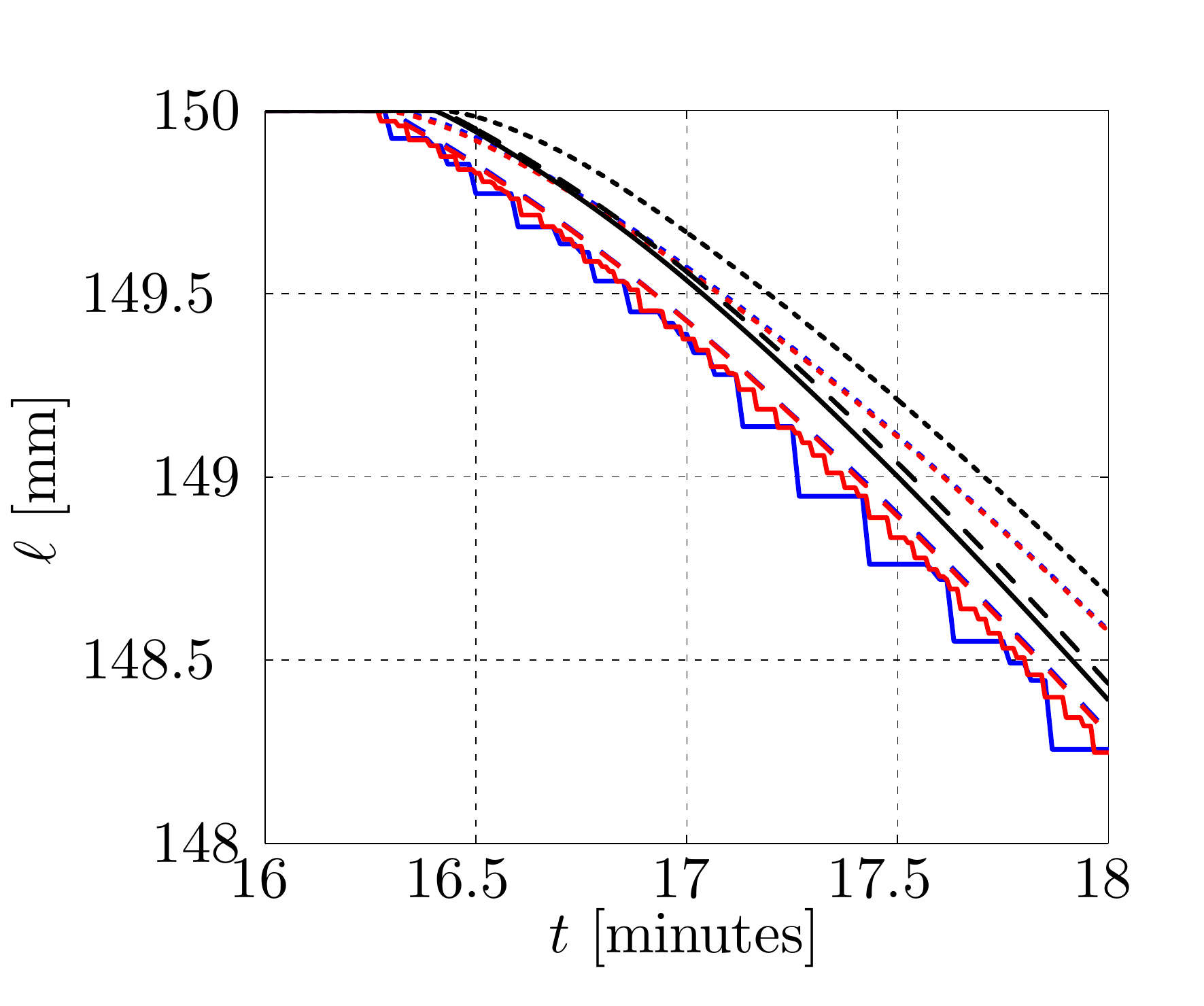}
  &
  \includegraphics[angle=0,width=6.4cm]{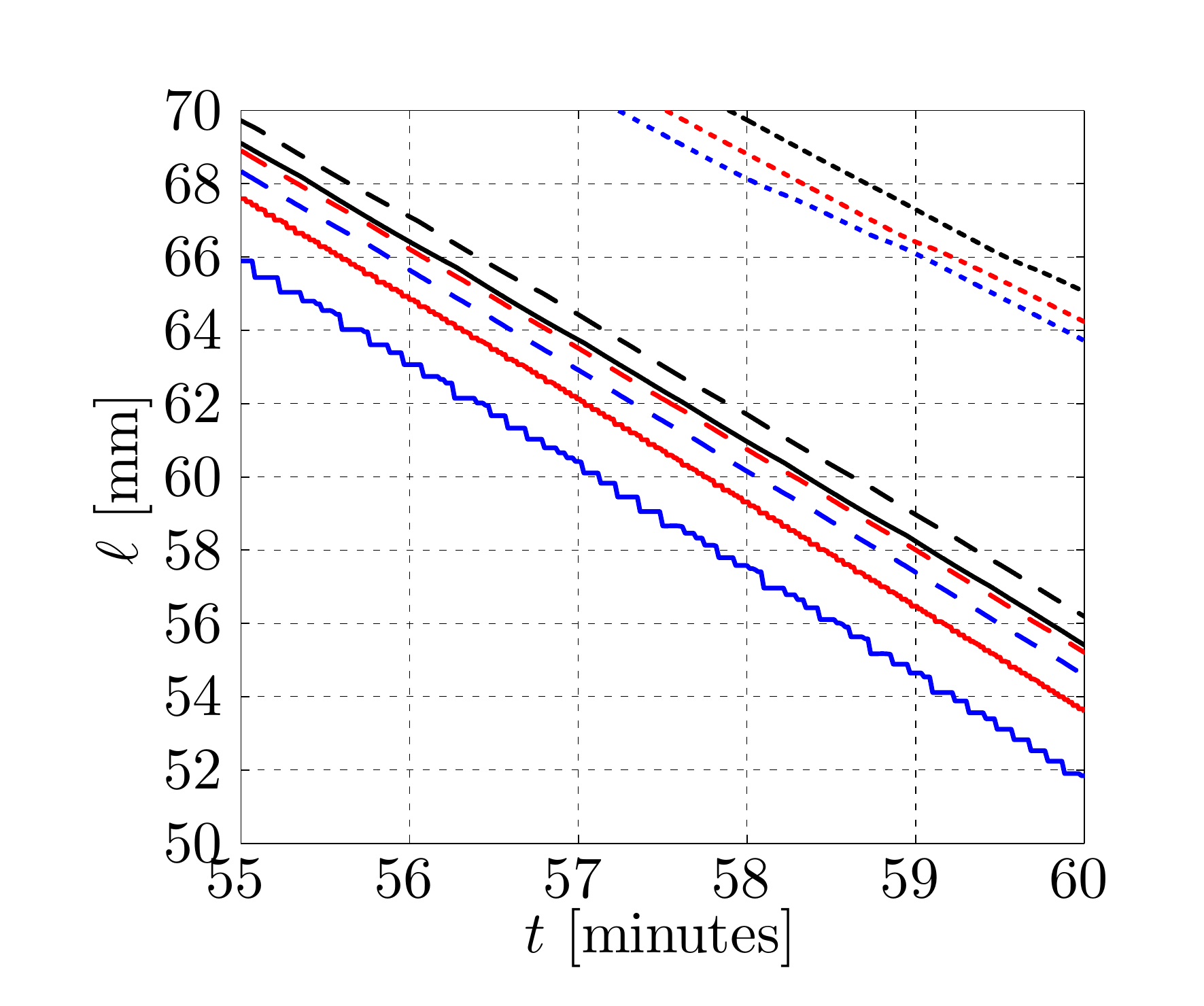}
\end{tabular}
\caption{Reduction of the specimen thickness due to concrete spalling for the analysed experiment \citep{Mindeguia2009} obtained by the present model}
\label{Fig_Example3_Thickness_Evolution}
\end{figure}

In this case, the spalling of concrete occurred during the test, which is also correctly simulated by the present model. The reduction of the specimen thickness obtained by the calculations is shown in Figure~\ref{Fig_Example3_Thickness_Evolution}.

\begin{figure}[h]
\centering
\begin{tabular}{cc}
  \\
  \includegraphics[angle=0,width=6.4cm]{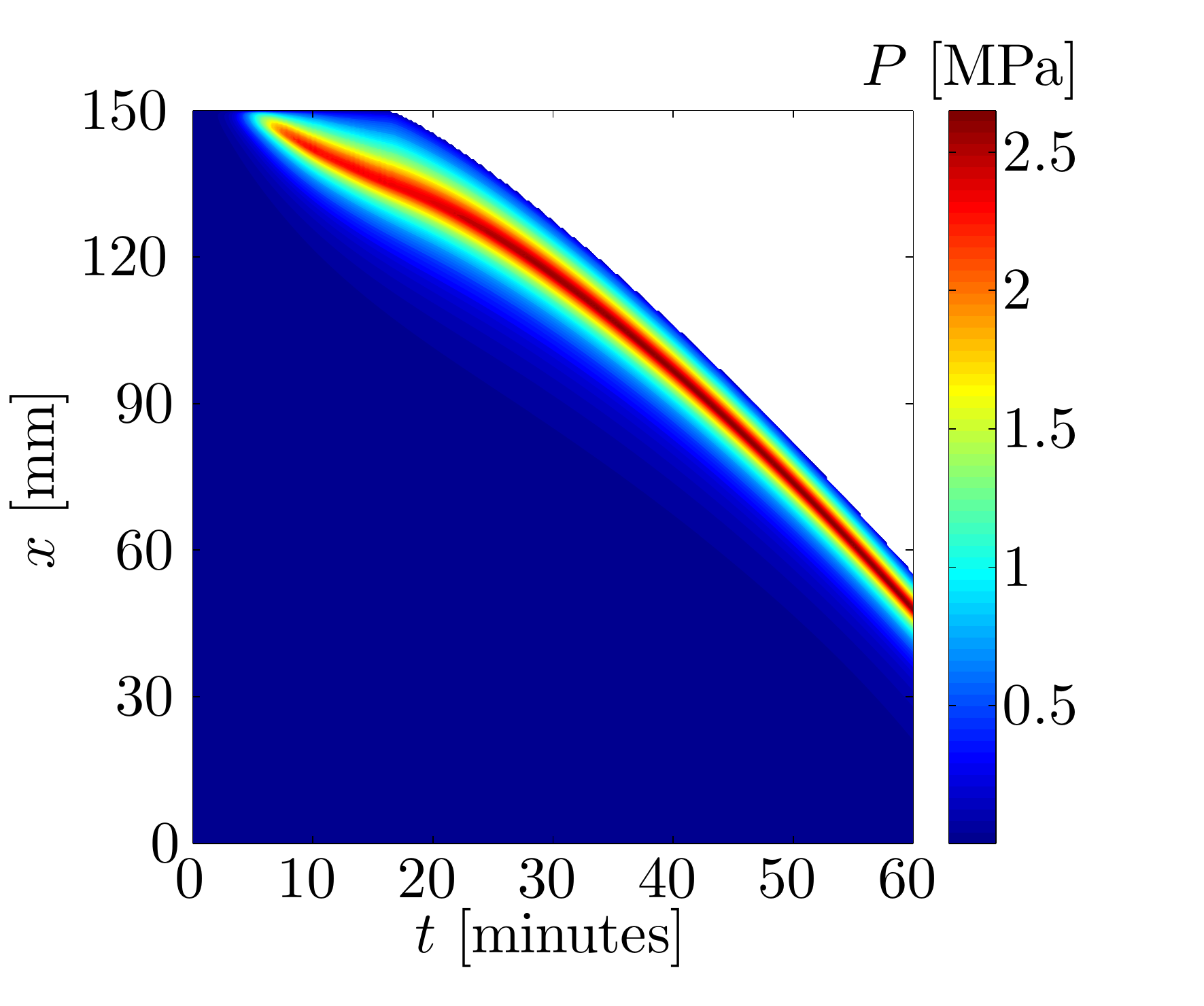}
  &
  \includegraphics[angle=0,width=6.4cm]{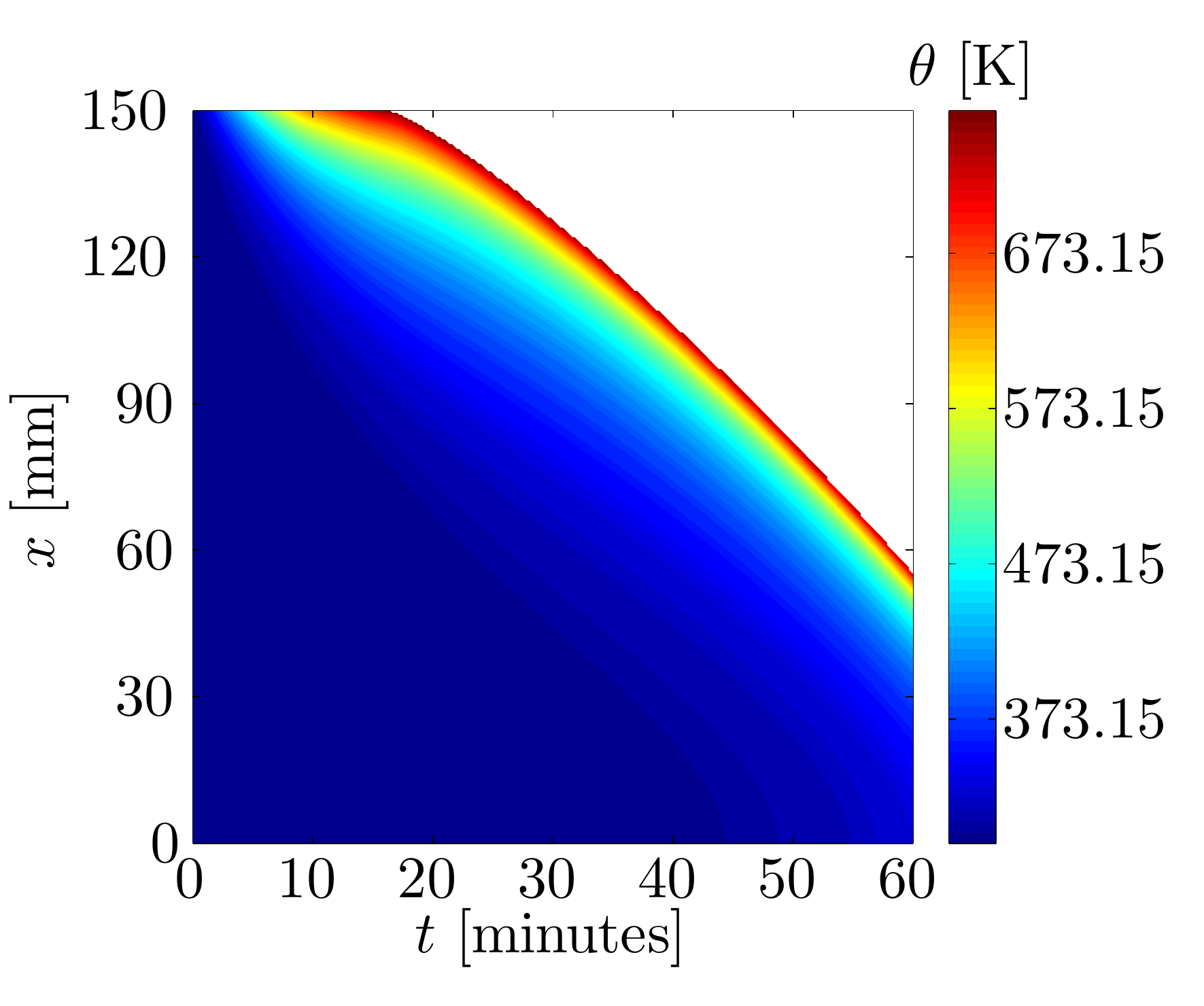}
  \\
  \includegraphics[angle=0,width=6.4cm]{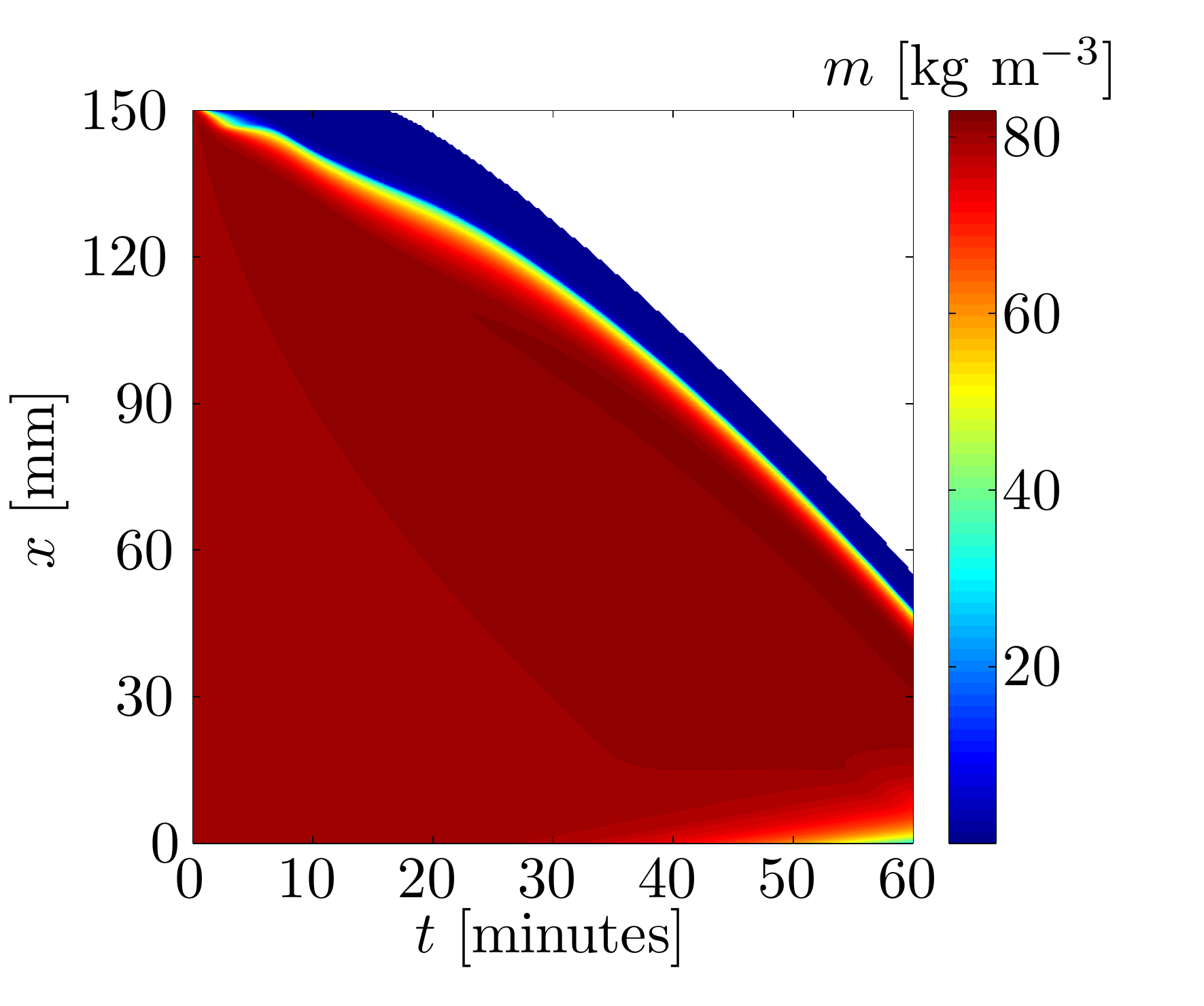}
  &
  \includegraphics[angle=0,width=6.4cm]{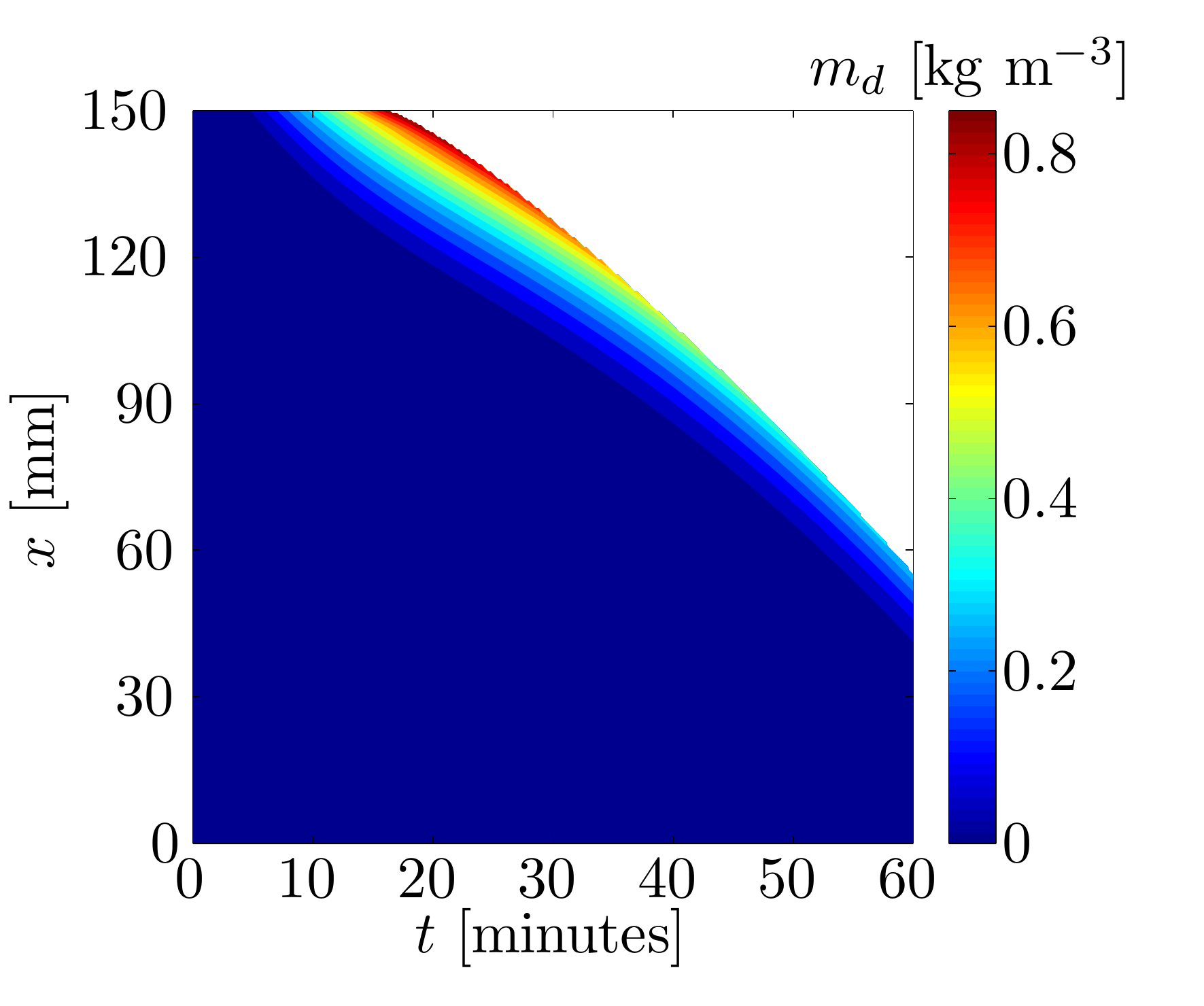}
\end{tabular}
\caption{Spatial and time distribution of the primary unknowns for the analysed experiment \citep{Mindeguia2009} obtained by the present model (for $\Delta t = 0.1~\rm{s}$ and $\gamma = 1~\rm{s}$)}
\label{Fig_Example3_Distribution}
\end{figure}

The spatial and time distributions of the primary unknowns of the present model are
shown in Figure~\ref{Fig_Example3_Distribution}.

It is obvious that the model predicted a large amount of concrete spalling -- the spalling depth at the end of heating reaches the values from $85~\rm{mm}$ to $99~\rm{mm}$, depending on $\Delta t$ and $\gamma$ assumed for the calculation. These values of spalling depths are about 2.5 to 4 times as high as measured by \citet[Table~36]{Mindeguia2009} the maximal spalling depths of $34~{\rm{mm}}$ and $23~{\rm{mm}}$ for the first and for the second test are reported, respectively.

{Moreover, the values of pore pressure determined by the simulation are also significantly higher than measured within the experiment. The maximal value of pore pressure obtained by the simulation is about $2.5~\rm{MPa}$ (see Figure~\ref{Fig_Example3_Distribution}); the maximal value of pore pressure measured by Mindeguia is $0.63~\rm{MPa}$ (see \citep[Table~38]{Mindeguia2009}).

The overestimation of pore pressures indicates that in such a rapid heating conditions, the specimen was damaged more severely (which leads to increase of the permeability of concrete) than has been predicted by the model. The fact that the increasing heating rate leads to decreasing pore pressures within the spacemen was observed by \citet[Section~3.2]{Mindeguia2009a}. In our simulations, this trend is not so evident. When comparing the pore pressure distributions within the specimen heated by a radiant heater of a temperature of $600~\rm{^{\circ}C}$ (see Figures~\ref{Fig_Example2_Comparison_Press},~\ref{Fig_Example2_Comparison_Press_yy}) with the pore pressure in the specimen made of the same concrete exposed to the ISO~834 fire (see Figure~\ref{Fig_Example3_Distribution}), we get practically identical values (about $2.5~\rm{MPa}$). In our opinion, there are three possible explanations of this fact:
\begin{itemize}
\item the Bary function \eqref{permeabiliy} adopted in our model from \citet[eq.~(13)]{Davie2012} is not valid for such a rapid heating conditions as the ISO~834 fire exposure;
\item the mechanical damage parameter influencing the increase of concrete permeability cannot be simply assumed to be equal to the failure parametr assumed in our spalling criterion, as we supposed in equation \eqref{damage_mechanical};
\item the value of reference permeability of undamaged concrete at the room temperature that has been set in our simulation of the spalling test is not correct. It has been adopted from the previous PTM test of small specimen made of the same concrete. Within the PTM test, concrete permeability has been fitted by a trial-and error method and hence, it is possible that it covers indirectly some of the factors related to the PTM experiment of small samples. However, these factors don't need to be relevant for the spalling test of large specimens.
\end{itemize}

The detailed investigation of this phenomenon needs further research and it is out of scope of the present paper.}

{The} overestimation of the specimen thickness reduction is {consequently caused by the overestimation of pore pressures as discussed in the previous paragraph, and probably also} by the assumption adopted in the model that the specimen is fully mechanically restrained in the plane perpendicular to its thickness. In the investigated experiment, the only restraint arose from the fixing the heated concrete by the unheated lateral parts and upper layers of the specimen {(see \citep{Jansson2013} and references therein)}. It should be noted that the overestimation of the spalling process has been supposed since the assumption of the fully restrained conditions is generally considered as conservative. Moreover, in the real conditions, the structure usually is mechanically restrained and hence, in such a case, the model can provide more accurate prediction.

{Nevertheless}, numerical experiments demonstrate a computational stability of the present model. From Figure \ref{Fig_Example3_Thickness_Evolution}, it is obvious that the model is not extremely sensitive on the value of time step. It is also evident that by increasing the characteristic time of spalling as well as by decreing the time step, the spalling proces becomes smoother.

\section{Conclusions}
\label{Sec_Conclusions}
In the paper, we proposed a one-dimensional coupled model for simulation of hygro-thermo-mechanical behaviour of concrete walls exposed to high temperatures including the prediction of concrete spalling. The transport-processes-related part of the model was derived from the multi-phase formulations of the laws of conservation of mass and energy. Based on the detailed quantitative parameter analysis, the less important transport phenomena were neglected in order to obtain a simplified but still robust and realistic model of transport processes in heated concrete, which can be expressed in terms of relatively small number of material parameters that can be easily obtained from material tests. The main simplifications employed in our approach consisted in (i) neglecting the diffusive mass flux of water vapour, (ii) removing the separate term describing the effect of diffusion of adsorbed water and including this effect into the liquid water relative permeability, $K_{rw}$, and (iii) ignoring the effects of variations of pressure of dry air. The permeability of concrete was supposed to be influenced both by the thermal and mechanical damage of the material, which can be expressed by a multiplicative total damage parameter.

The spalling of concrete was assumed to be caused by a combination of the hygro-thermal stress due to the pore pressure build-up and the thermo-mechanical stress resulting from the restrained thermal dilatation.
We employed a simplified mechanical approach based on the assumption that the wall is fully mechanically restrained and the effects of stresses resulting from the external mechanical load as well as the self-weight of the wall on the potential spalling are neglected. An evolution law for moving boundary due to spalling was proposed. It enables to simulate the instantaneous spalling of concrete as a continuous process, which contributes to the computational stability of the numerical solution.

For the model represented by a system of partial differential equations and appropriate boundary and initial conditions, the finite element discretization in space and the semi-implicit discretization in time were employed. The resulting numerical algorithm was incorporated into an in-house MATLAB code that was applied for numerical experiments in order to validate the present model.

The validity of the model was verified by comparing the results obtained by the numerical simulations with the data measured within real high-temperature experiments. Two types of tests were investigated -- the tests of hygro-thermal behaviour of high-strength concrete specimens $300 \times 300 \times 120~\rm{mm^3}$ heated by a radiant heater of temperature of $600~\rm{^{\circ}C}$ performed by \citet{Kalifa2000} and \citet{Mindeguia2009} and the spalling test of high-strength concrete specimens $700 \times 600 \times 150~\rm{mm^3}$  exposed to the ISO 834 fire performed by \citet{Mindeguia2009}.
All the material properties assumed for the simulations, except the reference intrinsic permeability of undamaged concrete at the room temperature, $K_{ref}$, were adopted from the data measured by \citet{Kalifa2000} and \citet{Mindeguia2009}, respectively. The values of $K_{ref}$ were not reported by the authors of the tests (\citet{Kalifa2000} and \citet{Mindeguia2009} reported only the values of the intrinsic permeability dried at $105~\rm{^{\circ}C}$ or $80~\rm{^{\circ}C}$, respectively, and then heated up to several temperature levels).

By comparing the calculated and measured data, the following findings can be drawn:
\begin{itemize}
\item{the model is able to simulate the hygro-thermal behaviour of heated concrete on an appropriate level of accuracy, the temperature and pore pressure distributions obtained by the present model are in close agreement with the data measured by \citet{Kalifa2000} and \citet{Mindeguia2009};}
\item{for the investigated examples, the model provided a correct qualitative prediction of concrete spalling (whether it occurred or not);}
\item{the quantitative prediction of spalling for the Mindeguia's spalling test \citep{Mindeguia2009} was overestimated (the values of spalling depths determined by the model at the end of the test period are about 2.5 to 4 times as high as measured by \citet{Mindeguia2009}). {This result arises from the overestimation of pore pressures (as discussed in detail in Section~\ref{Sec_Example3_Discussion}) and probably also from the assumption of the full mechanical restraint of the specimen. This assumption has been adopted in the model but it is probably not fulfilled in the investigated test setup by \citet{Mindeguia2009}. It can be supposed that in real structures (with some level of restraint), the model can provide more accurate prediction.}}
\end{itemize}

For future research, it could be recommended to focuss on:
\begin{itemize}
\item{determination of the reference intrinsic permeability of undamaged concrete at the room temperature};
\item{validation of the present model on a real structure exposed to fire or on the results of full-scale fire experiments;}
\item{extension of the present model for two-dimensional problems and improvement of the mechanical part of the model in order to describe various types of loading and restrain conditions.}
\end{itemize}

\appendix
\section{Physical mechanisms of transport processes in concrete at high temperatures}
\label{parameter_analysis}
In this Appendix we address, on the quantitative level, contributions of different transport mechanisms in concrete exposed to ambient high-temperatures. To this goal, we will first summarize modelling assumptions and state variables employed in the derivation of complete multi-phase formulations, together with the dependence of material constants on the state variables. Then, by expressing the thermodynamic fluxes in terms of the same state variables, we will obtain closed-form expressions for the corresponding transport coefficients. Their relative importance will be evaluated at the level of material point and these results will allow us to explore the domains of applicability and the connection of the presented simplified model against detailed multi-phase descriptions.

\subsection{Conservation of mass}
The local form of the macroscopic mass balance equation of the phase $\alpha$ is \citep{Gawin2003}
\begin{equation}\label{eq:mass_balance}
\frac{D^{\alpha}\rho^{\alpha}}{Dt}
+ \rho^{\alpha} \nabla \cdot \bfv_{\alpha}
= \mathcal{M}^{\alpha}.
\end{equation}
Here,
\begin{equation}\label{eq:density}
\rho^{\alpha} = \eta_{\alpha}\rho_{\alpha}
\end{equation}
represents the phase averaged density, $\eta_{\alpha}$ [-] is the volume fraction of the $\alpha$ phase and $\rho_{\alpha}$ stands for the intrinsic phase averaged density. Note that
\begin{equation}\label{porosity_01}
\sum_{\alpha} \eta_{\alpha}  = 1.
\end{equation}
Further, $\bfv_{\alpha}$ [m s$^{-1}$] is the velocity of $\alpha$ phase and $\mathcal{M}^{\alpha}$ is the volumetric mass source. Using \eqref{eq:density} the equation \eqref{eq:mass_balance} can be rewritten as
\begin{equation}\label{mass_balance_2}
\frac{\partial (\eta_{\alpha}\rho_{\alpha})}{\partial t}
+\nabla\cdot \left( \eta_{\alpha}\rho_{\alpha}
\bfv_{\alpha} \right)
=
\mathcal{M}^{\alpha}.
\end{equation}

Based on multi-phase modelling, the moist concrete is considered as a multi-phase porous material consisted of solid skeleton with pores filled by liquid water and gas, which is a mixture of dry air and water vapour.
Let us write the balance equation \eqref{mass_balance_2} for each phase of concrete mixture:

\emph{liquid water conservation equation:}
\begin{equation}\label{balance_liquid_water}
\frac{\partial (\eta_{w}\rho_{w})}{\partial t}
+
\nabla\cdot \left( \eta_w \rho_w\bfv_{w} \right)
=
-\frac{\partial m_{e}}{\partial t} + \frac{\partial m_{d}}{\partial t};
\end{equation}

\emph{water vapour conservation equation:}
\begin{equation}\label{balance_vapour}
\frac{\partial (\eta_v\rho_v)}{\partial t}
+
\nabla\cdot \left( \eta_v \rho_v\bfv_v \right)
=
\frac{\partial m_{e}}{\partial t};
\end{equation}

\emph{dry air conservation equation:}
\begin{equation}\label{balance_dry_air}
\frac{\partial (\eta_a\rho_a)}{\partial t}
+
\nabla\cdot \left( \eta_a \rho_a\bfv_a \right)
=
0;
\end{equation}

\emph{solid mass conservation equation:}
\begin{equation}\label{balance_solid}
\frac{\partial (\eta_s\rho_s)}{\partial t}
+
\nabla\cdot \left( \eta_s\rho_s \bfv_s \right)
=
-\frac{\partial m_{d}}{\partial t},
\end{equation}
where $m_d$ $[{\rm kg\, m^{-3}}]$ is the mass source term related to the dehydration process and  $m_e$ $[{\rm kg\, m^{-3}}]$ is the vapour mass source caused by the liquid water evaporation.

The mass balances of the liquid water and of the vapour, summed together to eliminate the source term related to phase changes (evaporation or condensation), form the mass balance equation of moisture (liquid water and vapour):
\begin{equation}\label{balance_moisture}
\frac{\partial }{\partial t}(\eta_{w}\rho_{w}+\eta_v\rho_v)
+
\nabla\cdot \left( \eta_w \rho_w\bfv_{w} + \eta_v \rho_v\bfv_v \right)
=
\frac{\partial m_{d}}{\partial t}.
\end{equation}

\subsection{Moisture flux}
The total moisture flux in moist concrete can be expressed as
\begin{equation}
\label{total_moisture_appendix}
\bfJ_M
=
\underbrace{\eta_w \rho_w\bfv_w}_{\bfJ_{w}}
+
\underbrace{\eta_g \rho_v\bfv_v}_{\bfJ_v} ,
\end{equation}
where $\bfJ_w$ represents the liquid (capillary and adsorbed) water mass flux and $\bfJ_v$ is the water vapour mass flux.

The chosen primary state variables are temperature $\theta$, water vapour pressure $P_v$ and dry air pressure $P_a$. We express the thermodynamic fluxes in terms of state variables $\theta$, $P_v$ and $P_a$ to obtain the corresponding transport coefficients related to different physical mechanisms of moisture transfer.

\paragraph{Liquid water mass flux}
Different mechanisms governing the liquid (free) water mass flux need to be distinguished. In particular, above the solid saturation point the transport of liquid water consists of capillary water flows driven by the capillary pressure gradient. Otherwise, the physically adsorbed water flows diffuses due to the saturation gradient. Consequently, mathematical expression of liquid water (adsorbed or capillary) mass flux reads as (cf. \citep{Davie2006})
\begin{equation}
\bfJ_w
=
- \left( 1-\frac{S_B}{S_w} \right) \rho_{w}\frac{K K_{rw}}{\mu_{w}}\nabla (P_g - P_c)
-
\frac{S_B}{S_w}
\rho_{w} \mathbb{D}_B \nabla S_B,
\end{equation}
where the degree of saturation with adsorbed water $S_B$ is defined as
\begin{equation}\label{}
S_B
=
\left\{
\begin{array}{lll}
S_w
&
\textmd{ for }
&
S_w \leq S_{ssp},
\\
\\
S_{ssp}
&
\textmd{ for }
&
S_w > S_{ssp},
\end{array}\right.
\end{equation}
$S_w$ represents the degree of saturation with liquid water and $S_{ssp}$ is the solid saturation point. $\mathbb{D}_b$ is the bound water diffusion tensor (see \citep[eq. (39)]{Gawin1999}).

Note that the saturation degree with liquid water $S_w$ is defined as
\begin{equation}
S_w = \frac{\eta_{w}}{\phi} \qquad [-].
\end{equation}
Here the volume fraction of liquid water $\eta_{w}$ in concrete can be calculated via the sorption isotherms (introduced by Ba\v{z}ant et al. \citep{BaCheTh1981,BaTh1978,BaTh1979}) as a function of temperature $\theta$ and vapour pressure $P_v$.

The equilibrium state of the capillary water with the water vapour is expressed in the form corresponding to the {Kelvin equation}
\begin{equation}\label{equilibrium meniscus}
P_c(P_v, \theta)
=
- \rho_w  \frac{R \theta}{M_w} \ln
\left(
\frac{P_v}{P_s}
\right),
\end{equation}
where $P_c$ denotes the capillary pressure and the water vapour saturation pressure $P_s$ can be calculated from the following formula as a function of temperature $\theta$ [K]
\begin{equation}\label{wvs}
P_s(\theta)
=
\exp\left( 23.5771 - \frac{4042.9}{\theta-37.58} \right).
\end{equation}

Finally, the liquid water mass fluxes can be expressed in the closed-forms with respect to gradients of primary state variables:

\medskip

{Adsorbed water diffusion ($ S_w \leq S_{ssp}$, $S_B = S_w$):}
\begin{eqnarray}\label{flow_adsorbed}
\eta_w \rho_w\bfv_w
&=& - \rho_{w} \mathbb{D}_B \nabla S_w
\nonumber
\\
&=&
- \rho_{w} \mathbb{D}_B \frac{\partial S_w}{\partial P_v}\nabla P_v
- \rho_{w} \mathbb{D}_B \frac{\partial S_w}{\partial \theta} \nabla \theta;
\end{eqnarray}

{Capillary water flow ($ S_w > S_{ssp}$, $S_B = S_{ssp}$):}
\begin{align}
&\eta_w \rho_w\bfv_w
=
-\left( 1-\frac{S_{B}}{S_w} \right)
\rho_{w}\frac{K K_{rw}}{\mu_{w}}\nabla P_g
+
\left( 1-\frac{S_{B}}{S_w} \right)
\rho_{w}\frac{K K_{rw}}{\mu_{w}}\nabla P_c
\nonumber
\\
&
\quad
=
-
\left( 1-\frac{S_{B}}{S_w} \right)
\rho_{w}\frac{K K_{rw}}{\mu_{w}}\nabla P_g
+
\left( 1-\frac{S_{B}}{S_w} \right)
\rho_{w}\frac{K K_{rw}}{\mu_{w}}
\left[
\frac{\partial P_c}{\partial P_v}  \nabla P_v
+
\frac{\partial P_c}{\partial \theta} \nabla \theta
\right]
\nonumber
\\
&
\quad
=
\left( 1-\frac{S_{B}}{S_w} \right)
\left(
-
\rho_{w}\frac{K K_{rw}}{\mu_{w}}
+
\rho_{w}\frac{K K_{rw}}{\mu_{w}} \frac{\partial P_c}{\partial P_v}
\right)\nabla P_v
\nonumber
\\
&
\quad\quad
-
\left( 1-\frac{S_{B}}{S_w} \right)
\rho_{w}\frac{K K_{rw}}{\mu_{w}}\nabla P_a
+
\left( 1-\frac{S_{B}}{S_w} \right)
\rho_{w}\frac{K K_{rw}}{\mu_{w}} \frac{\partial P_c}{\partial \theta} \nabla\theta.
\end{align}

\paragraph{Water vapour mass flux}
Decomposition of the water vapour velocity into the diffusional ($\bfv_v-\bfv_g$) and advectional ($\bfv_g$) components yields
\begin{equation}
\bfJ_v
=
\eta_g \rho_v \bfv_g
+  \eta_g \rho_v(\bfv_v-\bfv_g).
\end{equation}
The advective flux may be desribed by Darcy's law in the form (vapour flow)
\begin{eqnarray}
\eta_g \rho_v \bfv_g
&=&
- \rho_{v}
\frac{K K_{rg}}{\mu_g} \nabla P_g
\nonumber
\\
&=&
- \rho_{v}
\frac{K K_{rg}}{\mu_g} \nabla P_v
- \rho_{v}
\frac{K K_{rg}}{\mu_g} \nabla P_a.
\end{eqnarray}
For the diffusive mass flux of water vapour, Fick's law (vapour diffusion) is applied in the form
\begin{multline}
\bfJ_d^v = \eta_g \rho_v(\bfv_v-\bfv_g)
=
-\rho_g \frac{M_a M_w}{M^2_g}
D_{eff} \nabla \left( \frac{P_v}{P_ g}
\right)
\\
=
-
\frac{M_a M_w P_a}{\theta R (P_v M_w + P_a M_a)}D_{eff} \nabla P_v
+
\frac{M_a M_w P_v}{\theta R (P_v M_w + P_a M_a)}D_{eff} \nabla P_a.
\end{multline}

\paragraph{Total moisture flux}
The total moisture flux can be expressed in terms of gradients of state variables $P_v$, $P_a$ and $\theta$ to obtain
\begin{eqnarray}
\label{eq:total_flux}
\eta_w \rho_w\bfv_w + \eta_g \rho_v\bfv_v
&=&
\left(
\underbrace{
-\rho_{v}\frac{K K_{rg}}{\mu_g}}_{vapour\,flow}
\underbrace{
-
\left( 1-\frac{S_B}{S_w} \right)
\rho_{w}\frac{K K_{rw}}{\mu_{w}} \left(1-\frac{\partial P_c}{\partial P_v}\right)}_{liquid\, water\,flow}
\nonumber
\right.
\\
&& \left.
\underbrace{
- \frac{S_B}{S_w} \rho_{w} \mathbb{D}_B \frac{\partial S_B}{\partial P_v}}_{adsorbed \, water \, diffusion}
\underbrace{
-
\frac{M_a M_w P_a}{\theta R (P_v M_w + P_a M_a)}D_{eff}}_{vapour\,diffusion}
\right)\nabla P_v
\nonumber
\\
&&
\!\!\!\!\!\!\!\!\!\!\!\!\!\!\!\!\!\!\!
\!\!\!\!\!\!\!\!\!\!\!\!\!\!\!\!\!\!\!
+
\left(
\underbrace{
-\rho_{v}\frac{K K_{rg}}{\mu_g}}_{vapour\,flow}
\underbrace{
-\left( 1-\frac{S_B}{S_w} \right)\rho_{w}\frac{K K_{rw}}{\mu_{w}}}_{liquid\,water\,flow}
\underbrace{
+\frac{M_a M_w P_v}{\theta R (P_v M_w + P_a M_a)}D_{eff}}_{vapour\,diffusion}
\right)
\nabla P_a
\nonumber
\\
&&
+
\left(
\underbrace{
\left( 1-\frac{S_B}{S_w} \right)\rho_{w}\frac{K K_{rw}}{\mu_{w}}
\frac{\partial P_c}{\partial \theta} }_{liquid \, water \, flow}
\underbrace{
-
\frac{S_B}{S_w}\rho_{w} \mathbb{D}_B \frac{\partial S_B}{\partial \theta}}_{adsorbed \, water \, diffusion}
\right)
\nabla\theta    .
\nonumber
\\
\end{eqnarray}

\begin{figure}[h]
\centering
\begin{tabular}{cc}
  \\
  \includegraphics[angle=0,width=6.4cm]{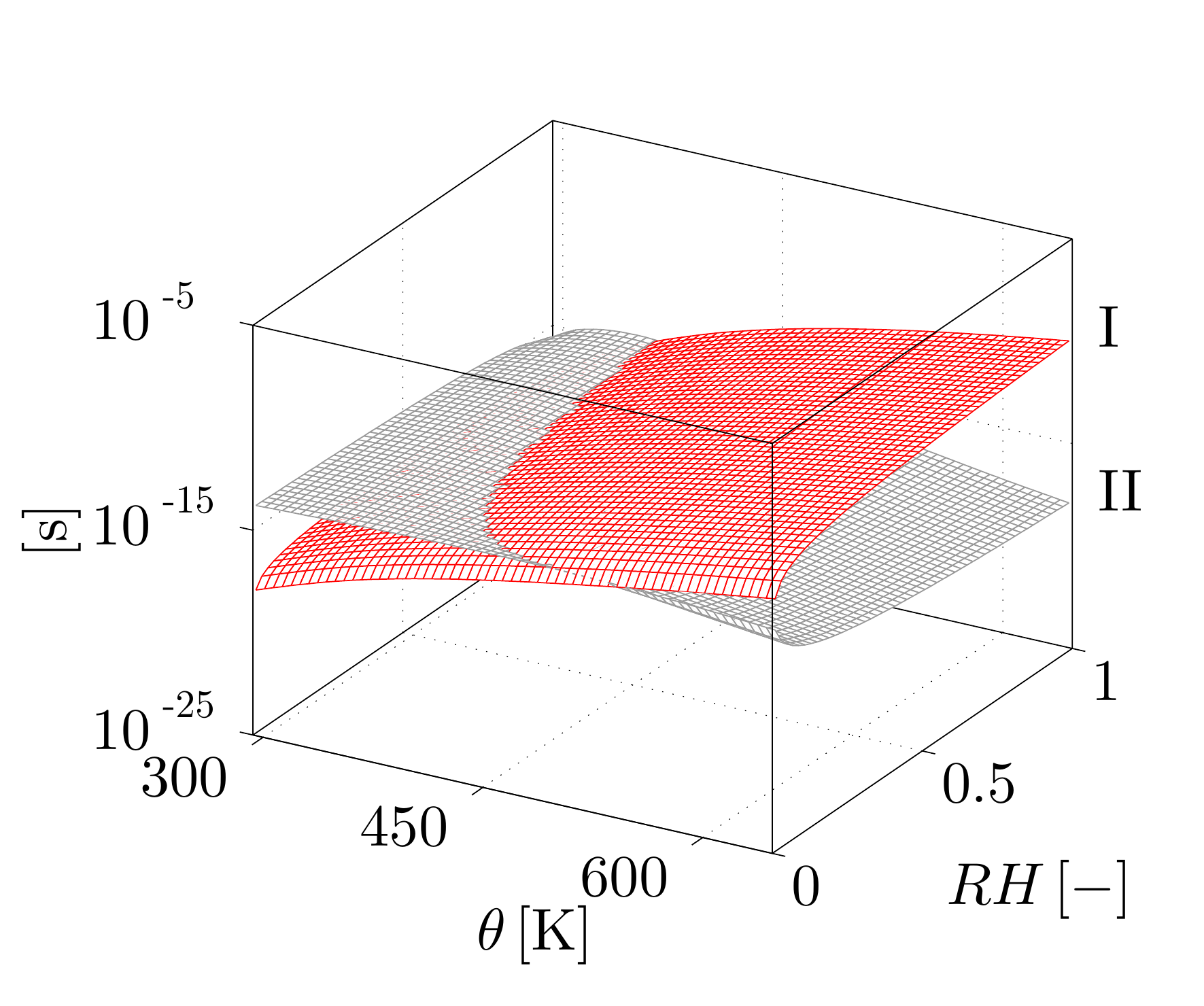}
  &
  \includegraphics[angle=0,width=6.4cm]{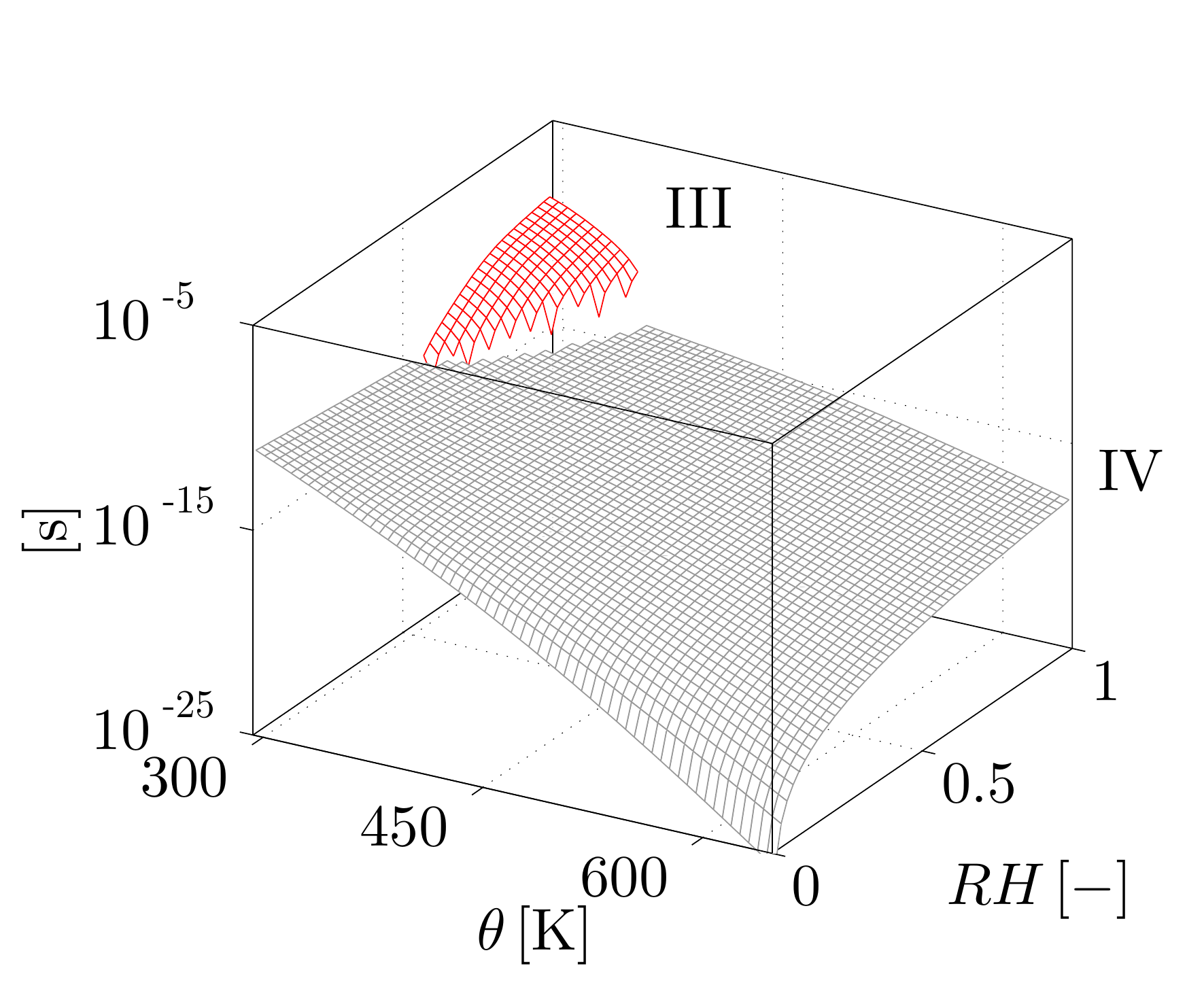}
\end{tabular}
\caption{Moisture fluxes related to $\nabla P_v$ representing the different moisture transfer mechanisms --
(I) vapour flow, (II) vapour diffusion, (III) liquid water flow, (IV) adsorbed water diffusion}
\label{fig:transp_coeff1}
\end{figure}

\begin{figure}[h]
\centering
\begin{tabular}{cc}
  \\
  \includegraphics[angle=0,width=6.4cm]{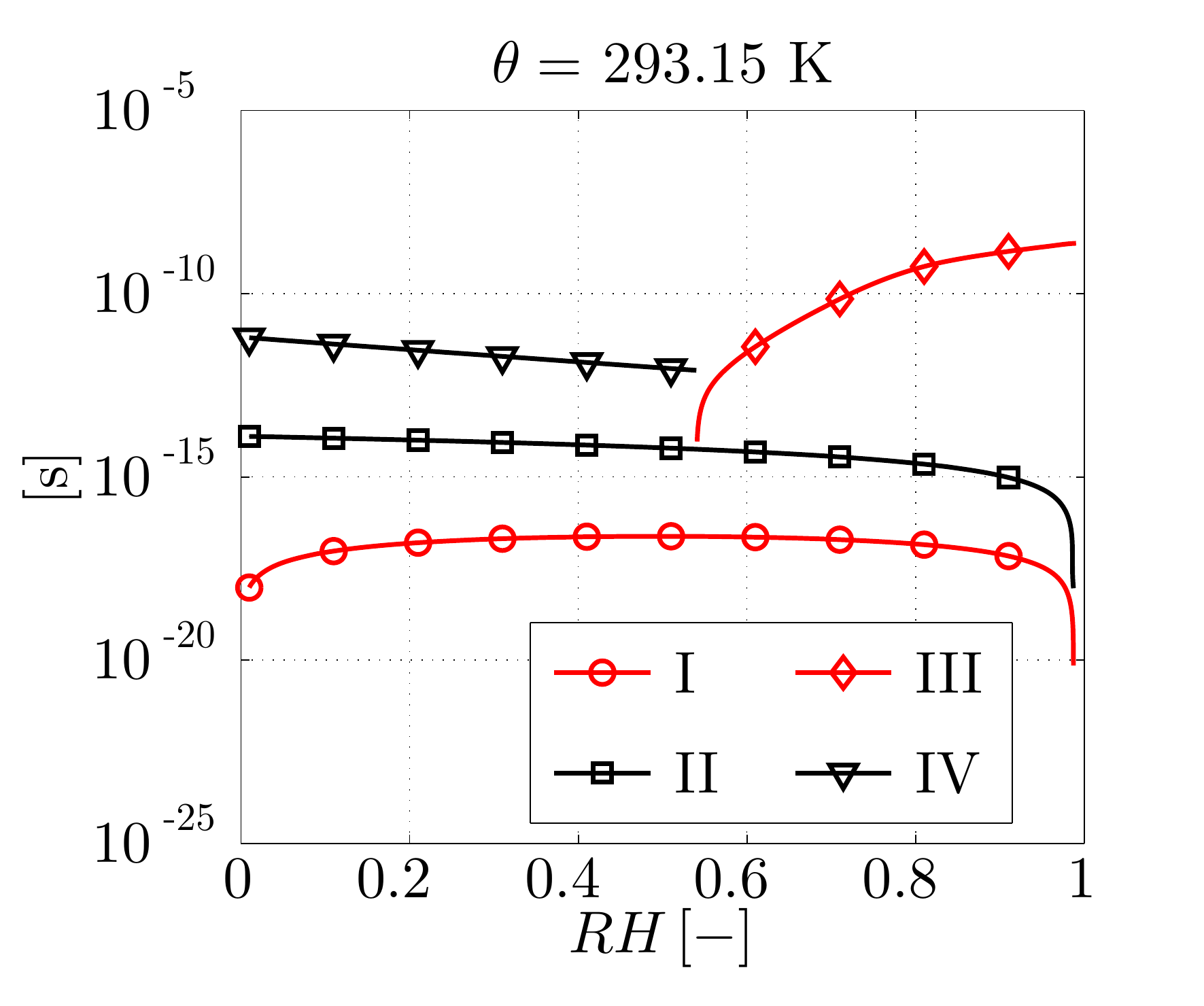}
  &
  \includegraphics[angle=0,width=6.4cm]{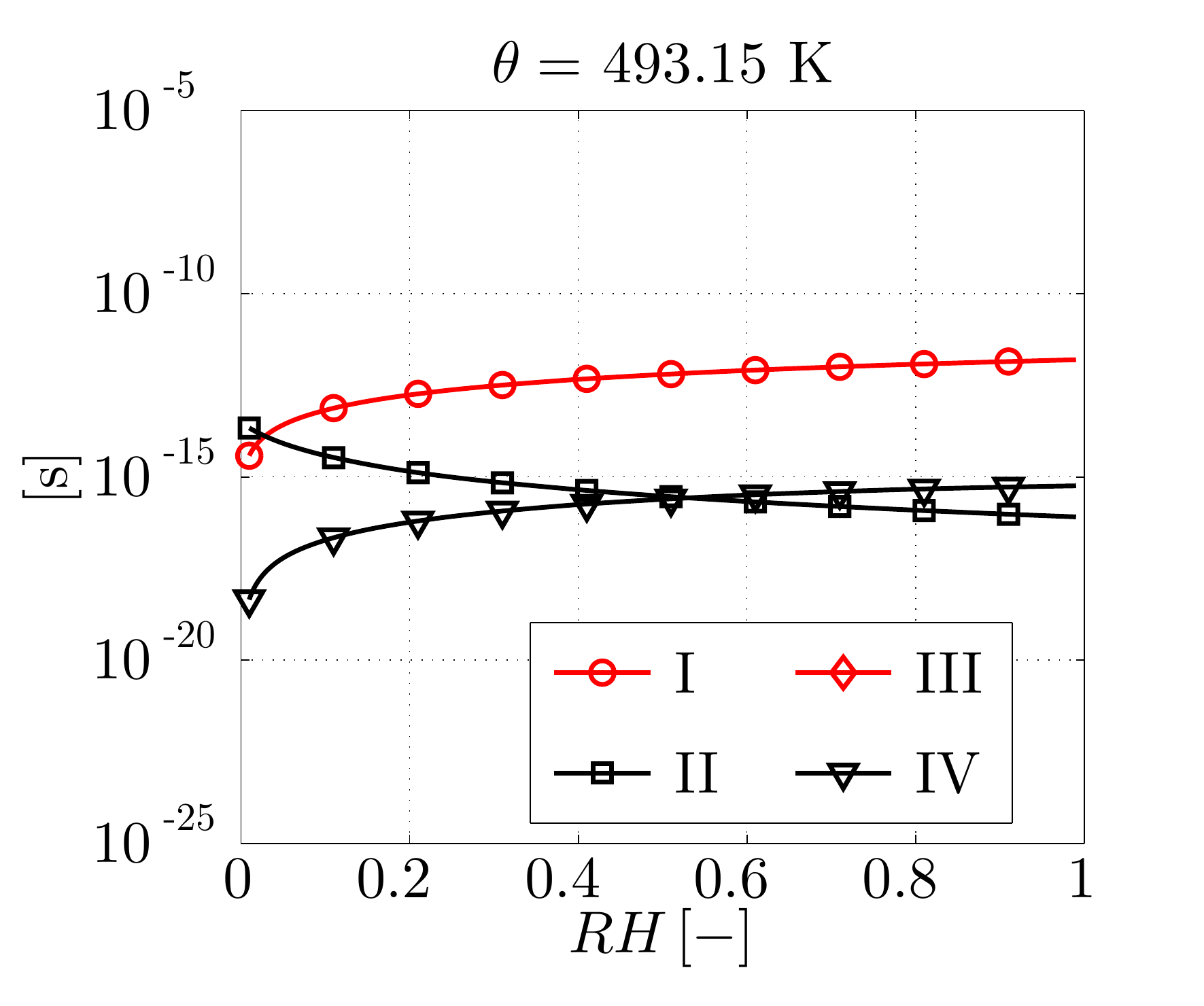}
  \\
  \includegraphics[angle=0,width=6.4cm]{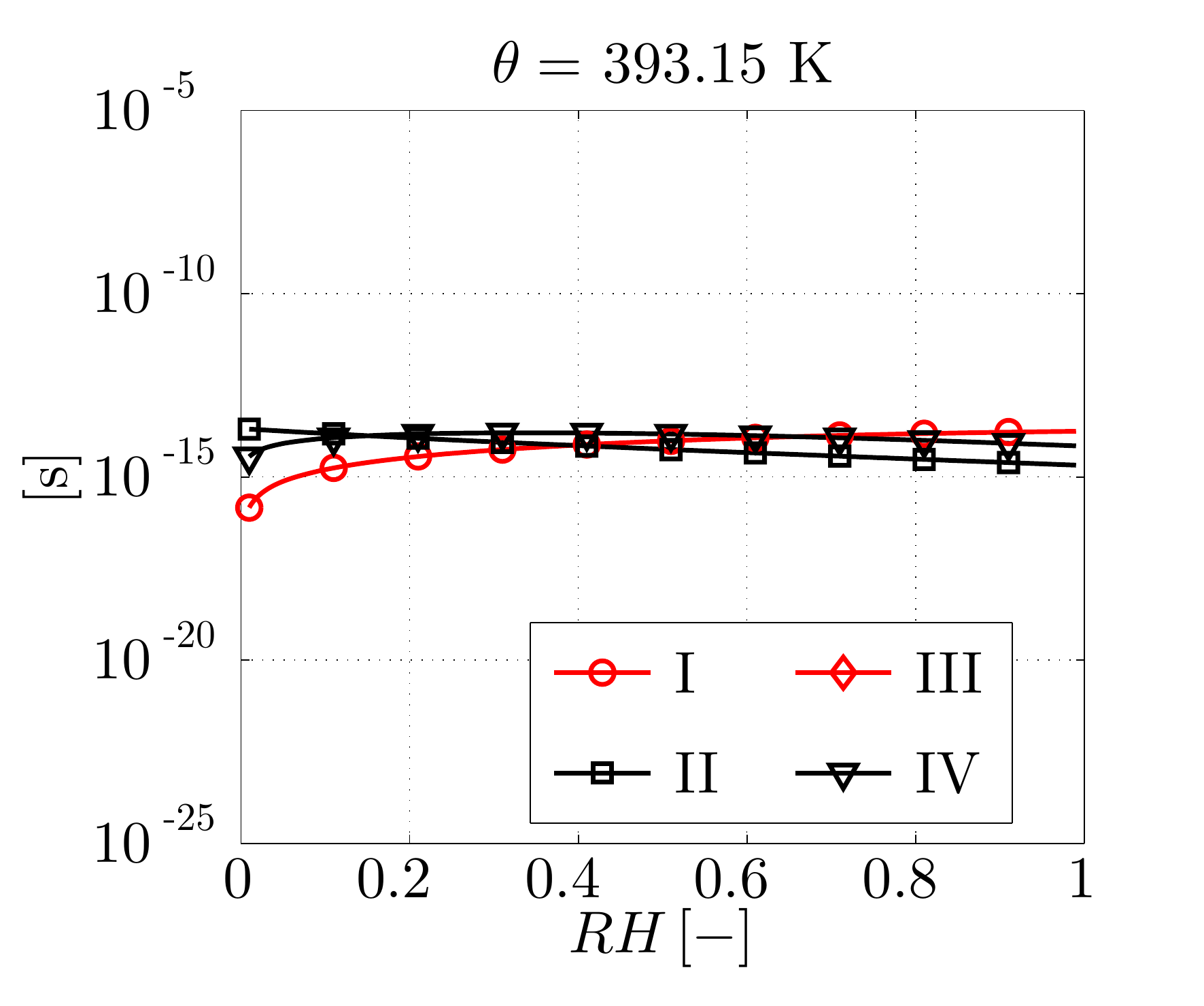}
  &
  \includegraphics[angle=0,width=6.4cm]{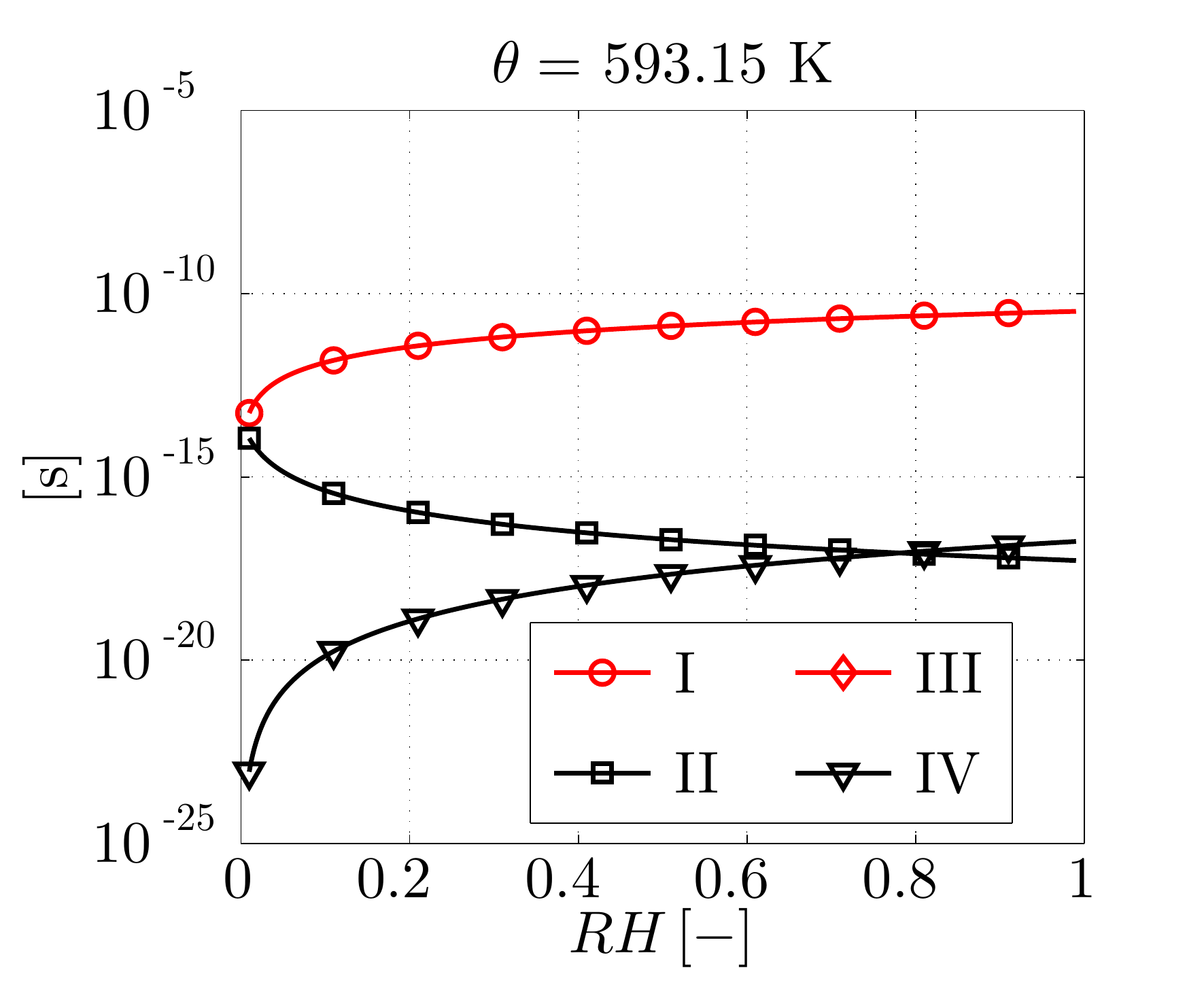}
\end{tabular}
\caption{Moisture fluxes related to $\nabla P_v$ representing the different moisture transfer mechanisms --
(I) vapour flow, (II) vapour diffusion, (III) liquid water flow, (IV) adsorbed water diffusion}
\label{fig:transp_coeff2}
\end{figure}

Figures \ref{fig:transp_coeff1} and \ref{fig:transp_coeff2} present the dependence of transport coefficients  upon the temperature and relative humidity at constant atmospheric pressure ($P_a = 101325$ Pa) related to the gradient of vapour pressure $\nabla P_v$. Material properties of concrete are taken from \citet{Gawin1999}. The moisture flux according to gradient of vapour pressure, as derived in \eqref{eq:total_flux}, consists of Darcian vapour flow, vapour diffusion and Darcian flow of capillary water (provided $S_w > S_{ssp}$) or diffusion of adsorbed water (provided $S_w \leq S_{ssp}$). Figures \ref{fig:transp_coeff1} and \ref{fig:transp_coeff2} show that with increasing temperature Darcian vapour flow plays a dominant role (approximately above $200 ^{\circ}$C). This observation is in accordance with \citet{DwaiKod2009} where it is assumed that only the vapour transport takes place during the temperature exposure of HSC. At lower temperatures, the adsorbed water diffusion or Darcian water flow (depending on degree of saturation) are of importance. Moreover, at the temperatures between $100 \sim 200~^{\circ}$C
in the relatively small $RH$-range, approximately up to $10~\%$, the vapour diffusion may be the dominant factor of moisture transport. However, following our observation based on numerical experiments, neglecting vapour diffusion has negligible effects on the spatial position of the amplitude of the peak and the peak value of the pore pressure.

In order to simplify the model as much as possible and minimize the number of input parameters, against complete multi-phase formulation, where the adsorbed and capillary water are assumed separately, we follow the approach introduced by \citet{ChungCon2005}, where the diffusion of adsorbed water is considered to be accounted for within the liquid water relative permeability $K_{rw}$ (see \citep{Davie2010}).
The main assumption in our model \eqref{eq:model_moisture}--\eqref{eq:ini_md} is that the moisture transport in HSC at high temperature related to the dry air pressure gradient is negligible when compared to the other causes, in particular, the vapour pressure and temperature gradients, respectively (see \citep{DwaiKod2009}).

\subsection{Heat energy conservation equation}
The local form of the macroscopic energy balance equation of the phase $\alpha$ is \citep{Gawin2003}
\begin{equation}\label{}
\rho^{\alpha}
c^{\alpha}_p
\frac{D^{\alpha}\theta^{\alpha}}{D t}
+
\nabla \cdot \bfq^{\alpha}_c
=
\mathcal{Q}^{\alpha}
+
\mathcal{E}^{\alpha}
-
H^{\alpha} \mathcal{M}^{\alpha}.
\end{equation}
Here, $c^{\alpha}_p$ is the specific isobaric heat, $\bfq^{\alpha}_c$ the heat flux due to conduction, $\mathcal{Q}^{\alpha}$ the volumetric heat source, $\mathcal{E}^{\alpha}$ represents energy exchange with other phases and $H^{\alpha}$ is the specific enthalpy of the phase $\alpha$.

Summing up the energy balances for all phases ($\alpha \equiv a,w,v,s$) of the multi-phase system, neglecting internal sources $\mathcal{Q}^{\alpha}$ and taking into consideration $\sum_{\alpha}\mathcal{E}^{\alpha} =0$ one arrives at the final form of~the

\emph{energy conservation equation for moist concrete as the multi-phase system:}
\begin{equation}\label{app:balance_energy}
(\rho c_p)
\frac{\partial\theta}{\partial t}
=
-
\nabla \cdot\bfq_c
-
(\rho c_p \bfv)
\cdot\nabla\theta
-
\frac{\partial m_{e}}{\partial t} h_{e}
-
\frac{\partial m_{d}}{\partial t} h_{d},
\end{equation}
where
\begin{eqnarray}
(\rho c_p)
&=&
c_p^{w}\, \rho_{w}\,\eta_{w}
+
c_p^{v}\, \rho_{v}\,\eta_{v}
+
c_p^{a}\, \rho_{a}\,\eta_{a}
+
c_p^{s}\, \rho_{s}\,\eta_{s},
\\
(\rho c_p \bfv)
&=&
c^w_p\rho_w\eta_w \bfv_{w}
+
c^v_p\rho_v\eta_v \bfv_v
+
c^a_p\rho_a\eta_a \bfv_a
+
c^s_p\rho_s\eta_s \bfv_s,
\\
\bfq_c
&=&
\bfq_c^{s} + \bfq_c^{w} + \bfq_c^{v} + \bfq_c^{a},
\\
h_{e}
&=&
H^v - H^{w},
\\
h_{d}
&=&
H^w - H^{s}.
\end{eqnarray}
Here, $c^w_p$, $c_p^{v}$, $c_p^{a}$ and $c_p^{s}$ $[{\rm J\, kg^{-1}\, K^{-1}}]$ are the specific heats at constant pressure of liquid water, water vapour, dry air and solid matrix, $\bfq_c^{w}$, $\bfq_c^{v}$, $\bfq_c^{a}$ and $\bfq_c^{s}$ represent heat fluxes due to conduction corresponding to liquid water, water vapour, dry air and solid matrix. Further, $h_e$ $[{\rm J \, kg^{-1}}]$ represents the enthalpy of evaporation per unit mass, while $h_d$ $[{\rm J \, kg^{-1}}]$  is the enthalpy of dehydration per unit mass. Finally, $H^{w}$, $H^v$  and $H^{s}$  $[{\rm J \, kg^{-1}}]$ are specific enthalpies of liquid water, water vapour and chemically bound water.

Neglecting the impact of diffusional flows of dry air and vapour in the gas mixture on heat transfer by convection yields
\begin{equation}
c^v_p\rho_v\eta_v \bfv_v
+
c^a_p\rho_a\eta_a \bfv_a
=
(c^v_p\rho_v\eta_v+c^a_p\rho_a\eta_a)\bfv_g
=
c^g_p\rho_g\eta_g\bfv_g,
\end{equation}
where $c^g_p$ is the specific heat at constant pressure of gas mixture.

\section{Nomenclature}
\label{nomenclature}
\begin{longtable}[l]{p{.13\textwidth} p{.18\textwidth} p{.62\textwidth}}
\textbf{Symbol}	        & \textbf{Unit} & \textbf{Description}
\\ $\alpha_c$	        & $[{\rm W\, m^{-2}\, K^{-1}}]$	& convective heat transfer coefficient
\\ $\beta_c$	        & $[{\rm m\, s^{-1}}]$	  & convective mass transfer coefficient
\\ $\epsilon_{tot}$     & $[-]$		              & total strain
\\ $\epsilon_{\theta}$	& $[-]$		              & free thermal strain
\\ $\epsilon_{\sigma}$	& $[-]$		              & instantaneous stress-related strain
\\ $\epsilon_{cr}$	    & $[-]$		              & creep strain
\\ $\epsilon_{tr}$ 	    & $[-]$		              & transient strain
\\ $\eta_s$	            & $[-]$	                  & volume fraction of the solid microstructure	
\\ $\eta_w$	            & $[-]$	                  & volume fraction of liquid phase
\\ $\eta_g$	            & $[-]$	                  & volume fraction of gas phase
\\ $\phi$	            & $[-]$		              & porosity
\\ $\rho_s$             & $[{\rm kg\, m^{-3}}]$   & density of the solid microstructure	
\\ $\rho_g$	            & $[{\rm kg\, m^{-3}}]$	  & gas phase density
\\ $\rho_a$	            & $[{\rm kg\, m^{-3}}]$	  & dry air phase density
\\ $\rho_v$	            & $[{\rm kg\, m^{-3}}]$	  & vapour phase density
\\ $\rho_{v\infty}$	    & $[{\rm kg\, m^{-3}}]$	  & ambient vapour phase density
\\ $\rho_w$	            & $[{\rm kg\, m^{-3}}]$	  & liquid phase density
\\ $\mu_{w}$	        & $[{\rm Pa \, s}]$		  & liquid water dynamic viscosity
\\ $\mu_g$  	        & $[{\rm Pa \, s]    }$	  & gas dynamic viscosity
\\ $\sigma$	            & $[{\rm Pa}]$		      & stress
\\ $\sigma_{ht}$	    & $[{\rm Pa}]$		      & stress caused by hygro-thermal processes
\\ $\sigma_{tm}$	    & $[{\rm Pa}]$		      & stress caused by thermo-mechanical processes
\\ $\sigma_{SB}$	    & $[{\rm W\, m^{-2} \, K^{-1}}]$	& Stefan-Boltzmann constant
\\ $\theta$	            & $[{\rm K}]$	          & absolute temperature
\\ $\theta_{\infty}$	& $[{\rm K}]$ 	          & ambient absolute temperature
\\ $\lambda_c$	        & $[{\rm W\, m^{-1}\, K^{-1}}]$  & effective thermal conductivity of moist concrete
\\ $\tau$	            & $[{\rm s}]$		      & characteristic time of mass loss
                                                    governing the asymptotic evolution of the dehydration process
\\ $c$                  & ${\rm kg\, m^{-3}}$     & mass of cement per m$^3$ of concrete
\\ $c_p^{w}$	        & $[{\rm J\, kg^{-1}\, K^{-1}}]$	& specific heat at constant pressure of liquid water
\\ $c_p^{v}$	        & $[{\rm J\, kg^{-1}\, K^{-1}}]$	& specific heat at constant pressure of water vapour
\\ $c_p^{a}$	        & $[{\rm J\, kg^{-1}\, K^{-1}}]$	& specific heat at constant pressure of dry air
\\ $c_p^{s}$	        & $[{\rm J\, kg^{-1}\, K^{-1}}]$	& specific heat at constant pressure of solid phase
\\ $c_p^{g}$	        & $[{\rm J\, kg^{-1}\, K^{-1}}]$	& specific heat at constant pressure of gas mixture
\\ $e$	                & $[-]$	                  & emissivity of the interface
\\ $f_c$                & $[{\rm Pa}]$	          & uniaxial compressive strength of concrete
\\ $f_t$                & $[{\rm Pa}]$	          & uniaxial tensile strength of concrete
\\ $F$	                & $[-]$		              & failure function
\\ $h_e$	            & $[{\rm J \, kg^{-1}}]$  & enthalpy of evaporation per unit mass
\\ $h_d$	            & $[{\rm J \, kg^{-1}}]$  & enthalpy of dehydration per unit mass
\\ $H^{w}$	            & $[{\rm J \, kg^{-1}}]$  & specific enthalpy of liquid water
\\ $H^v$                & $[{\rm J \, kg^{-1}}]$  & specific enthalpy of water vapour
\\ $H^{s}$	            & $[{\rm J \, kg^{-1}}]$  & specific enthalpy of chemically bound water
\\ $\bfJ_w$	            & $[{\rm kg \, m^{-2} \, s^{-1}}]$ & liquid water mass flux
\\ $\bfJ_v$	            & $[{\rm kg \, m^{-2} \, s^{-1}}]$ & water vapour mass flux
\\ $\bfJ_M$	            & $[{\rm kg \, m^{-2} \, s^{-1}}]$ & moisture flux
\\ $K$	                & $[{\rm m^2}]$		      & intrinsic permeability
\\ $K_{rw}$	            & $[-]$			          & relative permeability of liquid water
\\ $K_{rg}$	            & $[-]$			          & relative permeability of gas
\\ $\ell$               & $[{\rm m}]$			  & thickness of a concrete wall
\\ $m_d$	            & $[{\rm kg\, m^{-3}}]$  & mass source term related to the dehydration process
\\ $m_{d,eq}$	        & $[{\rm kg\, m^{-3}}]$  & mass of water released at the equilibrium
\\ $m_e$	            & $[{\rm kg\, m^{-3}}]$  & vapour mass source caused by the liquid water evaporation
\\ $M_w$	            & $[{\rm kg\,kmol^{-1}}]$         & molar mass of liquid water
\\ $M_a$	            & $[{\rm kg\,kmol^{-1}}]$		  & molar mass of dry air
\\ $P$                  & $[{\rm Pa}]$            & pore pressure due to water vapour ($P = P_v$)
\\ $P_c$                & $[{\rm Pa}]$	          & capillary pressure
\\ $P_v$                & $[{\rm Pa}]$	          & water vapour pressure
\\ $P_a$                & $[{\rm Pa}]$	          & dry air pressure
\\ $P_g$                & $[{\rm Pa}]$            & gas pressure
\\ $P_w$                & $[{\rm Pa}]$	          & pressure of liquid water
\\ $P_s$                & $[{\rm Pa}]$	          & water vapour saturation pressure
\\ $q_c$	            & $[{\rm W \, m^{-2}}]$	  &     heat flux vector
\\ $R$                  & $[{\rm J \, kmol^{-1} K^{-1}}]$	& gas constant
\\ $RH$                 & $[-]$                   & relative humidity ($RH = {P_v}/{P_s}$)
\\ $S_w$                & $[-]$	                  & degree of saturation with liquid water
\\ $S_B$                & $[-]$                   & degree of saturation with adsorbed water
\\ $S_{ssp}$            & $[-]$ 	              & solid saturation point
\\ $v_g$	            & $[{\rm m \, s^{-1}}]$	  & velocity of gaseous phase
\\ $v_w$	            & $[{\rm m \, s^{-1}}]$	  & velocity of liquid phase
\\ $v_a$	            & $[{\rm m \, s^{-1}}]$	  & velocity of dry air
\\ $v_v$	            & $[{\rm m \, s^{-1}}]$	  & velocity of vapour
\end{longtable}


\bigskip
\paragraph{Acknowledgement}
This research has been supported by the Czech Science Foundation, project GA\v{C}R 13-18652S (the first and the second author), and by the Grant Agency of the Czech Technical University in Prague, project SGS13/038/ OHK1/1T/11 (the second author). The support is gratefully acknowledged.

\section*{References}
\addcontentsline{toc}{section}{References} 
\phantomsection


\end{document}